\documentclass[aps,pra,twocolumn,superscriptaddress]{revtex4-2}
\usepackage[colorlinks,linkcolor=blue,urlcolor=blue,anchorcolor=blue,citecolor=blue]{hyperref}
\usepackage{amsmath}
\usepackage{graphicx}
\usepackage{dcolumn}
\usepackage{mathrsfs}
\usepackage{amssymb}
\usepackage{amsfonts,multirow}
\usepackage{bm,lineno}
\usepackage{color}
\usepackage[markup=underlined, defaultcolor=red]{changes}
\begin{document}
\title{Critical quantum metrology robust against dissipation and non-adiabaticity}
\author{Jia-Hao L\"{u}}
\thanks{These authors equally contributed to the work.}
\author{Wen Ning}\thanks{These authors equally contributed to the work.}
\author{Fan Wu}\thanks{These authors equally contributed to the work.}
\author{Ri-Hua Zheng}
\author{Ken Chen}
\author{Xin Zhu}
\affiliation{Fujian Key Laboratory of Quantum Information and Quantum Optics, College of Physics and Information Engineering, Fuzhou University, Fuzhou, Fujian 350108, China}
\author{Zhen-Biao Yang}\email{zbyang@fzu.edu.cn}
\affiliation{Fujian Key Laboratory of Quantum Information and Quantum Optics, College of Physics and Information Engineering, Fuzhou University, Fuzhou, Fujian 350108, China}
\affiliation{Hefei National Laboratory, Hefei 230088, China}
\author{Huai-Zhi Wu}\email{huaizhi.wu@fzu.edu.cn}
\affiliation{Fujian Key Laboratory of Quantum Information and Quantum Optics, College of Physics and Information Engineering, Fuzhou University, Fuzhou, Fujian 350108, China}
\author{Shi-Biao Zheng}\email{t96034@fzu.edu.cn}
\affiliation{Fujian Key Laboratory of Quantum Information and Quantum Optics, College of Physics and Information Engineering, Fuzhou University, Fuzhou, Fujian 350108, China}
\affiliation{Hefei National Laboratory, Hefei 230088, China}
\date{\today }

\begin{abstract}
Critical systems near quantum phase transitions were predicted to be useful for improvement of metrological precision, thanks to their ultra-sensitive response to tiny variations of the control Hamiltonian. Despite the promising perspective, realization of criticality-enhanced quantum metrology is an experimentally challenging task, mainly owing to decoherence and critical slowing down associated with the corresponding quantum state preparation. We here circumvent these problems by making use of the critical behaviors in the Jaynes-Cummings model, to which the signal field is coupled. The information is encoded in the qubit's excitation number, which displays a divergent changing rate at the critical point, and is extremely robust against decoherence and non-adiabatic effects. We demonstrate such a metrological protocol in a superconducting circuit, where an Xmon qubit, interacting with a resonator, is used as a probe for estimating the amplitude of a microwave field. The measured quantum Fisher information exhibits a critical quantum enhancement, confirming the potential of this system for quantum metrology.
\end{abstract}
\pacs{PACS number: }

\vskip0.5cm

\maketitle

\section{Teaser}

Ultrahigh precision measurement of a signal field is demonstrated by coupling it to an on-chip critical spin-boson system.

\section{Introduction}
 The capability of accurately measuring a weak signal is crucial for
advancing modern science and technology, exemplified by the detection of
gravitational waves \cite{PRL061102}, which not only provides a direct evidence for the
validity of general relativity, but also lays the experimental
foundation for gravitational-wave astronomy. Quantum metrology aims to
exploit quantum sources to improve the precision of measurement of special
physical quantities, including the amplitude of a magnetic or electric
field, the frequency of an oscillator, and the magnitude of a force \cite{RMP035002,RMP035006}.
In conventional quantum metrological protocols, the signal is encoded in a
quantum superposition of two components with highly distinct quantum numbers
for a single element \cite{Nature262,Nature86,NC4382}, e.g., Fock states $\left\vert 0\right\rangle $
and $\left\vert N\right\rangle $ of a bosonic mode, or in a highly entangled
state for many qubits \cite{Science726,Science1476,PRL130506,Nature472}. With such encoding protocols, the achievable
sensitivity can exhibit a scaling surpassing the standard quantum limit in
principle. However, the vulnerability of these nonclassical states restricts
their metrological practicalities, as the gain from the increase of the
system size is quickly cancelled out by the environment-induced decoherence
effects \cite{PNAS11459}.

Quantum critical phenomena represent an alternative quantum resource for
realizing quantum sensing. When a quantum system evolves near the critical
point of a phase transition \cite{BOOK1}, its properties exhibit dramatic changes in response to a slight variation of the governing Hamiltonian. These phenomena
are useful for realizing criticality-enhanced quantum sensors \cite{PRE031123,PRB064418,NJP063039,PRE022101,PRA042105,PRA021801,PRA042106,NJP053035,PRA023803,PRA022111,PRE052118,PRX021022,PRE052107,PRL020402,PRL120504,PRL210506,QST035010,chenken,NPJ170,NP1447,PRL150501,PRL173601,PRA013817,PRXQ010354,NPJ23,PRL010502,PRA062616,SCPMA250313}, which aim to exploit critical systems to amplify the effects of the signal fields. In conventional critical metrological protocols \cite{PRE031123,PRB064418,NJP063039,PRE022101,PRA042105,PRA021801,PRA042106,NJP053035,PRA023803,PRA022111,PRE052118,PRX021022,PRE052107,PRL020402,PRL120504,PRL210506,QST035010,chenken,NPJ170,NP1447,arxiv16931}, the physical quantity
is encoded in one specific eigenstate of the control Hamiltonian that is
adiabatically steered towards the critical point. Benefitting from the
adiabatic nature, this approach bears an intrinsic robustness against
noises. However, it is an experimental challenge to meet the adiabatic condition close to a quantum phase transition, owing to the critical
slowing down. To make a quantum system adiabatically follow one specific
eigenstate, the changing rate of the control parameter usually needs to be
much lower than the corresponding energy gaps. At the
phase transition, these gaps vanish, which indicate the probability
of leaking to other eigenstates is non-negligible when the system is
approaching the critical point within a limited time. This inherent state
leakage challenges the practical usefulness of such protocols. Recent
theoretical investigations show that the adiabatic condition can be removed
with dynamical approaches \cite{chenken,PRL010502,PRA062616,SCPMA250313}, where the parameter to be estimated
is encoded in the time-evolving state under a time-independent Hamiltonian, to which the
physical quantity of interest is coupled. However, such dynamical protocols
also require extremely long evolution time \cite{Quantum700}, furthermore, their
performances depend on precise timing.

We here propose and demonstrate a critical quantum metrological protocol,
where the process to bring the system towards the critical point is
significantly speeded up, but without loss of the robustness associated with
adiabatic evolution. The sensor is composed of a single qubit and a photonic
mode interacting with each other, referred to as the Jaynes-Cummings model
(JCM) \cite{IEEE89}. The signal field, whose amplitude is to be probed, couples to
the bosonic mode, producing a continuous drive. Below the critical point,
the system has a unique dark state, in which the signal is encoded. The
energy gaps between this dark state and the nearest bright states are
continually narrowed when approaching the critical point, where the average
excitation numbers for both the photonic mode and the qubit exhibit
diverging changing rates. The qubit excitation number, which serves as a robust indicator for estimating the amplitude of the signal field, is insensitive to state leakage. We demonstrate this criticality-enhanced metrology with a circuit quantum
electrodynamics (QED) system, where a superconducting qubit and a microwave resonator form the JCM. The
experimental results unambiguously demonstrate the robustness of the critical
quantum sensing protocol. 

\section{Results}
\subsection{Model}

\begin{figure}
	\includegraphics[width=0.5\textwidth]{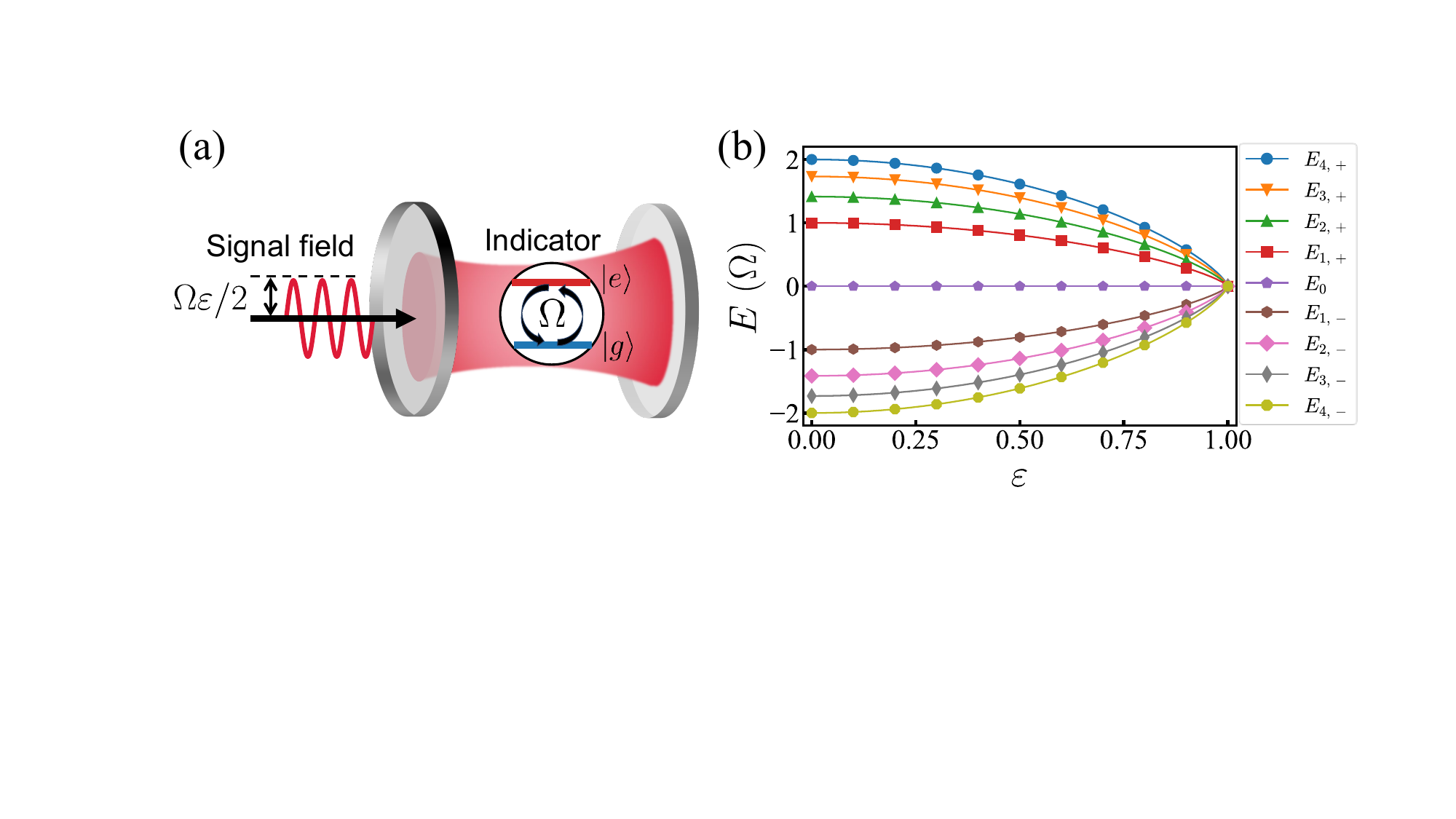}
     \caption{(color online). Critical quantum sensing protocol. (a) Theoretical model. The model comprises a cavity mode interacting with the transition between the lower ($\left\vert g\right\rangle $) and upper state ($\left\vert
e\right\rangle $) of a qubit, with the photonic swapping strength $\Omega $.
The signal field with an unknown rescaled amplitude $\varepsilon $ is
coupled to the resonator. (b) Quasi-energy spectrum. Such a model exhibits a highly nonlinear
quasi-energy spectrum, with $E_{0}=0$ and $E_{n,\pm }=\pm\sqrt{n}\Omega
(1-\varepsilon ^{2})^{3/4}$. When $\varepsilon \ll 1$, the excitation number
in the adiabatically evolved dark state is negligible, as a consequence of
the photon blockade. At $\varepsilon =1$ quasi-energy gaps vanish,
leading to the breakdown of the photon blockade. Near this critical point,
both the photon number and qubit excitation number exhibit a diverging
changing rate in response to the variation of $\varepsilon $.}\label{fig1}
\end{figure}

The theoretical model under consideration involves a photonic mode resonantly
interacting with a qubit and driven by a signal field, whose amplitude is
to be estimated, as schematically  shown in Fig. \ref{fig1}(a). In the interaction picture, the system dynamics is
described by the driven JCM (setting $\hbar=1$) \cite{PRA5135}%
\begin{equation}
H=\Omega \lbrack( a^{\dagger }\left\vert g\right\rangle \left\langle
e\right\vert +a\left\vert e\right\rangle \left\langle g\right\vert
)+\varepsilon (a^{\dagger }+a)/2], \label{eq1}
\end{equation}
where $a^{\dagger }$ and $a$ denote the photonic creation and annihilation
operators, $\left\vert g\right\rangle $ and $\left\vert e\right\rangle $ are
the ground and excited state of the qubit, $\Omega $ represents the
qubit-boson interaction strength, and $\varepsilon $ characterizes the
rescaled amplitude of the signal field coupled to the photonic mode. When $%
\varepsilon <1$, the system possesses discrete eigenenergies, given by%
\begin{eqnarray}
E_{0} &=&0,  \nonumber\\
E_{n,\pm } &=&\pm \sqrt{n}\Omega A^{3/4},
\end{eqnarray}%
where $A=1-\varepsilon ^{2}$, as depicted in Fig. \ref{fig1}(b). We note that the eigenenergies are derived in the framework rotating at the frequency of the signal field, and will refer to them as quasi-energies \cite{PRA5135}. The  corresponding eigenstates are%
\begin{eqnarray}
\left\vert \psi _{0}\right\rangle &=&S(r)\left\vert 0\right\rangle
\left\vert \phi _{0}\right\rangle , \nonumber\\
\left\vert \psi _{n,\pm }\right\rangle &=&S(r)D(\alpha _{n,\pm })(\left\vert
n-1\right\rangle \left\vert \phi _{1}\right\rangle \pm \left\vert
n\right\rangle \left\vert \phi _{0}\right\rangle )/\sqrt{2},
\end{eqnarray}
where $\left\vert n\right\rangle $ denotes the $n$-photon state, and $S(r)=\exp{[r(a^2-a^{\dagger2})/2]}$
and $D(\alpha _{n,\pm })=\exp{[\alpha _{n,\pm }(a^\dagger-a)]}$ are respectively the squeezing and displacement
operators for the photonic field, with $r=\frac{1}{4}\ln A$ and $\alpha
_{n,\pm }=\mp \sqrt{n}\varepsilon $.  $\left\vert \phi _{0,1 }\right\rangle $
denote the qubit parts, given by
\begin{eqnarray}
\left\vert \phi _{0}\right\rangle &=&c_{+}\left\vert g\right\rangle
-c_{-}\left\vert e\right\rangle , \nonumber\\
\left\vert \phi _{1}\right\rangle &=&c_{+}\left\vert e\right\rangle
-c_{-}\left\vert g\right\rangle ,
\end{eqnarray}
with $c_{\pm }=(1\pm \sqrt{A})^{1/2}/\sqrt{2}$. Hereafter, we will call $\left\vert \psi _{0}\right\rangle$ and $\left\vert \psi _{n,\pm }\right\rangle$ dark and bright eigenstates, respectively. The results show that when
the driving strength is increased, the quasi-energy splittings are continually
narrowed, all vanishing at the critical point $\varepsilon =1$, beyond which the Hamiltonian $H$ cannot be diagonalized so that the system does not possess a well-defined quasi-energy spectrum \cite{PRR023062}. The quasi-energy level
configuration can also be interpreted in terms of the competition between
the JC coupling and the external driving. When the JC coupling dominates
over the external driving, the system features the JC-like ladder,
manifested by the $\sqrt{n}$-dependent splitting. If the external driving is
dominant, the photonic mode is disentangled with the anharmonic qubit and
the spectrum becomes a continuum. This spectral feature is associated with
the dissipative-driven photon-blockade breakdown phase transition \cite{PRX031028,PRX011012}. We note these phase transitions cannot be described by the Ginzburg-Landau-Wilson symmetry-breaking paradigm, as the Hamiltonian itself does not possess the U(1) symmetry of the JCM, which is explicitly broken by the drive.

We consider the system behaviors below but close to the critical point $%
\varepsilon =1$. The associated average photon number, $\left\langle
N\right\rangle _{0}=\sinh ^{2}r$, displays a divergent behavior and thus can
be used as a critical sensing indicator in principle. However, it is
sensitive to leakage to the bright states $\left\vert \psi _{n,\pm
}\right\rangle $, whose average photon number, $\left\langle N\right\rangle
_{n}=2n\varepsilon ^{2}e^{-2r}+n\cosh{2r}-\frac{1}{2}$, is significantly
distinct from that of the dark state. This sensitivity is unfavorable to
improvement of the signal-to-noise ratio. In distinct contrast, the qubit's excitation number is insensitive to leakage to the bright states, which can
be caused either by decoherence or by the non-adiabatic effects. Due to this insensitivity, the qubit's excitation number can be used as a robust indicator for estimating
the amplitude of the external drive. We note that the qubit responds to the change of the control field by interacting with the photonic mode, which is in a nonclassical state.

\subsection{Numerical simulations}
\begin{figure}
	\includegraphics[width=0.5\textwidth]{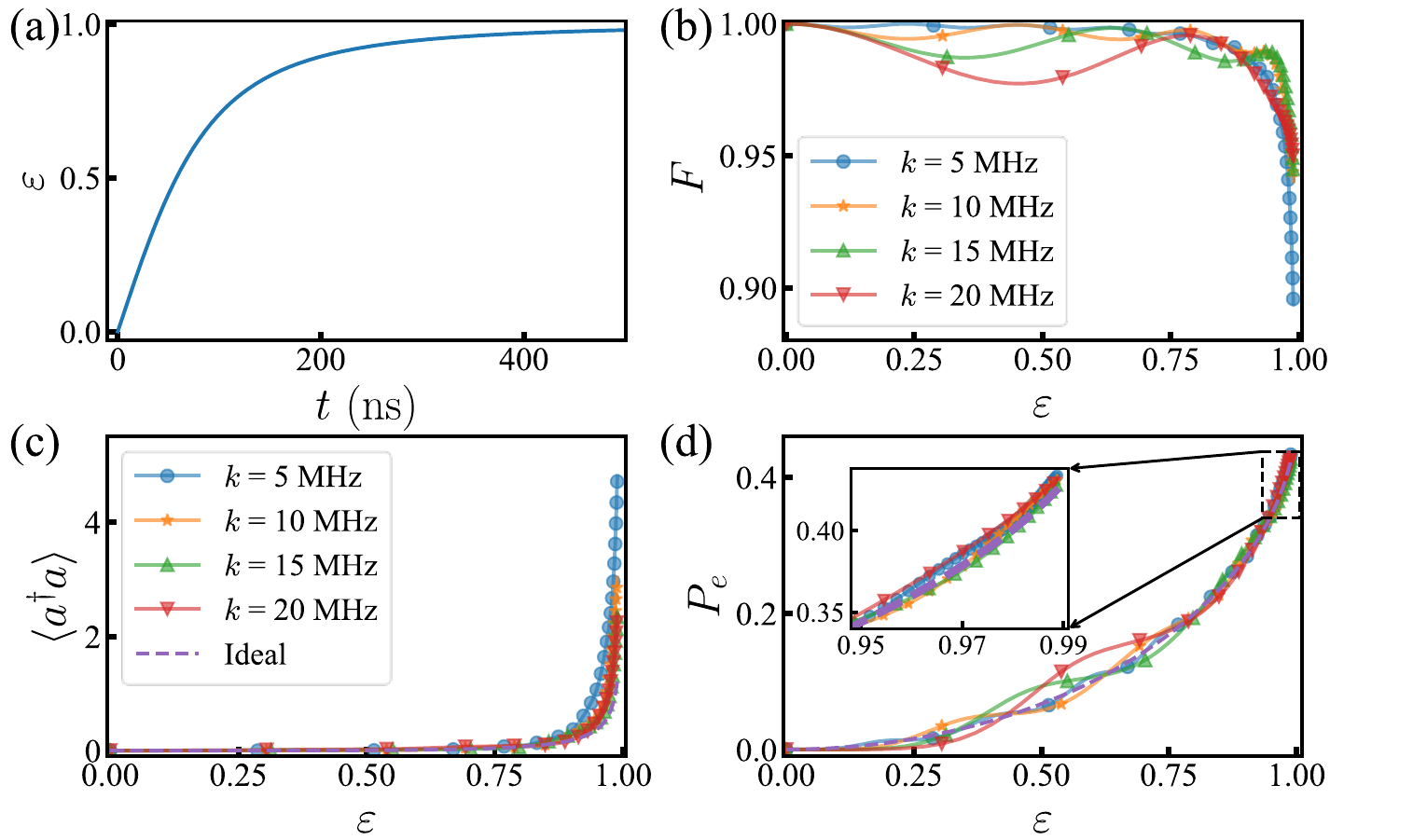}
     \caption{(color online). Simulations of the quench process. (a) $\varepsilon $
versus $t$. Here $\varepsilon $ is varied according to Eq. (\ref{epsilon}) with $k=10$ MHz. (b)
Fidelities of the qubit-resonator state to the ideal dark state as functions
of $\varepsilon $. In the simulation, the qubit-resonator coupling strength
is set to $\Omega =2\pi\times20.9$ MHz, and the dissipation rates for the qubit and resonator
are $0.05$ MHz and $0.08$ MHz, respectively. (c)
Average photon number of the resonator and (d) qubit's excitation number versus $\varepsilon$ for different $k$. The
dashed lines in (c) and (d) denote the results for the ideal dark state. The nice agreement between the results from simulation and the ideal dark state demonstrates that the qubit's excited-state population is insensitive to the dissipations for both the qubit and the photonic mode. }\label{fig2}
\end{figure}

In our sensing protocol, the signal is encoded in the time-evolving
eigenstate $\left\vert \psi _{0}\right\rangle $, which is separated from its nearest eigenstates by $E_{G,\min }=\Omega A^{3/4}$. The ratio between
this energy gap and the ramping rate  of the control parameter $%
\varepsilon $ determines how well the system can be kept in the dark state.
The lower this ramping rate, the smaller the leakage probability to the
bright eigenstates. However, a longer evolution time would result in more
serious decoherence effects. For a practical system, the attainable accuracy is limited by the non-adiabaticity and decoherence. As a compromise between these two effects, the dependence of $\varepsilon $ on $t$ is modeled as%
\begin{equation}
    \varepsilon =\sqrt{1-(k^{2}t^{2}+1)^{-1}}, \label{epsilon}
\end{equation}
where $k$ is the coefficient controlling the ramping velocity. Fig. \ref{fig2}(a) presents $\varepsilon$ plotted as a function of $t$, where $k=10$ MHz. To clearly show how well the system is restricted to the dark state during this quench process, we perform a simulation of the fidelity (${ F}$) as a function of $\varepsilon $, which is defined as ${ F}=\left\langle \psi _{0}\right\vert \rho \left\vert \psi_{0}\right\rangle $, where $\rho $ denotes the density operator for the system.  $\rho $ is calculated from the master equation, which includes the coherent dynamics governed by the Hamiltonian of Eq. (\ref{eq1}), as well as the incoherent dissipations for both the qubit and the resonator (see Supplemental Material \cite{supp}). The numerical result is displayed in Fig. \ref{fig2}(b), where the qubit-resonator coupling strength and the energy dissipative rates are set to be the same as those in our experimental system. As expected, the fidelity drops fast when approaching the critical point, as a consequence of the leakage to bright states caused by decoherence and non-adiabatic effect.

As analyzed above, the average photon number is very sensitive to the state leakage.
To provide quantitative evidence for this sensitivity, we simulate evolutions of the photon-number during quench process, defined as $\langle a^\dagger a\rangle={\rm Tr} (\rho a^\dagger a)$. The result is shown in Fig.
\ref{fig2}(c), which confirms that even a slight leakage to the bright states can cause a
significant deviation of the average photon number from that in the dark
state.
In contrast, the evolution of the qubit's $\left\vert e\right\rangle $-state population  ($P_{e}$) is robust against the state leakage,  as shown in Fig. \ref{fig2}(d).  As expected, when approaching the critical point, $P_{e}$
calculated with different ramping velocities converge to that of the ideal
dark eigenstate. This robustness is due to the fact that at the critical
point, each of the bright states is a product state, where the qubit part is
the same as in the dark state. Consequently, the evolution of $P_{e}$ is
robust against dissipation, non-adiabaticity, and imperfect timing. More
importantly, near the critical point $P_{e}$ almost displays the same
dependence on $\varepsilon $ for a quite wide range of $k$. This allows the
problem associated with critical slowing down to be largely compensated
by increasing $k$. For example, for the choice of $k=10$ MHz, the ramping of $%
\varepsilon $ from 0 to 0.99 is accomplished within a time $T\approx 700$ ns, which is much shorter than the energy relaxation times for the qubit and the photonic mode of the experimental system, $T^q_1=20$ $\mu$s and $T^p_1=12$ $\mu$s. With this choice, the dependence of $P_{e}$ on $\varepsilon $
well coincides with that for the ideal dark state within the regime $0.8<\varepsilon <0.99$.
 For the ideal dark state, $P_e$ becomes more sensitive to the change of $\varepsilon$ when $\varepsilon$ gets closer to 1, sharply increasing from 0.429 to 0.5 when $\varepsilon$ is varied from 0.99 to 1, and exhibiting an infinite sensitivity at the critical point. However, for the real system, the sensitivity cannot be infinitely improved owing to  decoherence and non-adiabatic effects, and consequently, a sharp increase of $P_e$ to $0.5$ at the critical point cannot be observed.

\subsection{Experimental setup and results}
The critical system is realized in a circuit QED architecture,
which involves a bus resonator with a fixed frequency $\omega _{r}/2\pi\simeq5.584$ GHz
and five frequency-tunable Xmon qubits, one of which is used as the test
qubit. The test qubit has an anharmonicity of $\chi \simeq2\pi\times0.245$ GHz. The on-resonance
photonic swapping rate between the qubit and the resonator is $\Omega \simeq2\pi\times 20.9$ MHz. The energy decaying rates for the qubit and the resonator are $%
\kappa _{q}\simeq 0.05$ MHz and $\kappa _{r}\simeq 0.08$ MHz, respectively. The dephasing rate of the qubit, measured at its idle frequency is around $1.25$ MHz. We note that the dephasing
noises are highly suppressed by resonantly coupling to the photonic field, as
a consequence of the dynamical decoupling \cite{PRL130501}.

\begin{figure}
	\includegraphics[width=0.5\textwidth]{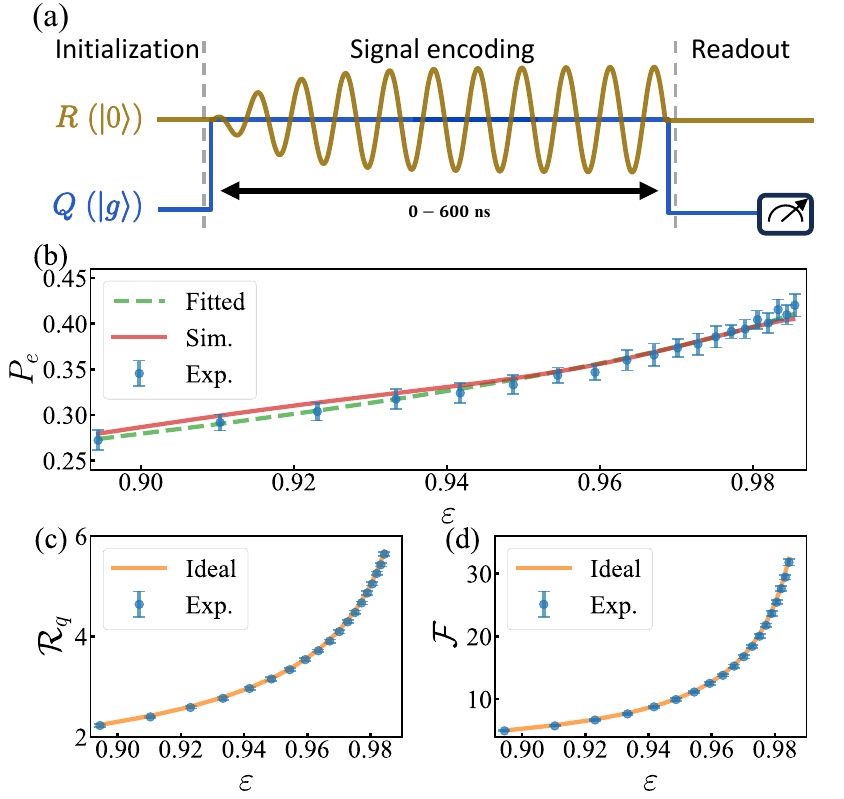}
     \caption{(color online). Performance of the criticality-enhanced sensing. (a) Pulse sequence. The qubit ($Q$) is tuned from its idle frequency to the resonator's frequency. At the same time the signal field with a rescaled amplitude $\varepsilon $ is coupled to the resonator. After the quench process, the qubit is biased back to its idle frequency for state readout. $\varepsilon$ is the time-dependent function shown in Fig. \ref{fig2}(a). (b) Qubit's $\left\vert e\right\rangle $-state population measured versus $\varepsilon $. The solid line denotes the simulation including the second-excited state of the Xmon, and the dashed line is the fitted function $P_{e}(\varepsilon)=C(1-\sqrt{A})/2$, with $C=0.9899$. (c) Signal-to-noise ratio (${\cal R}_{q}$) and (d)  Fisher information ($\cal F$) versus $\varepsilon$. The values ${\cal R}_{q}$ and $\cal F$ depend both on $P_{e}$ and on its derivative with respect to $\varepsilon$. The value of this derivative at each point is calculated with the fitted function $P_{e}(\varepsilon)$. The solid lines in (c) and (d) denote the results for the ideal dark state. 
}  \label{fig3}
\end{figure}

Before the experiment, the resonator is in the vacuum and the test qubit is initially in its ground state at
the idle frequency $\omega _{0}/2\pi=5.35$ GHz, where it is effectively decoupled from
the bus resonator since the corresponding detuning is much larger than $%
\Omega $. The interaction between this qubit and the resonator is activated
by tuning the qubit to the resonator's frequency. At the same time, the signal field with a rescaled amplitude $\varepsilon $ is coupled
to the resonator, continuously driving the photonic field. The experimental
pulse sequence is shown in Fig. \ref{fig3}(a).  The rescaled field
amplitude $\varepsilon $ versus $t$ is set to be the same as that shown in Fig. \ref{fig2}(a).

Due to the limited anharmonicity, the Xmon has a probability of leaking to levels higher than $\vert e\rangle$, which increases with the resonator's photon number. To suppress this leakage, we fine tune the detunings of the qubit and the resonator  (see
Supplemental Material). When $\varepsilon$ reaches a preset value, the qubit is biased back to its idle frequency for state readout. The qubit's $\vert e\rangle $-state populations, measured for different values
of $\varepsilon $, are displayed in Fig. \ref{fig3}(b).
The measured data well agree with the numerical simulation, where
the Xmon's second-excited state $\vert f\rangle$ is included.
The slight fluctuations are mainly caused by imperfect calibration of the amplitude of the signal field. In our experiment, $\varepsilon$ is calibrated by coupling the signal field to the resonator for a given time, and then reading out the resonator's average photon number with an ancilla qubit (see Supplemental Material). The errors in this procedure make the calibrated value of $\varepsilon$ deviate from the real value.
 The dashed line denotes the function $C(1-\sqrt{A})/2$, fitted with the data measured within the range $0.9<\varepsilon <0.985$, where the fitted parameter $C$ is $0.9899$, close to the ideal value $C=1$. This agreement confirms  that the critical behaviors of the qubit are well controlled.

The critical enhancement is manifested by the diverging changing rate of $%
P_{e}$ in the dark state $\left\vert \psi _{0}\right\rangle $, given by $%
dP_{e}/d\varepsilon =\varepsilon /(2\sqrt{A})$, which tends to $1/[2\sqrt{%
2(1-\varepsilon )}]$ when $1-\varepsilon \ll 1$. Consequently, in proximity
of the critical point, a tiny variation of $\varepsilon $ would result in
dramatic responses of the qubit. The signal-to-noise ratio is given by
\begin{equation}
    {\cal R}_{q}=\frac{dP_{e}/d\varepsilon }{\Delta P_{e}},
\end{equation}
where $\Delta P_{e}=\sqrt{P_{e}(1-P_e)}$ denotes the standard deviation for
measuring $P_{e}$ (see Supplemental Material). In proximity of the critical
point, ${\cal R}_{q}$ is approximately given by $1/\sqrt{2(1-\varepsilon )}
$. ${\cal R}_{q}$ versus $\varepsilon $  [inferred from $\varepsilon$-dependence of $P_e$ in Fig. \ref{fig3}(b)] is shown in Fig. \ref{fig3}(c). The performance of the
quantum sensor can be further quantified by the Fisher information, defined
as
\begin{equation}
    {\cal F}=\frac{1}{P_{e}}\left( \frac{dP_{e}}{d\varepsilon }\right) ^{2}+%
\frac{1}{P_{g}}\left( \frac{dP_{g}}{d\varepsilon }\right) ^{2}.
\end{equation}
As shown in the Supplemental Material, ${\cal F}$ is equal to the
quantum Fisher information, thereby saturating  the Cram\'{e}r-Rao bound. Moreover, $\mathcal{F}$ linearly scales with the square of the time, corresponding to the optimal Heisenberg scaling precision \cite{PRL120504}.  Fig. \ref{fig3}(d) presents the Fisher information evolution  obtained  from the measured $P_e$. Within the range $0.9<\varepsilon <0.985$, the results for ${\cal R}_{q}$ and $\cal F$ both agree well with  those calculated with the ideal dark state (solid lines). 

The most remarkable feature of our scheme is the incorporation of the intrinsic robustness against experimental imperfections with the high sensitivity near the critical point. This feature can be illustrated by comparing our protocol with the conventional Rabi measurement method \cite{RMP035002}, which aims to estimate the amplitude of the signal field by directly coupling it to a qubit. As detailed in the Supplemental Material, the measurement error is proportional to the timing error. In distinct contrast, the present protocol is robust against imperfect timing. As illustrated in Fig. \ref{fig2}(d), within the working regime the dependence of the probability $P_e$ on $\varepsilon$ is insensitive to $k$, which determines the ramping time $T$. We further confirm this robustness by directly displaying $P_e$ as a function of $T$ for different values of $\varepsilon$, simulated with the master equation (see Sec. S8 of the Supplemental Material). We also note that $P_{e}$ is insensitive to the frequency fluctuation of the signal field, as shown in Fig. S2(b).
\section{Discussion}
In conclusion, we have theoretically proposed and experimentally
demonstrated a scheme for realizing a criticality-enhanced quantum sensing
protocol. The critical system comprises a qubit and a photonic field stored
in a resonator. Under the driving of the signal field, the system has a
unique dark eigenstate, where the qubit's excitation number is used as the
indicator for estimating the signal field amplitude. The performance of the
sensor is insensitive to leakage to the bright states, caused by
non-adiabaticity and dissipation, as the
qubit is in the same state at the critical point, no matter the system is in
the dark or bright states. We demonstrate our critical sensing
protocol with a circuit QED architecture.   It should be noted that the criticality-enhanced sensitivity is achieved only when the field amplitude is within a narrow region below the critical point. For the general case, the critical sensing needs to be realized in a two-step procedure, where a rough estimation is first made, following which the field amplitude is shifted to the desired critical region for fine measurement \cite{NPJ170}. The working region of the sensor can be broadened by using a tunable coupler \cite{PRAD054062}, with which the qubit-resonator coupling is tunable so that the critical sensor can be adapted for a broader range of the field amplitude. Our approach can be directly applied to other spin-boson systems, e.g., ion traps (see Sec. S10), where the qubit can be better controlled and measured with a higher fidelity \cite{RMP1103}.

High-precision estimation of the amplitude of a driving field is crucial for implementation of high-fidelity single-qubit gates (see Sec. S11), which are realized by driving the relevant qubits with external fields with well controlled amplitudes.
  These gates are essential for implementation of quantum algorithms \cite{book} and for multi-qubit quantum state reconstructions  \cite{PRL180511}. High-precision calibration of the driving amplitude is also crucial for continuous-variable-based quantum information processing protocols with photonic fields \cite{science607}, where high-precision calibration of external driving fields is crucial for high-fidelity manipulation and readout of the photonic qubits.

\bibliography{ref}

\begin{thebibliography}{62}%
\makeatletter
\providecommand \@ifxundefined [1]{%
 \@ifx{#1\undefined}
}%
\providecommand \@ifnum [1]{%
 \ifnum #1\expandafter \@firstoftwo
 \else \expandafter \@secondoftwo
 \fi
}%
\providecommand \@ifx [1]{%
 \ifx #1\expandafter \@firstoftwo
 \else \expandafter \@secondoftwo
 \fi
}%
\providecommand \natexlab [1]{#1}%
\providecommand \enquote  [1]{``#1''}%
\providecommand \bibnamefont  [1]{#1}%
\providecommand \bibfnamefont [1]{#1}%
\providecommand \citenamefont [1]{#1}%
\providecommand \href@noop [0]{\@secondoftwo}%
\providecommand \href [0]{\begingroup \@sanitize@url \@href}%
\providecommand \@href[1]{\@@startlink{#1}\@@href}%
\providecommand \@@href[1]{\endgroup#1\@@endlink}%
\providecommand \@sanitize@url [0]{\catcode `\\12\catcode `\$12\catcode
  `\&12\catcode `\#12\catcode `\^12\catcode `\_12\catcode `\%12\relax}%
\providecommand \@@startlink[1]{}%
\providecommand \@@endlink[0]{}%
\providecommand \url  [0]{\begingroup\@sanitize@url \@url }%
\providecommand \@url [1]{\endgroup\@href {#1}{\urlprefix }}%
\providecommand \urlprefix  [0]{URL }%
\providecommand \Eprint [0]{\href }%
\providecommand \doibase [0]{https://doi.org/}%
\providecommand \selectlanguage [0]{\@gobble}%
\providecommand \bibinfo  [0]{\@secondoftwo}%
\providecommand \bibfield  [0]{\@secondoftwo}%
\providecommand \translation [1]{[#1]}%
\providecommand \BibitemOpen [0]{}%
\providecommand \bibitemStop [0]{}%
\providecommand \bibitemNoStop [0]{.\EOS\space}%
\providecommand \EOS [0]{\spacefactor3000\relax}%
\providecommand \BibitemShut  [1]{\csname bibitem#1\endcsname}%
\let\auto@bib@innerbib\@empty
\bibitem [{\citenamefont {et~al.}(2016)}]{PRL061102}%
  \BibitemOpen
  \bibfield  {author} {\bibinfo {author} {\bibfnamefont {B.~P.~A.}\
  \bibnamefont {et~al.}} (\bibinfo {collaboration} {LIGO Scientific and Virgo
  Collaborations}),\ }\bibfield  {title} {\bibinfo {title} {Observation of
  gravitational waves from a binary black hole merger},\ }\href
  {https://doi.org/10.1103/PhysRevLett.116.061102} {\bibfield  {journal}
  {\bibinfo  {journal} {Phys. Rev. Lett.}\ }\textbf {\bibinfo {volume} {116}},\
  \bibinfo {pages} {061102} (\bibinfo {year} {2016})}\BibitemShut {NoStop}%
\bibitem [{\citenamefont {Degen}\ \emph {et~al.}(2017)\citenamefont {Degen},
  \citenamefont {Reinhard},\ and\ \citenamefont {Cappellaro}}]{RMP035002}%
  \BibitemOpen
  \bibfield  {author} {\bibinfo {author} {\bibfnamefont {C.~L.}\ \bibnamefont
  {Degen}}, \bibinfo {author} {\bibfnamefont {F.}~\bibnamefont {Reinhard}},\
  and\ \bibinfo {author} {\bibfnamefont {P.}~\bibnamefont {Cappellaro}},\
  }\bibfield  {title} {\bibinfo {title} {Quantum sensing},\ }\href
  {https://doi.org/10.1103/RevModPhys.89.035002} {\bibfield  {journal}
  {\bibinfo  {journal} {Rev. Mod. Phys.}\ }\textbf {\bibinfo {volume} {89}},\
  \bibinfo {pages} {035002} (\bibinfo {year} {2017})}\BibitemShut {NoStop}%
\bibitem [{\citenamefont {Braun}\ \emph {et~al.}(2018)\citenamefont {Braun},
  \citenamefont {Adesso}, \citenamefont {Benatti}, \citenamefont {Floreanini},
  \citenamefont {Marzolino}, \citenamefont {Mitchell},\ and\ \citenamefont
  {Pirandola}}]{RMP035006}%
  \BibitemOpen
  \bibfield  {author} {\bibinfo {author} {\bibfnamefont {D.}~\bibnamefont
  {Braun}}, \bibinfo {author} {\bibfnamefont {G.}~\bibnamefont {Adesso}},
  \bibinfo {author} {\bibfnamefont {F.}~\bibnamefont {Benatti}}, \bibinfo
  {author} {\bibfnamefont {R.}~\bibnamefont {Floreanini}}, \bibinfo {author}
  {\bibfnamefont {U.}~\bibnamefont {Marzolino}}, \bibinfo {author}
  {\bibfnamefont {M.~W.}\ \bibnamefont {Mitchell}},\ and\ \bibinfo {author}
  {\bibfnamefont {S.}~\bibnamefont {Pirandola}},\ }\bibfield  {title} {\bibinfo
  {title} {Quantum-enhanced measurements without entanglement},\ }\href
  {https://doi.org/10.1103/RevModPhys.90.035006} {\bibfield  {journal}
  {\bibinfo  {journal} {Rev. Mod. Phys.}\ }\textbf {\bibinfo {volume} {90}},\
  \bibinfo {pages} {035006} (\bibinfo {year} {2018})}\BibitemShut {NoStop}%
\bibitem [{\citenamefont {Facon}\ \emph {et~al.}(2016)\citenamefont {Facon},
  \citenamefont {Dietsche}, \citenamefont {Grosso}, \citenamefont {Haroche},
  \citenamefont {Raimond}, \citenamefont {Brune},\ and\ \citenamefont
  {Gleyzes}}]{Nature262}%
  \BibitemOpen
  \bibfield  {author} {\bibinfo {author} {\bibfnamefont {A.}~\bibnamefont
  {Facon}}, \bibinfo {author} {\bibfnamefont {E.-K.}\ \bibnamefont {Dietsche}},
  \bibinfo {author} {\bibfnamefont {D.}~\bibnamefont {Grosso}}, \bibinfo
  {author} {\bibfnamefont {S.}~\bibnamefont {Haroche}}, \bibinfo {author}
  {\bibfnamefont {J.-M.}\ \bibnamefont {Raimond}}, \bibinfo {author}
  {\bibfnamefont {M.}~\bibnamefont {Brune}},\ and\ \bibinfo {author}
  {\bibfnamefont {S.}~\bibnamefont {Gleyzes}},\ }\bibfield  {title} {\bibinfo
  {title} {A sensitive electrometer based on a {Rydberg} atom in a
  {Schrödinger}-cat state},\ }\href {https://doi.org/10.1038/nature18327}
  {\bibfield  {journal} {\bibinfo  {journal} {Nature (London)}\ }\textbf
  {\bibinfo {volume} {535}},\ \bibinfo {pages} {262} (\bibinfo {year}
  {2016})}\BibitemShut {NoStop}%
\bibitem [{\citenamefont {McCormick}\ \emph {et~al.}(2019)\citenamefont
  {McCormick}, \citenamefont {Keller}, \citenamefont {Burd}, \citenamefont
  {Wineland}, \citenamefont {Wilson},\ and\ \citenamefont
  {Leibfried}}]{Nature86}%
  \BibitemOpen
  \bibfield  {author} {\bibinfo {author} {\bibfnamefont {K.~C.}\ \bibnamefont
  {McCormick}}, \bibinfo {author} {\bibfnamefont {J.}~\bibnamefont {Keller}},
  \bibinfo {author} {\bibfnamefont {S.~C.}\ \bibnamefont {Burd}}, \bibinfo
  {author} {\bibfnamefont {D.~J.}\ \bibnamefont {Wineland}}, \bibinfo {author}
  {\bibfnamefont {A.~C.}\ \bibnamefont {Wilson}},\ and\ \bibinfo {author}
  {\bibfnamefont {D.}~\bibnamefont {Leibfried}},\ }\bibfield  {title} {\bibinfo
  {title} {Quantum-enhanced sensing of a single-ion mechanical oscillator},\
  }\href {https://doi.org/10.1038/s41586-019-1421-y} {\bibfield  {journal}
  {\bibinfo  {journal} {Nature (London)}\ }\textbf {\bibinfo {volume} {572}},\
  \bibinfo {pages} {86} (\bibinfo {year} {2019})}\BibitemShut {NoStop}%
\bibitem [{\citenamefont {Wang}\ \emph {et~al.}(2019)\citenamefont {Wang},
  \citenamefont {Wu}, \citenamefont {Ma}, \citenamefont {Cai}, \citenamefont
  {Hu}, \citenamefont {Mu}, \citenamefont {Xu}, \citenamefont {Chen},
  \citenamefont {Wang}, \citenamefont {Song}, \citenamefont {Yuan},
  \citenamefont {Zou}, \citenamefont {Duan},\ and\ \citenamefont
  {Sun}}]{NC4382}%
  \BibitemOpen
  \bibfield  {author} {\bibinfo {author} {\bibfnamefont {W.}~\bibnamefont
  {Wang}}, \bibinfo {author} {\bibfnamefont {Y.}~\bibnamefont {Wu}}, \bibinfo
  {author} {\bibfnamefont {Y.}~\bibnamefont {Ma}}, \bibinfo {author}
  {\bibfnamefont {W.}~\bibnamefont {Cai}}, \bibinfo {author} {\bibfnamefont
  {L.}~\bibnamefont {Hu}}, \bibinfo {author} {\bibfnamefont {X.}~\bibnamefont
  {Mu}}, \bibinfo {author} {\bibfnamefont {Y.}~\bibnamefont {Xu}}, \bibinfo
  {author} {\bibfnamefont {Z.-J.}\ \bibnamefont {Chen}}, \bibinfo {author}
  {\bibfnamefont {H.}~\bibnamefont {Wang}}, \bibinfo {author} {\bibfnamefont
  {Y.~P.}\ \bibnamefont {Song}}, \bibinfo {author} {\bibfnamefont
  {H.}~\bibnamefont {Yuan}}, \bibinfo {author} {\bibfnamefont {C.-L.}\
  \bibnamefont {Zou}}, \bibinfo {author} {\bibfnamefont {L.-M.}\ \bibnamefont
  {Duan}},\ and\ \bibinfo {author} {\bibfnamefont {L.}~\bibnamefont {Sun}},\
  }\bibfield  {title} {\bibinfo {title} {Heisenberg-limited single-mode quantum
  metrology in a superconducting circuit},\ }\href
  {https://doi.org/10.1038/s41467-019-12290-7} {\bibfield  {journal} {\bibinfo
  {journal} {Nat. Commun.}\ }\textbf {\bibinfo {volume} {10}},\ \bibinfo
  {pages} {4382} (\bibinfo {year} {2019})}\BibitemShut {NoStop}%
\bibitem [{\citenamefont {Nagata}\ \emph {et~al.}(2007)\citenamefont {Nagata},
  \citenamefont {Okamoto}, \citenamefont {O'Brien}, \citenamefont {Sasaki},\
  and\ \citenamefont {Takeuchi}}]{Science726}%
  \BibitemOpen
  \bibfield  {author} {\bibinfo {author} {\bibfnamefont {T.}~\bibnamefont
  {Nagata}}, \bibinfo {author} {\bibfnamefont {R.}~\bibnamefont {Okamoto}},
  \bibinfo {author} {\bibfnamefont {J.~L.}\ \bibnamefont {O'Brien}}, \bibinfo
  {author} {\bibfnamefont {K.}~\bibnamefont {Sasaki}},\ and\ \bibinfo {author}
  {\bibfnamefont {S.}~\bibnamefont {Takeuchi}},\ }\bibfield  {title} {\bibinfo
  {title} {Beating the standard quantum limit with four-entangled photons},\
  }\href {https://doi.org/10.1126/science.1138007} {\bibfield  {journal}
  {\bibinfo  {journal} {Science}\ }\textbf {\bibinfo {volume} {316}},\ \bibinfo
  {pages} {726} (\bibinfo {year} {2007})}\BibitemShut {NoStop}%
\bibitem [{\citenamefont {Leibfried}\ \emph {et~al.}(2004)\citenamefont
  {Leibfried}, \citenamefont {Barrett}, \citenamefont {Schaetz}, \citenamefont
  {Britton}, \citenamefont {Chiaverini}, \citenamefont {Itano}, \citenamefont
  {Jost}, \citenamefont {Langer},\ and\ \citenamefont
  {Wineland}}]{Science1476}%
  \BibitemOpen
  \bibfield  {author} {\bibinfo {author} {\bibfnamefont {D.}~\bibnamefont
  {Leibfried}}, \bibinfo {author} {\bibfnamefont {M.~D.}\ \bibnamefont
  {Barrett}}, \bibinfo {author} {\bibfnamefont {T.}~\bibnamefont {Schaetz}},
  \bibinfo {author} {\bibfnamefont {J.}~\bibnamefont {Britton}}, \bibinfo
  {author} {\bibfnamefont {J.}~\bibnamefont {Chiaverini}}, \bibinfo {author}
  {\bibfnamefont {W.~M.}\ \bibnamefont {Itano}}, \bibinfo {author}
  {\bibfnamefont {J.~D.}\ \bibnamefont {Jost}}, \bibinfo {author}
  {\bibfnamefont {C.}~\bibnamefont {Langer}},\ and\ \bibinfo {author}
  {\bibfnamefont {D.~J.}\ \bibnamefont {Wineland}},\ }\bibfield  {title}
  {\bibinfo {title} {Toward heisenberg-limited spectroscopy with multiparticle
  entangled states},\ }\href {https://doi.org/10.1126/science.1097576}
  {\bibfield  {journal} {\bibinfo  {journal} {Science}\ }\textbf {\bibinfo
  {volume} {304}},\ \bibinfo {pages} {1476} (\bibinfo {year}
  {2004})}\BibitemShut {NoStop}%
\bibitem [{\citenamefont {Monz}\ \emph {et~al.}(2011)\citenamefont {Monz},
  \citenamefont {Schindler}, \citenamefont {Barreiro}, \citenamefont {Chwalla},
  \citenamefont {Nigg}, \citenamefont {Coish}, \citenamefont {Harlander},
  \citenamefont {H\"ansel}, \citenamefont {Hennrich},\ and\ \citenamefont
  {Blatt}}]{PRL130506}%
  \BibitemOpen
  \bibfield  {author} {\bibinfo {author} {\bibfnamefont {T.}~\bibnamefont
  {Monz}}, \bibinfo {author} {\bibfnamefont {P.}~\bibnamefont {Schindler}},
  \bibinfo {author} {\bibfnamefont {J.~T.}\ \bibnamefont {Barreiro}}, \bibinfo
  {author} {\bibfnamefont {M.}~\bibnamefont {Chwalla}}, \bibinfo {author}
  {\bibfnamefont {D.}~\bibnamefont {Nigg}}, \bibinfo {author} {\bibfnamefont
  {W.~A.}\ \bibnamefont {Coish}}, \bibinfo {author} {\bibfnamefont
  {M.}~\bibnamefont {Harlander}}, \bibinfo {author} {\bibfnamefont
  {W.}~\bibnamefont {H\"ansel}}, \bibinfo {author} {\bibfnamefont
  {M.}~\bibnamefont {Hennrich}},\ and\ \bibinfo {author} {\bibfnamefont
  {R.}~\bibnamefont {Blatt}},\ }\bibfield  {title} {\bibinfo {title} {14-qubit
  entanglement: Creation and coherence},\ }\href
  {https://doi.org/10.1103/PhysRevLett.106.130506} {\bibfield  {journal}
  {\bibinfo  {journal} {Phys. Rev. Lett.}\ }\textbf {\bibinfo {volume} {106}},\
  \bibinfo {pages} {130506} (\bibinfo {year} {2011})}\BibitemShut {NoStop}%
\bibitem [{\citenamefont {Greve}\ \emph {et~al.}(2022)\citenamefont {Greve},
  \citenamefont {Luo}, \citenamefont {Wu},\ and\ \citenamefont
  {Thompson}}]{Nature472}%
  \BibitemOpen
  \bibfield  {author} {\bibinfo {author} {\bibfnamefont {G.~P.}\ \bibnamefont
  {Greve}}, \bibinfo {author} {\bibfnamefont {C.}~\bibnamefont {Luo}}, \bibinfo
  {author} {\bibfnamefont {B.}~\bibnamefont {Wu}},\ and\ \bibinfo {author}
  {\bibfnamefont {J.~K.}\ \bibnamefont {Thompson}},\ }\bibfield  {title}
  {\bibinfo {title} {Entanglement-enhanced matter-wave interferometry in a
  high-finesse cavity},\ }\href {https://doi.org/10.1038/s41586-022-05197-9}
  {\bibfield  {journal} {\bibinfo  {journal} {Nature (London)}\ }\textbf
  {\bibinfo {volume} {610}},\ \bibinfo {pages} {472} (\bibinfo {year}
  {2022})}\BibitemShut {NoStop}%
\bibitem [{\citenamefont {Pezzè}\ \emph {et~al.}(2016)\citenamefont {Pezzè},
  \citenamefont {Li}, \citenamefont {Li},\ and\ \citenamefont
  {Smerzi}}]{PNAS11459}%
  \BibitemOpen
  \bibfield  {author} {\bibinfo {author} {\bibfnamefont {L.}~\bibnamefont
  {Pezzè}}, \bibinfo {author} {\bibfnamefont {Y.}~\bibnamefont {Li}}, \bibinfo
  {author} {\bibfnamefont {W.}~\bibnamefont {Li}},\ and\ \bibinfo {author}
  {\bibfnamefont {A.}~\bibnamefont {Smerzi}},\ }\bibfield  {title} {\bibinfo
  {title} {Witnessing entanglement without entanglement witness operators},\
  }\href {https://doi.org/10.1073/pnas.1603346113} {\bibfield  {journal}
  {\bibinfo  {journal} {Proc. Natl. Acad. Sci. U.S.A.}\ }\textbf {\bibinfo
  {volume} {113}},\ \bibinfo {pages} {11459} (\bibinfo {year}
  {2016})}\BibitemShut {NoStop}%
\bibitem [{\citenamefont {Sachdev}(2011)}]{BOOK1}%
  \BibitemOpen
  \bibfield  {author} {\bibinfo {author} {\bibfnamefont {S.}~\bibnamefont
  {Sachdev}},\ }\href@noop {} {\emph {\bibinfo {title} {Quantum Phase
  Transitions}}},\ \bibinfo {edition} {2nd}\ ed.\ (\bibinfo  {publisher}
  {Cambridge University Press},\ \bibinfo {year} {2011})\BibitemShut {NoStop}%
\bibitem [{\citenamefont {Zanardi}\ and\ \citenamefont
  {Paunkovi\ifmmode~\acute{c}\else \'{c}\fi{}}(2006)}]{PRE031123}%
  \BibitemOpen
  \bibfield  {author} {\bibinfo {author} {\bibfnamefont {P.}~\bibnamefont
  {Zanardi}}\ and\ \bibinfo {author} {\bibfnamefont {N.}~\bibnamefont
  {Paunkovi\ifmmode~\acute{c}\else \'{c}\fi{}}},\ }\bibfield  {title} {\bibinfo
  {title} {Ground state overlap and quantum phase transitions},\ }\href
  {https://doi.org/10.1103/PhysRevE.74.031123} {\bibfield  {journal} {\bibinfo
  {journal} {Phys. Rev. E}\ }\textbf {\bibinfo {volume} {74}},\ \bibinfo
  {pages} {031123} (\bibinfo {year} {2006})}\BibitemShut {NoStop}%
\bibitem [{\citenamefont {Albuquerque}\ \emph {et~al.}(2010)\citenamefont
  {Albuquerque}, \citenamefont {Alet}, \citenamefont {Sire},\ and\
  \citenamefont {Capponi}}]{PRB064418}%
  \BibitemOpen
  \bibfield  {author} {\bibinfo {author} {\bibfnamefont {A.~F.}\ \bibnamefont
  {Albuquerque}}, \bibinfo {author} {\bibfnamefont {F.}~\bibnamefont {Alet}},
  \bibinfo {author} {\bibfnamefont {C.}~\bibnamefont {Sire}},\ and\ \bibinfo
  {author} {\bibfnamefont {S.}~\bibnamefont {Capponi}},\ }\bibfield  {title}
  {\bibinfo {title} {Quantum critical scaling of fidelity susceptibility},\
  }\href {https://doi.org/10.1103/PhysRevB.81.064418} {\bibfield  {journal}
  {\bibinfo  {journal} {Phys. Rev. B}\ }\textbf {\bibinfo {volume} {81}},\
  \bibinfo {pages} {064418} (\bibinfo {year} {2010})}\BibitemShut {NoStop}%
\bibitem [{\citenamefont {Wang}\ \emph {et~al.}(2014)\citenamefont {Wang},
  \citenamefont {Wu}, \citenamefont {Yang}, \citenamefont {Jin}, \citenamefont
  {Lambert},\ and\ \citenamefont {Nori}}]{NJP063039}%
  \BibitemOpen
  \bibfield  {author} {\bibinfo {author} {\bibfnamefont {T.-L.}\ \bibnamefont
  {Wang}}, \bibinfo {author} {\bibfnamefont {L.-N.}\ \bibnamefont {Wu}},
  \bibinfo {author} {\bibfnamefont {W.}~\bibnamefont {Yang}}, \bibinfo {author}
  {\bibfnamefont {G.-R.}\ \bibnamefont {Jin}}, \bibinfo {author} {\bibfnamefont
  {N.}~\bibnamefont {Lambert}},\ and\ \bibinfo {author} {\bibfnamefont
  {F.}~\bibnamefont {Nori}},\ }\bibfield  {title} {\bibinfo {title} {Quantum
  fisher information as a signature of the superradiant quantum phase
  transition},\ }\href {https://doi.org/10.1088/1367-2630/16/6/063039}
  {\bibfield  {journal} {\bibinfo  {journal} {New J. Phys.}\ }\textbf {\bibinfo
  {volume} {16}},\ \bibinfo {pages} {063039} (\bibinfo {year}
  {2014})}\BibitemShut {NoStop}%
\bibitem [{\citenamefont {You}\ \emph {et~al.}(2007)\citenamefont {You},
  \citenamefont {Li},\ and\ \citenamefont {Gu}}]{PRE022101}%
  \BibitemOpen
  \bibfield  {author} {\bibinfo {author} {\bibfnamefont {W.-L.}\ \bibnamefont
  {You}}, \bibinfo {author} {\bibfnamefont {Y.-W.}\ \bibnamefont {Li}},\ and\
  \bibinfo {author} {\bibfnamefont {S.-J.}\ \bibnamefont {Gu}},\ }\bibfield
  {title} {\bibinfo {title} {Fidelity, dynamic structure factor, and
  susceptibility in critical phenomena},\ }\href
  {https://doi.org/10.1103/PhysRevE.76.022101} {\bibfield  {journal} {\bibinfo
  {journal} {Phys. Rev. E}\ }\textbf {\bibinfo {volume} {76}},\ \bibinfo
  {pages} {022101} (\bibinfo {year} {2007})}\BibitemShut {NoStop}%
\bibitem [{\citenamefont {Zanardi}\ \emph {et~al.}(2008)\citenamefont
  {Zanardi}, \citenamefont {Paris},\ and\ \citenamefont
  {Campos~Venuti}}]{PRA042105}%
  \BibitemOpen
  \bibfield  {author} {\bibinfo {author} {\bibfnamefont {P.}~\bibnamefont
  {Zanardi}}, \bibinfo {author} {\bibfnamefont {M.~G.~A.}\ \bibnamefont
  {Paris}},\ and\ \bibinfo {author} {\bibfnamefont {L.}~\bibnamefont
  {Campos~Venuti}},\ }\bibfield  {title} {\bibinfo {title} {Quantum criticality
  as a resource for quantum estimation},\ }\href
  {https://doi.org/10.1103/PhysRevA.78.042105} {\bibfield  {journal} {\bibinfo
  {journal} {Phys. Rev. A}\ }\textbf {\bibinfo {volume} {78}},\ \bibinfo
  {pages} {042105} (\bibinfo {year} {2008})}\BibitemShut {NoStop}%
\bibitem [{\citenamefont {Tsang}(2013)}]{PRA021801}%
  \BibitemOpen
  \bibfield  {author} {\bibinfo {author} {\bibfnamefont {M.}~\bibnamefont
  {Tsang}},\ }\bibfield  {title} {\bibinfo {title} {Quantum transition-edge
  detectors},\ }\href {https://doi.org/10.1103/PhysRevA.88.021801} {\bibfield
  {journal} {\bibinfo  {journal} {Phys. Rev. A}\ }\textbf {\bibinfo {volume}
  {88}},\ \bibinfo {pages} {021801} (\bibinfo {year} {2013})}\BibitemShut
  {NoStop}%
\bibitem [{\citenamefont {Invernizzi}\ \emph {et~al.}(2008)\citenamefont
  {Invernizzi}, \citenamefont {Korbman}, \citenamefont {Campos~Venuti},\ and\
  \citenamefont {Paris}}]{PRA042106}%
  \BibitemOpen
  \bibfield  {author} {\bibinfo {author} {\bibfnamefont {C.}~\bibnamefont
  {Invernizzi}}, \bibinfo {author} {\bibfnamefont {M.}~\bibnamefont {Korbman}},
  \bibinfo {author} {\bibfnamefont {L.}~\bibnamefont {Campos~Venuti}},\ and\
  \bibinfo {author} {\bibfnamefont {M.~G.~A.}\ \bibnamefont {Paris}},\
  }\bibfield  {title} {\bibinfo {title} {Optimal quantum estimation in spin
  systems at criticality},\ }\href {https://doi.org/10.1103/PhysRevA.78.042106}
  {\bibfield  {journal} {\bibinfo  {journal} {Phys. Rev. A}\ }\textbf {\bibinfo
  {volume} {78}},\ \bibinfo {pages} {042106} (\bibinfo {year}
  {2008})}\BibitemShut {NoStop}%
\bibitem [{\citenamefont {Gammelmark}\ and\ \citenamefont
  {Mølmer}(2011)}]{NJP053035}%
  \BibitemOpen
  \bibfield  {author} {\bibinfo {author} {\bibfnamefont {S.}~\bibnamefont
  {Gammelmark}}\ and\ \bibinfo {author} {\bibfnamefont {K.}~\bibnamefont
  {Mølmer}},\ }\bibfield  {title} {\bibinfo {title} {Phase transitions and
  heisenberg limited metrology in an ising chain interacting with a single-mode
  cavity field},\ }\href {https://doi.org/10.1088/1367-2630/13/5/053035}
  {\bibfield  {journal} {\bibinfo  {journal} {New J. Phys.}\ }\textbf {\bibinfo
  {volume} {13}},\ \bibinfo {pages} {053035} (\bibinfo {year}
  {2011})}\BibitemShut {NoStop}%
\bibitem [{\citenamefont {Ivanov}\ and\ \citenamefont
  {Porras}(2013)}]{PRA023803}%
  \BibitemOpen
  \bibfield  {author} {\bibinfo {author} {\bibfnamefont {P.~A.}\ \bibnamefont
  {Ivanov}}\ and\ \bibinfo {author} {\bibfnamefont {D.}~\bibnamefont
  {Porras}},\ }\bibfield  {title} {\bibinfo {title} {Adiabatic quantum
  metrology with strongly correlated quantum optical systems},\ }\href
  {https://doi.org/10.1103/PhysRevA.88.023803} {\bibfield  {journal} {\bibinfo
  {journal} {Phys. Rev. A}\ }\textbf {\bibinfo {volume} {88}},\ \bibinfo
  {pages} {023803} (\bibinfo {year} {2013})}\BibitemShut {NoStop}%
\bibitem [{\citenamefont {Salvatori}\ \emph {et~al.}(2014)\citenamefont
  {Salvatori}, \citenamefont {Mandarino},\ and\ \citenamefont
  {Paris}}]{PRA022111}%
  \BibitemOpen
  \bibfield  {author} {\bibinfo {author} {\bibfnamefont {G.}~\bibnamefont
  {Salvatori}}, \bibinfo {author} {\bibfnamefont {A.}~\bibnamefont
  {Mandarino}},\ and\ \bibinfo {author} {\bibfnamefont {M.~G.~A.}\ \bibnamefont
  {Paris}},\ }\bibfield  {title} {\bibinfo {title} {Quantum metrology in
  lipkin-meshkov-glick critical systems},\ }\href
  {https://doi.org/10.1103/PhysRevA.90.022111} {\bibfield  {journal} {\bibinfo
  {journal} {Phys. Rev. A}\ }\textbf {\bibinfo {volume} {90}},\ \bibinfo
  {pages} {022111} (\bibinfo {year} {2014})}\BibitemShut {NoStop}%
\bibitem [{\citenamefont {Bina}\ \emph {et~al.}(2016)\citenamefont {Bina},
  \citenamefont {Amelio},\ and\ \citenamefont {Paris}}]{PRE052118}%
  \BibitemOpen
  \bibfield  {author} {\bibinfo {author} {\bibfnamefont {M.}~\bibnamefont
  {Bina}}, \bibinfo {author} {\bibfnamefont {I.}~\bibnamefont {Amelio}},\ and\
  \bibinfo {author} {\bibfnamefont {M.~G.~A.}\ \bibnamefont {Paris}},\
  }\bibfield  {title} {\bibinfo {title} {Dicke coupling by feasible local
  measurements at the superradiant quantum phase transition},\ }\href
  {https://doi.org/10.1103/PhysRevE.93.052118} {\bibfield  {journal} {\bibinfo
  {journal} {Phys. Rev. E}\ }\textbf {\bibinfo {volume} {93}},\ \bibinfo
  {pages} {052118} (\bibinfo {year} {2016})}\BibitemShut {NoStop}%
\bibitem [{\citenamefont {Rams}\ \emph {et~al.}(2018)\citenamefont {Rams},
  \citenamefont {Sierant}, \citenamefont {Dutta}, \citenamefont {Horodecki},\
  and\ \citenamefont {Zakrzewski}}]{PRX021022}%
  \BibitemOpen
  \bibfield  {author} {\bibinfo {author} {\bibfnamefont {M.~M.}\ \bibnamefont
  {Rams}}, \bibinfo {author} {\bibfnamefont {P.}~\bibnamefont {Sierant}},
  \bibinfo {author} {\bibfnamefont {O.}~\bibnamefont {Dutta}}, \bibinfo
  {author} {\bibfnamefont {P.}~\bibnamefont {Horodecki}},\ and\ \bibinfo
  {author} {\bibfnamefont {J.}~\bibnamefont {Zakrzewski}},\ }\bibfield  {title}
  {\bibinfo {title} {At the limits of criticality-based quantum metrology:
  Apparent super-heisenberg scaling revisited},\ }\href
  {https://doi.org/10.1103/PhysRevX.8.021022} {\bibfield  {journal} {\bibinfo
  {journal} {Phys. Rev. X}\ }\textbf {\bibinfo {volume} {8}},\ \bibinfo {pages}
  {021022} (\bibinfo {year} {2018})}\BibitemShut {NoStop}%
\bibitem [{\citenamefont {Wald}\ \emph {et~al.}(2020)\citenamefont {Wald},
  \citenamefont {Moreira},\ and\ \citenamefont {Semi\~ao}}]{PRE052107}%
  \BibitemOpen
  \bibfield  {author} {\bibinfo {author} {\bibfnamefont {S.}~\bibnamefont
  {Wald}}, \bibinfo {author} {\bibfnamefont {S.~V.}\ \bibnamefont {Moreira}},\
  and\ \bibinfo {author} {\bibfnamefont {F.~L.}\ \bibnamefont {Semi\~ao}},\
  }\bibfield  {title} {\bibinfo {title} {In- and out-of-equilibrium quantum
  metrology with mean-field quantum criticality},\ }\href
  {https://doi.org/10.1103/PhysRevE.101.052107} {\bibfield  {journal} {\bibinfo
   {journal} {Phys. Rev. E}\ }\textbf {\bibinfo {volume} {101}},\ \bibinfo
  {pages} {052107} (\bibinfo {year} {2020})}\BibitemShut {NoStop}%
\bibitem [{\citenamefont {Fr\'erot}\ and\ \citenamefont
  {Roscilde}(2018)}]{PRL020402}%
  \BibitemOpen
  \bibfield  {author} {\bibinfo {author} {\bibfnamefont {I.}~\bibnamefont
  {Fr\'erot}}\ and\ \bibinfo {author} {\bibfnamefont {T.}~\bibnamefont
  {Roscilde}},\ }\bibfield  {title} {\bibinfo {title} {Quantum critical
  metrology},\ }\href {https://doi.org/10.1103/PhysRevLett.121.020402}
  {\bibfield  {journal} {\bibinfo  {journal} {Phys. Rev. Lett.}\ }\textbf
  {\bibinfo {volume} {121}},\ \bibinfo {pages} {020402} (\bibinfo {year}
  {2018})}\BibitemShut {NoStop}%
\bibitem [{\citenamefont {Garbe}\ \emph {et~al.}(2020)\citenamefont {Garbe},
  \citenamefont {Bina}, \citenamefont {Keller}, \citenamefont {Paris},\ and\
  \citenamefont {Felicetti}}]{PRL120504}%
  \BibitemOpen
  \bibfield  {author} {\bibinfo {author} {\bibfnamefont {L.}~\bibnamefont
  {Garbe}}, \bibinfo {author} {\bibfnamefont {M.}~\bibnamefont {Bina}},
  \bibinfo {author} {\bibfnamefont {A.}~\bibnamefont {Keller}}, \bibinfo
  {author} {\bibfnamefont {M.~G.~A.}\ \bibnamefont {Paris}},\ and\ \bibinfo
  {author} {\bibfnamefont {S.}~\bibnamefont {Felicetti}},\ }\bibfield  {title}
  {\bibinfo {title} {Critical quantum metrology with a finite-component quantum
  phase transition},\ }\href {https://doi.org/10.1103/PhysRevLett.124.120504}
  {\bibfield  {journal} {\bibinfo  {journal} {Phys. Rev. Lett.}\ }\textbf
  {\bibinfo {volume} {124}},\ \bibinfo {pages} {120504} (\bibinfo {year}
  {2020})}\BibitemShut {NoStop}%
\bibitem [{\citenamefont {Niezgoda}\ and\ \citenamefont
  {Chwede\ifmmode~\acute{n}\else \'{n}\fi{}czuk}(2021)}]{PRL210506}%
  \BibitemOpen
  \bibfield  {author} {\bibinfo {author} {\bibfnamefont {A.}~\bibnamefont
  {Niezgoda}}\ and\ \bibinfo {author} {\bibfnamefont {J.}~\bibnamefont
  {Chwede\ifmmode~\acute{n}\else \'{n}\fi{}czuk}},\ }\bibfield  {title}
  {\bibinfo {title} {Many-body nonlocality as a resource for quantum-enhanced
  metrology},\ }\href {https://doi.org/10.1103/PhysRevLett.126.210506}
  {\bibfield  {journal} {\bibinfo  {journal} {Phys. Rev. Lett.}\ }\textbf
  {\bibinfo {volume} {126}},\ \bibinfo {pages} {210506} (\bibinfo {year}
  {2021})}\BibitemShut {NoStop}%
\bibitem [{\citenamefont {Garbe}\ \emph {et~al.}(2022)\citenamefont {Garbe},
  \citenamefont {Abah}, \citenamefont {Felicetti},\ and\ \citenamefont
  {Puebla}}]{QST035010}%
  \BibitemOpen
  \bibfield  {author} {\bibinfo {author} {\bibfnamefont {L.}~\bibnamefont
  {Garbe}}, \bibinfo {author} {\bibfnamefont {O.}~\bibnamefont {Abah}},
  \bibinfo {author} {\bibfnamefont {S.}~\bibnamefont {Felicetti}},\ and\
  \bibinfo {author} {\bibfnamefont {R.}~\bibnamefont {Puebla}},\ }\bibfield
  {title} {\bibinfo {title} {Critical quantum metrology with fully-connected
  models: from heisenberg to kibble–zurek scaling},\ }\href
  {https://doi.org/10.1088/2058-9565/ac6ca5} {\bibfield  {journal} {\bibinfo
  {journal} {Quantum Sci. Technol.}\ }\textbf {\bibinfo {volume} {7}},\
  \bibinfo {pages} {035010} (\bibinfo {year} {2022})}\BibitemShut {NoStop}%
\bibitem [{\citenamefont {Chen}\ \emph {et~al.}(2024)\citenamefont {Chen},
  \citenamefont {Lü}, \citenamefont {Zhu}, \citenamefont {Zhang},
  \citenamefont {Ning}, \citenamefont {Yang},\ and\ \citenamefont
  {Zheng}}]{chenken}%
  \BibitemOpen
  \bibfield  {author} {\bibinfo {author} {\bibfnamefont {K.}~\bibnamefont
  {Chen}}, \bibinfo {author} {\bibfnamefont {J.}~\bibnamefont {Lü}}, \bibinfo
  {author} {\bibfnamefont {X.}~\bibnamefont {Zhu}}, \bibinfo {author}
  {\bibfnamefont {H.}~\bibnamefont {Zhang}}, \bibinfo {author} {\bibfnamefont
  {W.}~\bibnamefont {Ning}}, \bibinfo {author} {\bibfnamefont {Z.}~\bibnamefont
  {Yang}},\ and\ \bibinfo {author} {\bibfnamefont {S.}~\bibnamefont {Zheng}},\
  }\bibfield  {title} {\bibinfo {title} {Critical {Sensing} with a {Single}
  {Bosonic} {Mode} {Without} {Boson}–{Boson} {Interactions}},\ }\href
  {https://doi.org/10.1002/qute.202400105} {\bibfield  {journal} {\bibinfo
  {journal} {Adv. Quantum Technol.}\ ,\ \bibinfo {pages} {2400105}} (\bibinfo
  {year} {2024})}\BibitemShut {NoStop}%
\bibitem [{\citenamefont {Liu}\ \emph {et~al.}(2021)\citenamefont {Liu},
  \citenamefont {Chen}, \citenamefont {Jiang}, \citenamefont {Yang},
  \citenamefont {Wu}, \citenamefont {Li}, \citenamefont {Yuan}, \citenamefont
  {Peng},\ and\ \citenamefont {Du}}]{NPJ170}%
  \BibitemOpen
  \bibfield  {author} {\bibinfo {author} {\bibfnamefont {R.}~\bibnamefont
  {Liu}}, \bibinfo {author} {\bibfnamefont {Y.}~\bibnamefont {Chen}}, \bibinfo
  {author} {\bibfnamefont {M.}~\bibnamefont {Jiang}}, \bibinfo {author}
  {\bibfnamefont {X.}~\bibnamefont {Yang}}, \bibinfo {author} {\bibfnamefont
  {Z.}~\bibnamefont {Wu}}, \bibinfo {author} {\bibfnamefont {Y.}~\bibnamefont
  {Li}}, \bibinfo {author} {\bibfnamefont {H.}~\bibnamefont {Yuan}}, \bibinfo
  {author} {\bibfnamefont {X.}~\bibnamefont {Peng}},\ and\ \bibinfo {author}
  {\bibfnamefont {J.}~\bibnamefont {Du}},\ }\bibfield  {title} {\bibinfo
  {title} {Experimental critical quantum metrology with the {Heisenberg}
  scaling},\ }\href {https://doi.org/10.1038/s41534-021-00507-x} {\bibfield
  {journal} {\bibinfo  {journal} {npj Quantum Inf.}\ }\textbf {\bibinfo
  {volume} {7}},\ \bibinfo {pages} {170} (\bibinfo {year} {2021})}\BibitemShut
  {NoStop}%
\bibitem [{\citenamefont {Ding}\ \emph {et~al.}(2022)\citenamefont {Ding},
  \citenamefont {Liu}, \citenamefont {Shi}, \citenamefont {Guo}, \citenamefont
  {Mølmer},\ and\ \citenamefont {Adams}}]{NP1447}%
  \BibitemOpen
  \bibfield  {author} {\bibinfo {author} {\bibfnamefont {D.-S.}\ \bibnamefont
  {Ding}}, \bibinfo {author} {\bibfnamefont {Z.-K.}\ \bibnamefont {Liu}},
  \bibinfo {author} {\bibfnamefont {B.-S.}\ \bibnamefont {Shi}}, \bibinfo
  {author} {\bibfnamefont {G.-C.}\ \bibnamefont {Guo}}, \bibinfo {author}
  {\bibfnamefont {K.}~\bibnamefont {Mølmer}},\ and\ \bibinfo {author}
  {\bibfnamefont {C.~S.}\ \bibnamefont {Adams}},\ }\bibfield  {title} {\bibinfo
  {title} {Enhanced metrology at the critical point of a many-body {Rydberg}
  atomic system},\ }\href {https://doi.org/10.1038/s41567-022-01777-8}
  {\bibfield  {journal} {\bibinfo  {journal} {Nat. Phys.}\ }\textbf {\bibinfo
  {volume} {18}},\ \bibinfo {pages} {1447} (\bibinfo {year}
  {2022})}\BibitemShut {NoStop}%
\bibitem [{\citenamefont {Raghunandan}\ \emph {et~al.}(2018)\citenamefont
  {Raghunandan}, \citenamefont {Wrachtrup},\ and\ \citenamefont
  {Weimer}}]{PRL150501}%
  \BibitemOpen
  \bibfield  {author} {\bibinfo {author} {\bibfnamefont {M.}~\bibnamefont
  {Raghunandan}}, \bibinfo {author} {\bibfnamefont {J.}~\bibnamefont
  {Wrachtrup}},\ and\ \bibinfo {author} {\bibfnamefont {H.}~\bibnamefont
  {Weimer}},\ }\bibfield  {title} {\bibinfo {title} {High-density quantum
  sensing with dissipative first order transitions},\ }\href
  {https://doi.org/10.1103/PhysRevLett.120.150501} {\bibfield  {journal}
  {\bibinfo  {journal} {Phys. Rev. Lett.}\ }\textbf {\bibinfo {volume} {120}},\
  \bibinfo {pages} {150501} (\bibinfo {year} {2018})}\BibitemShut {NoStop}%
\bibitem [{\citenamefont {Heugel}\ \emph {et~al.}(2019)\citenamefont {Heugel},
  \citenamefont {Biondi}, \citenamefont {Zilberberg},\ and\ \citenamefont
  {Chitra}}]{PRL173601}%
  \BibitemOpen
  \bibfield  {author} {\bibinfo {author} {\bibfnamefont {T.~L.}\ \bibnamefont
  {Heugel}}, \bibinfo {author} {\bibfnamefont {M.}~\bibnamefont {Biondi}},
  \bibinfo {author} {\bibfnamefont {O.}~\bibnamefont {Zilberberg}},\ and\
  \bibinfo {author} {\bibfnamefont {R.}~\bibnamefont {Chitra}},\ }\bibfield
  {title} {\bibinfo {title} {Quantum transducer using a parametric
  driven-dissipative phase transition},\ }\href
  {https://doi.org/10.1103/PhysRevLett.123.173601} {\bibfield  {journal}
  {\bibinfo  {journal} {Phys. Rev. Lett.}\ }\textbf {\bibinfo {volume} {123}},\
  \bibinfo {pages} {173601} (\bibinfo {year} {2019})}\BibitemShut {NoStop}%
\bibitem [{\citenamefont {Fern\'andez-Lorenzo}\ and\ \citenamefont
  {Porras}(2017)}]{PRA013817}%
  \BibitemOpen
  \bibfield  {author} {\bibinfo {author} {\bibfnamefont {S.}~\bibnamefont
  {Fern\'andez-Lorenzo}}\ and\ \bibinfo {author} {\bibfnamefont
  {D.}~\bibnamefont {Porras}},\ }\bibfield  {title} {\bibinfo {title} {Quantum
  sensing close to a dissipative phase transition: Symmetry breaking and
  criticality as metrological resources},\ }\href
  {https://doi.org/10.1103/PhysRevA.96.013817} {\bibfield  {journal} {\bibinfo
  {journal} {Phys. Rev. A}\ }\textbf {\bibinfo {volume} {96}},\ \bibinfo
  {pages} {013817} (\bibinfo {year} {2017})}\BibitemShut {NoStop}%
\bibitem [{\citenamefont {Ilias}\ \emph {et~al.}(2022)\citenamefont {Ilias},
  \citenamefont {Yang}, \citenamefont {Huelga},\ and\ \citenamefont
  {Plenio}}]{PRXQ010354}%
  \BibitemOpen
  \bibfield  {author} {\bibinfo {author} {\bibfnamefont {T.}~\bibnamefont
  {Ilias}}, \bibinfo {author} {\bibfnamefont {D.}~\bibnamefont {Yang}},
  \bibinfo {author} {\bibfnamefont {S.~F.}\ \bibnamefont {Huelga}},\ and\
  \bibinfo {author} {\bibfnamefont {M.~B.}\ \bibnamefont {Plenio}},\ }\bibfield
   {title} {\bibinfo {title} {Criticality-enhanced quantum sensing via
  continuous measurement},\ }\href
  {https://doi.org/10.1103/PRXQuantum.3.010354} {\bibfield  {journal} {\bibinfo
   {journal} {PRX Quantum}\ }\textbf {\bibinfo {volume} {3}},\ \bibinfo {pages}
  {010354} (\bibinfo {year} {2022})}\BibitemShut {NoStop}%
\bibitem [{\citenamefont {Di~Candia}\ \emph {et~al.}(2023)\citenamefont
  {Di~Candia}, \citenamefont {Minganti}, \citenamefont {Petrovnin},
  \citenamefont {Paraoanu},\ and\ \citenamefont {Felicetti}}]{NPJ23}%
  \BibitemOpen
  \bibfield  {author} {\bibinfo {author} {\bibfnamefont {R.}~\bibnamefont
  {Di~Candia}}, \bibinfo {author} {\bibfnamefont {F.}~\bibnamefont {Minganti}},
  \bibinfo {author} {\bibfnamefont {K.~V.}\ \bibnamefont {Petrovnin}}, \bibinfo
  {author} {\bibfnamefont {G.~S.}\ \bibnamefont {Paraoanu}},\ and\ \bibinfo
  {author} {\bibfnamefont {S.}~\bibnamefont {Felicetti}},\ }\bibfield  {title}
  {\bibinfo {title} {Critical parametric quantum sensing},\ }\href
  {https://doi.org/10.1038/s41534-023-00690-z} {\bibfield  {journal} {\bibinfo
  {journal} {npj Quantum Inf.}\ }\textbf {\bibinfo {volume} {9}},\ \bibinfo
  {pages} {23} (\bibinfo {year} {2023})}\BibitemShut {NoStop}%
\bibitem [{\citenamefont {Chu}\ \emph {et~al.}(2021)\citenamefont {Chu},
  \citenamefont {Zhang}, \citenamefont {Yu},\ and\ \citenamefont
  {Cai}}]{PRL010502}%
  \BibitemOpen
  \bibfield  {author} {\bibinfo {author} {\bibfnamefont {Y.}~\bibnamefont
  {Chu}}, \bibinfo {author} {\bibfnamefont {S.}~\bibnamefont {Zhang}}, \bibinfo
  {author} {\bibfnamefont {B.}~\bibnamefont {Yu}},\ and\ \bibinfo {author}
  {\bibfnamefont {J.}~\bibnamefont {Cai}},\ }\bibfield  {title} {\bibinfo
  {title} {Dynamic framework for criticality-enhanced quantum sensing},\ }\href
  {https://doi.org/10.1103/PhysRevLett.126.010502} {\bibfield  {journal}
  {\bibinfo  {journal} {Phys. Rev. Lett.}\ }\textbf {\bibinfo {volume} {126}},\
  \bibinfo {pages} {010502} (\bibinfo {year} {2021})}\BibitemShut {NoStop}%
\bibitem [{\citenamefont {L\"u}\ \emph {et~al.}(2022)\citenamefont {L\"u},
  \citenamefont {Ning}, \citenamefont {Zhu}, \citenamefont {Wu}, \citenamefont
  {Shen}, \citenamefont {Yang},\ and\ \citenamefont {Zheng}}]{PRA062616}%
  \BibitemOpen
  \bibfield  {author} {\bibinfo {author} {\bibfnamefont {J.-H.}\ \bibnamefont
  {L\"u}}, \bibinfo {author} {\bibfnamefont {W.}~\bibnamefont {Ning}}, \bibinfo
  {author} {\bibfnamefont {X.}~\bibnamefont {Zhu}}, \bibinfo {author}
  {\bibfnamefont {F.}~\bibnamefont {Wu}}, \bibinfo {author} {\bibfnamefont
  {L.-T.}\ \bibnamefont {Shen}}, \bibinfo {author} {\bibfnamefont {Z.-B.}\
  \bibnamefont {Yang}},\ and\ \bibinfo {author} {\bibfnamefont {S.-B.}\
  \bibnamefont {Zheng}},\ }\bibfield  {title} {\bibinfo {title} {Critical
  quantum sensing based on the jaynes-cummings model with a squeezing drive},\
  }\href {https://doi.org/10.1103/PhysRevA.106.062616} {\bibfield  {journal}
  {\bibinfo  {journal} {Phys. Rev. A}\ }\textbf {\bibinfo {volume} {106}},\
  \bibinfo {pages} {062616} (\bibinfo {year} {2022})}\BibitemShut {NoStop}%
\bibitem [{\citenamefont {Zhu}\ \emph {et~al.}(2023)\citenamefont {Zhu},
  \citenamefont {Lü}, \citenamefont {Ning}, \citenamefont {Wu}, \citenamefont
  {Shen}, \citenamefont {Yang},\ and\ \citenamefont {Zheng}}]{SCPMA250313}%
  \BibitemOpen
  \bibfield  {author} {\bibinfo {author} {\bibfnamefont {X.}~\bibnamefont
  {Zhu}}, \bibinfo {author} {\bibfnamefont {J.-H.}\ \bibnamefont {Lü}},
  \bibinfo {author} {\bibfnamefont {W.}~\bibnamefont {Ning}}, \bibinfo {author}
  {\bibfnamefont {F.}~\bibnamefont {Wu}}, \bibinfo {author} {\bibfnamefont
  {L.-T.}\ \bibnamefont {Shen}}, \bibinfo {author} {\bibfnamefont {Z.-B.}\
  \bibnamefont {Yang}},\ and\ \bibinfo {author} {\bibfnamefont {S.-B.}\
  \bibnamefont {Zheng}},\ }\bibfield  {title} {\bibinfo {title}
  {Criticality-enhanced quantum sensing in the anisotropic quantum {Rabi}
  model},\ }\href {https://doi.org/10.1007/s11433-022-2073-9} {\bibfield
  {journal} {\bibinfo  {journal} {Sci. China Phys. Mech. Astron.}\ }\textbf
  {\bibinfo {volume} {66}},\ \bibinfo {pages} {250313} (\bibinfo {year}
  {2023})}\BibitemShut {NoStop}%
\bibitem [{\citenamefont {Mihailescu}\ \emph {et~al.}()\citenamefont
  {Mihailescu}, \citenamefont {Bayat}, \citenamefont {Campbell},\ and\
  \citenamefont {Mitchell}}]{arxiv16931}%
  \BibitemOpen
  \bibfield  {author} {\bibinfo {author} {\bibfnamefont {G.}~\bibnamefont
  {Mihailescu}}, \bibinfo {author} {\bibfnamefont {A.}~\bibnamefont {Bayat}},
  \bibinfo {author} {\bibfnamefont {S.}~\bibnamefont {Campbell}},\ and\
  \bibinfo {author} {\bibfnamefont {A.~K.}\ \bibnamefont {Mitchell}},\
  }\href@noop {} {\bibinfo {title} {Multiparameter critical quantum metrology
  with impurity probes}},\ \Eprint {https://arxiv.org/abs/2311.16931}
  {arXiv:2311.16931} \BibitemShut {NoStop}%
\bibitem [{\citenamefont {Gietka}\ \emph {et~al.}(2022)\citenamefont {Gietka},
  \citenamefont {Ruks},\ and\ \citenamefont {Busch}}]{Quantum700}%
  \BibitemOpen
  \bibfield  {author} {\bibinfo {author} {\bibfnamefont {K.}~\bibnamefont
  {Gietka}}, \bibinfo {author} {\bibfnamefont {L.}~\bibnamefont {Ruks}},\ and\
  \bibinfo {author} {\bibfnamefont {T.}~\bibnamefont {Busch}},\ }\bibfield
  {title} {\bibinfo {title} {Understanding and {Improving} {Critical}
  {Metrology}. {Quenching} {Superradiant} {Light}-{Matter} {Systems} {Beyond}
  the {Critical} {Point}},\ }\href {https://doi.org/10.22331/q-2022-04-27-700}
  {\bibfield  {journal} {\bibinfo  {journal} {Quantum}\ }\textbf {\bibinfo
  {volume} {6}},\ \bibinfo {pages} {700} (\bibinfo {year} {2022})}\BibitemShut
  {NoStop}%
\bibitem [{\citenamefont {Jaynes}\ and\ \citenamefont
  {Cummings}(1963)}]{IEEE89}%
  \BibitemOpen
  \bibfield  {author} {\bibinfo {author} {\bibfnamefont {E.}~\bibnamefont
  {Jaynes}}\ and\ \bibinfo {author} {\bibfnamefont {F.}~\bibnamefont
  {Cummings}},\ }\bibfield  {title} {\bibinfo {title} {Comparison of quantum
  and semiclassical radiation theories with application to the beam maser},\
  }\href {https://doi.org/10.1109/PROC.1963.1664} {\bibfield  {journal}
  {\bibinfo  {journal} {Proc. IEEE}\ }\textbf {\bibinfo {volume} {51}},\
  \bibinfo {pages} {89} (\bibinfo {year} {1963})}\BibitemShut {NoStop}%
\bibitem [{\citenamefont {Alsing}\ \emph {et~al.}(1992)\citenamefont {Alsing},
  \citenamefont {Guo},\ and\ \citenamefont {Carmichael}}]{PRA5135}%
  \BibitemOpen
  \bibfield  {author} {\bibinfo {author} {\bibfnamefont {P.}~\bibnamefont
  {Alsing}}, \bibinfo {author} {\bibfnamefont {D.-S.}\ \bibnamefont {Guo}},\
  and\ \bibinfo {author} {\bibfnamefont {H.~J.}\ \bibnamefont {Carmichael}},\
  }\bibfield  {title} {\bibinfo {title} {Dynamic stark effect for the
  jaynes-cummings system},\ }\href {https://doi.org/10.1103/PhysRevA.45.5135}
  {\bibfield  {journal} {\bibinfo  {journal} {Phys. Rev. A}\ }\textbf {\bibinfo
  {volume} {45}},\ \bibinfo {pages} {5135} (\bibinfo {year}
  {1992})}\BibitemShut {NoStop}%
\bibitem [{\citenamefont {Curtis}\ \emph {et~al.}(2021)\citenamefont {Curtis},
  \citenamefont {Boettcher}, \citenamefont {Young}, \citenamefont {Maghrebi},
  \citenamefont {Carmichael}, \citenamefont {Gorshkov},\ and\ \citenamefont
  {Foss-Feig}}]{PRR023062}%
  \BibitemOpen
  \bibfield  {author} {\bibinfo {author} {\bibfnamefont {J.~B.}\ \bibnamefont
  {Curtis}}, \bibinfo {author} {\bibfnamefont {I.}~\bibnamefont {Boettcher}},
  \bibinfo {author} {\bibfnamefont {J.~T.}\ \bibnamefont {Young}}, \bibinfo
  {author} {\bibfnamefont {M.~F.}\ \bibnamefont {Maghrebi}}, \bibinfo {author}
  {\bibfnamefont {H.}~\bibnamefont {Carmichael}}, \bibinfo {author}
  {\bibfnamefont {A.~V.}\ \bibnamefont {Gorshkov}},\ and\ \bibinfo {author}
  {\bibfnamefont {M.}~\bibnamefont {Foss-Feig}},\ }\bibfield  {title} {\bibinfo
  {title} {Critical theory for the breakdown of photon blockade},\ }\href
  {https://doi.org/10.1103/PhysRevResearch.3.023062} {\bibfield  {journal}
  {\bibinfo  {journal} {Phys. Rev. Res.}\ }\textbf {\bibinfo {volume} {3}},\
  \bibinfo {pages} {023062} (\bibinfo {year} {2021})}\BibitemShut {NoStop}%
\bibitem [{\citenamefont {Carmichael}(2015)}]{PRX031028}%
  \BibitemOpen
  \bibfield  {author} {\bibinfo {author} {\bibfnamefont {H.~J.}\ \bibnamefont
  {Carmichael}},\ }\bibfield  {title} {\bibinfo {title} {Breakdown of photon
  blockade: A dissipative quantum phase transition in zero dimensions},\ }\href
  {https://doi.org/10.1103/PhysRevX.5.031028} {\bibfield  {journal} {\bibinfo
  {journal} {Phys. Rev. X}\ }\textbf {\bibinfo {volume} {5}},\ \bibinfo {pages}
  {031028} (\bibinfo {year} {2015})}\BibitemShut {NoStop}%
\bibitem [{\citenamefont {Fink}\ \emph {et~al.}(2017)\citenamefont {Fink},
  \citenamefont {Dombi}, \citenamefont {Vukics}, \citenamefont {Wallraff},\
  and\ \citenamefont {Domokos}}]{PRX011012}%
  \BibitemOpen
  \bibfield  {author} {\bibinfo {author} {\bibfnamefont {J.~M.}\ \bibnamefont
  {Fink}}, \bibinfo {author} {\bibfnamefont {A.}~\bibnamefont {Dombi}},
  \bibinfo {author} {\bibfnamefont {A.}~\bibnamefont {Vukics}}, \bibinfo
  {author} {\bibfnamefont {A.}~\bibnamefont {Wallraff}},\ and\ \bibinfo
  {author} {\bibfnamefont {P.}~\bibnamefont {Domokos}},\ }\bibfield  {title}
  {\bibinfo {title} {Observation of the photon-blockade breakdown phase
  transition},\ }\href {https://doi.org/10.1103/PhysRevX.7.011012} {\bibfield
  {journal} {\bibinfo  {journal} {Phys. Rev. X}\ }\textbf {\bibinfo {volume}
  {7}},\ \bibinfo {pages} {011012} (\bibinfo {year} {2017})}\BibitemShut
  {NoStop}%
\bibitem [{sup()}]{supp}%
  \BibitemOpen
  \href@noop {} {}\bibinfo {note} {See Supplemental Material for theoretical
  and experimental details, which includes Refs.
  [2,27,51-53,55-62].}\BibitemShut {Stop}%
\bibitem [{\citenamefont {Guo}\ \emph {et~al.}(2018)\citenamefont {Guo},
  \citenamefont {Zheng}, \citenamefont {Wang}, \citenamefont {Song},
  \citenamefont {Zhang}, \citenamefont {Li}, \citenamefont {Liu}, \citenamefont
  {Deng}, \citenamefont {Huang}, \citenamefont {Zheng}, \citenamefont {Zhu},
  \citenamefont {Wang}, \citenamefont {Lu},\ and\ \citenamefont
  {Pan}}]{PRL130501}%
  \BibitemOpen
  \bibfield  {author} {\bibinfo {author} {\bibfnamefont {Q.}~\bibnamefont
  {Guo}}, \bibinfo {author} {\bibfnamefont {S.-B.}\ \bibnamefont {Zheng}},
  \bibinfo {author} {\bibfnamefont {J.}~\bibnamefont {Wang}}, \bibinfo {author}
  {\bibfnamefont {C.}~\bibnamefont {Song}}, \bibinfo {author} {\bibfnamefont
  {P.}~\bibnamefont {Zhang}}, \bibinfo {author} {\bibfnamefont
  {K.}~\bibnamefont {Li}}, \bibinfo {author} {\bibfnamefont {W.}~\bibnamefont
  {Liu}}, \bibinfo {author} {\bibfnamefont {H.}~\bibnamefont {Deng}}, \bibinfo
  {author} {\bibfnamefont {K.}~\bibnamefont {Huang}}, \bibinfo {author}
  {\bibfnamefont {D.}~\bibnamefont {Zheng}}, \bibinfo {author} {\bibfnamefont
  {X.}~\bibnamefont {Zhu}}, \bibinfo {author} {\bibfnamefont {H.}~\bibnamefont
  {Wang}}, \bibinfo {author} {\bibfnamefont {C.-Y.}\ \bibnamefont {Lu}},\ and\
  \bibinfo {author} {\bibfnamefont {J.-W.}\ \bibnamefont {Pan}},\ }\bibfield
  {title} {\bibinfo {title} {Dephasing-insensitive quantum information storage
  and processing with superconducting qubits},\ }\href
  {https://doi.org/10.1103/PhysRevLett.121.130501} {\bibfield  {journal}
  {\bibinfo  {journal} {Phys. Rev. Lett.}\ }\textbf {\bibinfo {volume} {121}},\
  \bibinfo {pages} {130501} (\bibinfo {year} {2018})}\BibitemShut {NoStop}%
\bibitem [{\citenamefont {Yan}\ \emph {et~al.}(2018)\citenamefont {Yan},
  \citenamefont {Krantz}, \citenamefont {Sung}, \citenamefont {Kjaergaard},
  \citenamefont {Campbell}, \citenamefont {Orlando}, \citenamefont
  {Gustavsson},\ and\ \citenamefont {Oliver}}]{PRAD054062}%
  \BibitemOpen
  \bibfield  {author} {\bibinfo {author} {\bibfnamefont {F.}~\bibnamefont
  {Yan}}, \bibinfo {author} {\bibfnamefont {P.}~\bibnamefont {Krantz}},
  \bibinfo {author} {\bibfnamefont {Y.}~\bibnamefont {Sung}}, \bibinfo {author}
  {\bibfnamefont {M.}~\bibnamefont {Kjaergaard}}, \bibinfo {author}
  {\bibfnamefont {D.~L.}\ \bibnamefont {Campbell}}, \bibinfo {author}
  {\bibfnamefont {T.~P.}\ \bibnamefont {Orlando}}, \bibinfo {author}
  {\bibfnamefont {S.}~\bibnamefont {Gustavsson}},\ and\ \bibinfo {author}
  {\bibfnamefont {W.~D.}\ \bibnamefont {Oliver}},\ }\bibfield  {title}
  {\bibinfo {title} {Tunable coupling scheme for implementing high-fidelity
  two-qubit gates},\ }\href {https://doi.org/10.1103/PhysRevApplied.10.054062}
  {\bibfield  {journal} {\bibinfo  {journal} {Phys. Rev. Appl.}\ }\textbf
  {\bibinfo {volume} {10}},\ \bibinfo {pages} {054062} (\bibinfo {year}
  {2018})}\BibitemShut {NoStop}%
\bibitem [{\citenamefont {Wineland}(2013)}]{RMP1103}%
  \BibitemOpen
  \bibfield  {author} {\bibinfo {author} {\bibfnamefont {D.~J.}\ \bibnamefont
  {Wineland}},\ }\bibfield  {title} {\bibinfo {title} {Nobel lecture:
  Superposition, entanglement, and raising schr\"odinger's cat},\ }\href
  {https://doi.org/10.1103/RevModPhys.85.1103} {\bibfield  {journal} {\bibinfo
  {journal} {Rev. Mod. Phys.}\ }\textbf {\bibinfo {volume} {85}},\ \bibinfo
  {pages} {1103} (\bibinfo {year} {2013})}\BibitemShut {NoStop}%
\bibitem [{\citenamefont {Nielsen}\ and\ \citenamefont {Chuang}(2011)}]{book}%
  \BibitemOpen
  \bibfield  {author} {\bibinfo {author} {\bibfnamefont {M.~A.}\ \bibnamefont
  {Nielsen}}\ and\ \bibinfo {author} {\bibfnamefont {I.~L.}\ \bibnamefont
  {Chuang}},\ }\href@noop {} {\emph {\bibinfo {title} {Quantum Computation and
  Quantum Information: 10th Anniversary Edition}}},\ \bibinfo {edition} {10th}\
  ed.\ (\bibinfo  {publisher} {Cambridge University Press},\ \bibinfo {address}
  {USA},\ \bibinfo {year} {2011})\BibitemShut {NoStop}%
\bibitem [{\citenamefont {Song}\ \emph
  {et~al.}(2017{\natexlab{a}})\citenamefont {Song}, \citenamefont {Xu},
  \citenamefont {Liu}, \citenamefont {Yang}, \citenamefont {Zheng},
  \citenamefont {Deng}, \citenamefont {Xie}, \citenamefont {Huang},
  \citenamefont {Guo}, \citenamefont {Zhang}, \citenamefont {Zhang},
  \citenamefont {Xu}, \citenamefont {Zheng}, \citenamefont {Zhu}, \citenamefont
  {Wang}, \citenamefont {Chen}, \citenamefont {Lu}, \citenamefont {Han},\ and\
  \citenamefont {Pan}}]{PRL180511}%
  \BibitemOpen
  \bibfield  {author} {\bibinfo {author} {\bibfnamefont {C.}~\bibnamefont
  {Song}}, \bibinfo {author} {\bibfnamefont {K.}~\bibnamefont {Xu}}, \bibinfo
  {author} {\bibfnamefont {W.}~\bibnamefont {Liu}}, \bibinfo {author}
  {\bibfnamefont {C.-p.}\ \bibnamefont {Yang}}, \bibinfo {author}
  {\bibfnamefont {S.-B.}\ \bibnamefont {Zheng}}, \bibinfo {author}
  {\bibfnamefont {H.}~\bibnamefont {Deng}}, \bibinfo {author} {\bibfnamefont
  {Q.}~\bibnamefont {Xie}}, \bibinfo {author} {\bibfnamefont {K.}~\bibnamefont
  {Huang}}, \bibinfo {author} {\bibfnamefont {Q.}~\bibnamefont {Guo}}, \bibinfo
  {author} {\bibfnamefont {L.}~\bibnamefont {Zhang}}, \bibinfo {author}
  {\bibfnamefont {P.}~\bibnamefont {Zhang}}, \bibinfo {author} {\bibfnamefont
  {D.}~\bibnamefont {Xu}}, \bibinfo {author} {\bibfnamefont {D.}~\bibnamefont
  {Zheng}}, \bibinfo {author} {\bibfnamefont {X.}~\bibnamefont {Zhu}}, \bibinfo
  {author} {\bibfnamefont {H.}~\bibnamefont {Wang}}, \bibinfo {author}
  {\bibfnamefont {Y.-A.}\ \bibnamefont {Chen}}, \bibinfo {author}
  {\bibfnamefont {C.-Y.}\ \bibnamefont {Lu}}, \bibinfo {author} {\bibfnamefont
  {S.}~\bibnamefont {Han}},\ and\ \bibinfo {author} {\bibfnamefont {J.-W.}\
  \bibnamefont {Pan}},\ }\bibfield  {title} {\bibinfo {title} {10-qubit
  entanglement and parallel logic operations with a superconducting circuit},\
  }\href {https://doi.org/10.1103/PhysRevLett.119.180511} {\bibfield  {journal}
  {\bibinfo  {journal} {Phys. Rev. Lett.}\ }\textbf {\bibinfo {volume} {119}},\
  \bibinfo {pages} {180511} (\bibinfo {year} {2017}{\natexlab{a}})}\BibitemShut
  {NoStop}%
\bibitem [{\citenamefont {Vlastakis}\ \emph {et~al.}(2013)\citenamefont
  {Vlastakis}, \citenamefont {Kirchmair}, \citenamefont {Leghtas},
  \citenamefont {Nigg}, \citenamefont {Frunzio}, \citenamefont {Girvin},
  \citenamefont {Mirrahimi}, \citenamefont {Devoret},\ and\ \citenamefont
  {Schoelkopf}}]{science607}%
  \BibitemOpen
  \bibfield  {author} {\bibinfo {author} {\bibfnamefont {B.}~\bibnamefont
  {Vlastakis}}, \bibinfo {author} {\bibfnamefont {G.}~\bibnamefont
  {Kirchmair}}, \bibinfo {author} {\bibfnamefont {Z.}~\bibnamefont {Leghtas}},
  \bibinfo {author} {\bibfnamefont {S.~E.}\ \bibnamefont {Nigg}}, \bibinfo
  {author} {\bibfnamefont {L.}~\bibnamefont {Frunzio}}, \bibinfo {author}
  {\bibfnamefont {S.~M.}\ \bibnamefont {Girvin}}, \bibinfo {author}
  {\bibfnamefont {M.}~\bibnamefont {Mirrahimi}}, \bibinfo {author}
  {\bibfnamefont {M.~H.}\ \bibnamefont {Devoret}},\ and\ \bibinfo {author}
  {\bibfnamefont {R.~J.}\ \bibnamefont {Schoelkopf}},\ }\bibfield  {title}
  {\bibinfo {title} {Deterministically {Encoding} {Quantum} {Information}
  {Using} 100-{Photon} {Schrödinger} {Cat} {States}},\ }\href
  {https://doi.org/10.1126/science.1243289} {\bibfield  {journal} {\bibinfo
  {journal} {Science}\ }\textbf {\bibinfo {volume} {342}},\ \bibinfo {pages}
  {607} (\bibinfo {year} {2013})}\BibitemShut {NoStop}%
\bibitem [{\citenamefont {Song}\ \emph
  {et~al.}(2017{\natexlab{b}})\citenamefont {Song}, \citenamefont {Zheng},
  \citenamefont {Zhang}, \citenamefont {Xu}, \citenamefont {Zhang},
  \citenamefont {Guo}, \citenamefont {Liu}, \citenamefont {Xu}, \citenamefont
  {Deng}, \citenamefont {Huang}, \citenamefont {Zheng}, \citenamefont {Zhu},\
  and\ \citenamefont {Wang}}]{2}%
  \BibitemOpen
  \bibfield  {author} {\bibinfo {author} {\bibfnamefont {C.}~\bibnamefont
  {Song}}, \bibinfo {author} {\bibfnamefont {S.-B.}\ \bibnamefont {Zheng}},
  \bibinfo {author} {\bibfnamefont {P.}~\bibnamefont {Zhang}}, \bibinfo
  {author} {\bibfnamefont {K.}~\bibnamefont {Xu}}, \bibinfo {author}
  {\bibfnamefont {L.}~\bibnamefont {Zhang}}, \bibinfo {author} {\bibfnamefont
  {Q.}~\bibnamefont {Guo}}, \bibinfo {author} {\bibfnamefont {W.}~\bibnamefont
  {Liu}}, \bibinfo {author} {\bibfnamefont {D.}~\bibnamefont {Xu}}, \bibinfo
  {author} {\bibfnamefont {H.}~\bibnamefont {Deng}}, \bibinfo {author}
  {\bibfnamefont {K.}~\bibnamefont {Huang}}, \bibinfo {author} {\bibfnamefont
  {D.}~\bibnamefont {Zheng}}, \bibinfo {author} {\bibfnamefont
  {X.}~\bibnamefont {Zhu}},\ and\ \bibinfo {author} {\bibfnamefont
  {H.}~\bibnamefont {Wang}},\ }\bibfield  {title} {\bibinfo {title}
  {Continuous-variable geometric phase and its manipulation for quantum
  computation in a superconducting circuit},\ }\href
  {https://doi.org/10.1038/s41467-017-01156-5} {\bibfield  {journal} {\bibinfo
  {journal} {Nat. Commun.}\ }\textbf {\bibinfo {volume} {8}},\ \bibinfo {pages}
  {1061} (\bibinfo {year} {2017}{\natexlab{b}})}\BibitemShut {NoStop}%
\bibitem [{\citenamefont {Ning}\ \emph {et~al.}(2019)\citenamefont {Ning},
  \citenamefont {Huang}, \citenamefont {Han}, \citenamefont {Li}, \citenamefont
  {Deng}, \citenamefont {Yang}, \citenamefont {Zhong}, \citenamefont {Xia},
  \citenamefont {Xu}, \citenamefont {Zheng},\ and\ \citenamefont {Zheng}}]{3}%
  \BibitemOpen
  \bibfield  {author} {\bibinfo {author} {\bibfnamefont {W.}~\bibnamefont
  {Ning}}, \bibinfo {author} {\bibfnamefont {X.-J.}\ \bibnamefont {Huang}},
  \bibinfo {author} {\bibfnamefont {P.-R.}\ \bibnamefont {Han}}, \bibinfo
  {author} {\bibfnamefont {H.}~\bibnamefont {Li}}, \bibinfo {author}
  {\bibfnamefont {H.}~\bibnamefont {Deng}}, \bibinfo {author} {\bibfnamefont
  {Z.-B.}\ \bibnamefont {Yang}}, \bibinfo {author} {\bibfnamefont {Z.-R.}\
  \bibnamefont {Zhong}}, \bibinfo {author} {\bibfnamefont {Y.}~\bibnamefont
  {Xia}}, \bibinfo {author} {\bibfnamefont {K.}~\bibnamefont {Xu}}, \bibinfo
  {author} {\bibfnamefont {D.}~\bibnamefont {Zheng}},\ and\ \bibinfo {author}
  {\bibfnamefont {S.-B.}\ \bibnamefont {Zheng}},\ }\bibfield  {title} {\bibinfo
  {title} {Deterministic entanglement swapping in a superconducting circuit},\
  }\href {https://doi.org/10.1103/PhysRevLett.123.060502} {\bibfield  {journal}
  {\bibinfo  {journal} {Phys. Rev. Lett.}\ }\textbf {\bibinfo {volume} {123}},\
  \bibinfo {pages} {060502} (\bibinfo {year} {2019})}\BibitemShut {NoStop}%
\bibitem [{\citenamefont {Zheng}\ \emph {et~al.}(2023)\citenamefont {Zheng},
  \citenamefont {Ning}, \citenamefont {Chen}, \citenamefont {L\"u},
  \citenamefont {Shen}, \citenamefont {Xu}, \citenamefont {Zhang},
  \citenamefont {Xu}, \citenamefont {Li}, \citenamefont {Xia}, \citenamefont
  {Wu}, \citenamefont {Yang}, \citenamefont {Miranowicz}, \citenamefont
  {Lambert}, \citenamefont {Zheng}, \citenamefont {Fan}, \citenamefont {Nori},\
  and\ \citenamefont {Zheng}}]{PRL113601}%
  \BibitemOpen
  \bibfield  {author} {\bibinfo {author} {\bibfnamefont {R.-H.}\ \bibnamefont
  {Zheng}}, \bibinfo {author} {\bibfnamefont {W.}~\bibnamefont {Ning}},
  \bibinfo {author} {\bibfnamefont {Y.-H.}\ \bibnamefont {Chen}}, \bibinfo
  {author} {\bibfnamefont {J.-H.}\ \bibnamefont {L\"u}}, \bibinfo {author}
  {\bibfnamefont {L.-T.}\ \bibnamefont {Shen}}, \bibinfo {author}
  {\bibfnamefont {K.}~\bibnamefont {Xu}}, \bibinfo {author} {\bibfnamefont
  {Y.-R.}\ \bibnamefont {Zhang}}, \bibinfo {author} {\bibfnamefont
  {D.}~\bibnamefont {Xu}}, \bibinfo {author} {\bibfnamefont {H.}~\bibnamefont
  {Li}}, \bibinfo {author} {\bibfnamefont {Y.}~\bibnamefont {Xia}}, \bibinfo
  {author} {\bibfnamefont {F.}~\bibnamefont {Wu}}, \bibinfo {author}
  {\bibfnamefont {Z.-B.}\ \bibnamefont {Yang}}, \bibinfo {author}
  {\bibfnamefont {A.}~\bibnamefont {Miranowicz}}, \bibinfo {author}
  {\bibfnamefont {N.}~\bibnamefont {Lambert}}, \bibinfo {author} {\bibfnamefont
  {D.}~\bibnamefont {Zheng}}, \bibinfo {author} {\bibfnamefont
  {H.}~\bibnamefont {Fan}}, \bibinfo {author} {\bibfnamefont {F.}~\bibnamefont
  {Nori}},\ and\ \bibinfo {author} {\bibfnamefont {S.-B.}\ \bibnamefont
  {Zheng}},\ }\bibfield  {title} {\bibinfo {title} {Observation of a
  superradiant phase transition with emergent cat states},\ }\href
  {https://doi.org/10.1103/PhysRevLett.131.113601} {\bibfield  {journal}
  {\bibinfo  {journal} {Phys. Rev. Lett.}\ }\textbf {\bibinfo {volume} {131}},\
  \bibinfo {pages} {113601} (\bibinfo {year} {2023})}\BibitemShut {NoStop}%
\bibitem [{\citenamefont {Xu}\ \emph {et~al.}(2020)\citenamefont {Xu},
  \citenamefont {Sun}, \citenamefont {Liu}, \citenamefont {Zhang},
  \citenamefont {Li}, \citenamefont {Dong}, \citenamefont {Ren}, \citenamefont
  {Zhang}, \citenamefont {Nori}, \citenamefont {Zheng}, \citenamefont {Fan},\
  and\ \citenamefont {Wang}}]{5}%
  \BibitemOpen
  \bibfield  {author} {\bibinfo {author} {\bibfnamefont {K.}~\bibnamefont
  {Xu}}, \bibinfo {author} {\bibfnamefont {Z.-H.}\ \bibnamefont {Sun}},
  \bibinfo {author} {\bibfnamefont {W.}~\bibnamefont {Liu}}, \bibinfo {author}
  {\bibfnamefont {Y.-R.}\ \bibnamefont {Zhang}}, \bibinfo {author}
  {\bibfnamefont {H.}~\bibnamefont {Li}}, \bibinfo {author} {\bibfnamefont
  {H.}~\bibnamefont {Dong}}, \bibinfo {author} {\bibfnamefont {W.}~\bibnamefont
  {Ren}}, \bibinfo {author} {\bibfnamefont {P.}~\bibnamefont {Zhang}}, \bibinfo
  {author} {\bibfnamefont {F.}~\bibnamefont {Nori}}, \bibinfo {author}
  {\bibfnamefont {D.}~\bibnamefont {Zheng}}, \bibinfo {author} {\bibfnamefont
  {H.}~\bibnamefont {Fan}},\ and\ \bibinfo {author} {\bibfnamefont
  {H.}~\bibnamefont {Wang}},\ }\bibfield  {title} {\bibinfo {title} {Probing
  dynamical phase transitions with a superconducting quantum simulator},\
  }\href {https://doi.org/10.1126/sciadv.aba4935} {\bibfield  {journal}
  {\bibinfo  {journal} {Sci. Adv.}\ }\textbf {\bibinfo {volume} {6}},\ \bibinfo
  {pages} {eaba4935} (\bibinfo {year} {2020})}\BibitemShut {NoStop}%
\bibitem [{\citenamefont {Hofheinz}\ \emph {et~al.}(2009)\citenamefont
  {Hofheinz}, \citenamefont {Wang}, \citenamefont {Ansmann}, \citenamefont
  {Bialczak}, \citenamefont {Lucero}, \citenamefont {Neeley}, \citenamefont
  {O'Connell}, \citenamefont {Sank}, \citenamefont {Wenner}, \citenamefont
  {Martinis},\ and\ \citenamefont {Cleland}}]{6}%
  \BibitemOpen
  \bibfield  {author} {\bibinfo {author} {\bibfnamefont {M.}~\bibnamefont
  {Hofheinz}}, \bibinfo {author} {\bibfnamefont {H.}~\bibnamefont {Wang}},
  \bibinfo {author} {\bibfnamefont {M.}~\bibnamefont {Ansmann}}, \bibinfo
  {author} {\bibfnamefont {R.~C.}\ \bibnamefont {Bialczak}}, \bibinfo {author}
  {\bibfnamefont {E.}~\bibnamefont {Lucero}}, \bibinfo {author} {\bibfnamefont
  {M.}~\bibnamefont {Neeley}}, \bibinfo {author} {\bibfnamefont {A.~D.}\
  \bibnamefont {O'Connell}}, \bibinfo {author} {\bibfnamefont {D.}~\bibnamefont
  {Sank}}, \bibinfo {author} {\bibfnamefont {J.}~\bibnamefont {Wenner}},
  \bibinfo {author} {\bibfnamefont {J.~M.}\ \bibnamefont {Martinis}},\ and\
  \bibinfo {author} {\bibfnamefont {A.~N.}\ \bibnamefont {Cleland}},\
  }\bibfield  {title} {\bibinfo {title} {Synthesizing arbitrary quantum states
  in a superconducting resonator},\ }\href
  {https://doi.org/10.1038/nature08005} {\bibfield  {journal} {\bibinfo
  {journal} {Nature}\ }\textbf {\bibinfo {volume} {459}},\ \bibinfo {pages}
  {546} (\bibinfo {year} {2009})}\BibitemShut {NoStop}%
\bibitem [{\citenamefont {Monroe}\ \emph {et~al.}(1995)\citenamefont {Monroe},
  \citenamefont {Meekhof}, \citenamefont {King}, \citenamefont {Itano},\ and\
  \citenamefont {Wineland}}]{PRL4714}%
  \BibitemOpen
  \bibfield  {author} {\bibinfo {author} {\bibfnamefont {C.}~\bibnamefont
  {Monroe}}, \bibinfo {author} {\bibfnamefont {D.~M.}\ \bibnamefont {Meekhof}},
  \bibinfo {author} {\bibfnamefont {B.~E.}\ \bibnamefont {King}}, \bibinfo
  {author} {\bibfnamefont {W.~M.}\ \bibnamefont {Itano}},\ and\ \bibinfo
  {author} {\bibfnamefont {D.~J.}\ \bibnamefont {Wineland}},\ }\bibfield
  {title} {\bibinfo {title} {Demonstration of a fundamental quantum logic
  gate},\ }\href {https://doi.org/10.1103/PhysRevLett.75.4714} {\bibfield
  {journal} {\bibinfo  {journal} {Phys. Rev. Lett.}\ }\textbf {\bibinfo
  {volume} {75}},\ \bibinfo {pages} {4714} (\bibinfo {year}
  {1995})}\BibitemShut {NoStop}%
\bibitem [{\citenamefont {Meekhof}\ \emph {et~al.}(1996)\citenamefont
  {Meekhof}, \citenamefont {Monroe}, \citenamefont {King}, \citenamefont
  {Itano},\ and\ \citenamefont {Wineland}}]{PRL1796}%
  \BibitemOpen
  \bibfield  {author} {\bibinfo {author} {\bibfnamefont {D.~M.}\ \bibnamefont
  {Meekhof}}, \bibinfo {author} {\bibfnamefont {C.}~\bibnamefont {Monroe}},
  \bibinfo {author} {\bibfnamefont {B.~E.}\ \bibnamefont {King}}, \bibinfo
  {author} {\bibfnamefont {W.~M.}\ \bibnamefont {Itano}},\ and\ \bibinfo
  {author} {\bibfnamefont {D.~J.}\ \bibnamefont {Wineland}},\ }\bibfield
  {title} {\bibinfo {title} {Generation of nonclassical motional states of a
  trapped atom},\ }\href {https://doi.org/10.1103/PhysRevLett.76.1796}
  {\bibfield  {journal} {\bibinfo  {journal} {Phys. Rev. Lett.}\ }\textbf
  {\bibinfo {volume} {76}},\ \bibinfo {pages} {1796} (\bibinfo {year}
  {1996})}\BibitemShut {NoStop}%
\bibitem [{\citenamefont {Mohseni}\ \emph {et~al.}(2008)\citenamefont
  {Mohseni}, \citenamefont {Rezakhani},\ and\ \citenamefont
  {Lidar}}]{PRA032322}%
  \BibitemOpen
  \bibfield  {author} {\bibinfo {author} {\bibfnamefont {M.}~\bibnamefont
  {Mohseni}}, \bibinfo {author} {\bibfnamefont {A.~T.}\ \bibnamefont
  {Rezakhani}},\ and\ \bibinfo {author} {\bibfnamefont {D.~A.}\ \bibnamefont
  {Lidar}},\ }\bibfield  {title} {\bibinfo {title} {Quantum-process tomography:
  Resource analysis of different strategies},\ }\href
  {https://doi.org/10.1103/PhysRevA.77.032322} {\bibfield  {journal} {\bibinfo
  {journal} {Phys. Rev. A}\ }\textbf {\bibinfo {volume} {77}},\ \bibinfo
  {pages} {032322} (\bibinfo {year} {2008})}\BibitemShut {NoStop}%
\end{thebibliography}%


\begin{thebibliography}{13}%
\makeatletter
\providecommand \@ifxundefined [1]{%
 \@ifx{#1\undefined}
}%
\providecommand \@ifnum [1]{%
 \ifnum #1\expandafter \@firstoftwo
 \else \expandafter \@secondoftwo
 \fi
}%
\providecommand \@ifx [1]{%
 \ifx #1\expandafter \@firstoftwo
 \else \expandafter \@secondoftwo
 \fi
}%
\providecommand \natexlab [1]{#1}%
\providecommand \enquote  [1]{``#1''}%
\providecommand \bibnamefont  [1]{#1}%
\providecommand \bibfnamefont [1]{#1}%
\providecommand \citenamefont [1]{#1}%
\providecommand \href@noop [0]{\@secondoftwo}%
\providecommand \href [0]{\begingroup \@sanitize@url \@href}%
\providecommand \@href[1]{\@@startlink{#1}\@@href}%
\providecommand \@@href[1]{\endgroup#1\@@endlink}%
\providecommand \@sanitize@url [0]{\catcode `\\12\catcode `\$12\catcode
  `\&12\catcode `\#12\catcode `\^12\catcode `\_12\catcode `\%12\relax}%
\providecommand \@@startlink[1]{}%
\providecommand \@@endlink[0]{}%
\providecommand \url  [0]{\begingroup\@sanitize@url \@url }%
\providecommand \@url [1]{\endgroup\@href {#1}{\urlprefix }}%
\providecommand \urlprefix  [0]{URL }%
\providecommand \Eprint [0]{\href }%
\providecommand \doibase [0]{https://doi.org/}%
\providecommand \selectlanguage [0]{\@gobble}%
\providecommand \bibinfo  [0]{\@secondoftwo}%
\providecommand \bibfield  [0]{\@secondoftwo}%
\providecommand \translation [1]{[#1]}%
\providecommand \BibitemOpen [0]{}%
\providecommand \bibitemStop [0]{}%
\providecommand \bibitemNoStop [0]{.\EOS\space}%
\providecommand \EOS [0]{\spacefactor3000\relax}%
\providecommand \BibitemShut  [1]{\csname bibitem#1\endcsname}%
\let\auto@bib@innerbib\@empty
\bibitem [{\citenamefont {Garbe}\ \emph {et~al.}(2020)\citenamefont {Garbe},
  \citenamefont {Bina}, \citenamefont {Keller}, \citenamefont {Paris},\ and\
  \citenamefont {Felicetti}}]{1}%
  \BibitemOpen
  \bibfield  {author} {\bibinfo {author} {\bibfnamefont {L.}~\bibnamefont
  {Garbe}}, \bibinfo {author} {\bibfnamefont {M.}~\bibnamefont {Bina}},
  \bibinfo {author} {\bibfnamefont {A.}~\bibnamefont {Keller}}, \bibinfo
  {author} {\bibfnamefont {M.~G.~A.}\ \bibnamefont {Paris}},\ and\ \bibinfo
  {author} {\bibfnamefont {S.}~\bibnamefont {Felicetti}},\ }\bibfield  {title}
  {\bibinfo {title} {Critical quantum metrology with a finite-component quantum
  phase transition},\ }\href {https://doi.org/10.1103/PhysRevLett.124.120504}
  {\bibfield  {journal} {\bibinfo  {journal} {Phys. Rev. Lett.}\ }\textbf
  {\bibinfo {volume} {124}},\ \bibinfo {pages} {120504} (\bibinfo {year}
  {2020})}\BibitemShut {NoStop}%
\bibitem [{\citenamefont {Song}\ \emph
  {et~al.}(2017{\natexlab{a}})\citenamefont {Song}, \citenamefont {Zheng},
  \citenamefont {Zhang}, \citenamefont {Xu}, \citenamefont {Zhang},
  \citenamefont {Guo}, \citenamefont {Liu}, \citenamefont {Xu}, \citenamefont
  {Deng}, \citenamefont {Huang}, \citenamefont {Zheng}, \citenamefont {Zhu},\
  and\ \citenamefont {Wang}}]{2}%
  \BibitemOpen
  \bibfield  {author} {\bibinfo {author} {\bibfnamefont {C.}~\bibnamefont
  {Song}}, \bibinfo {author} {\bibfnamefont {S.-B.}\ \bibnamefont {Zheng}},
  \bibinfo {author} {\bibfnamefont {P.}~\bibnamefont {Zhang}}, \bibinfo
  {author} {\bibfnamefont {K.}~\bibnamefont {Xu}}, \bibinfo {author}
  {\bibfnamefont {L.}~\bibnamefont {Zhang}}, \bibinfo {author} {\bibfnamefont
  {Q.}~\bibnamefont {Guo}}, \bibinfo {author} {\bibfnamefont {W.}~\bibnamefont
  {Liu}}, \bibinfo {author} {\bibfnamefont {D.}~\bibnamefont {Xu}}, \bibinfo
  {author} {\bibfnamefont {H.}~\bibnamefont {Deng}}, \bibinfo {author}
  {\bibfnamefont {K.}~\bibnamefont {Huang}}, \bibinfo {author} {\bibfnamefont
  {D.}~\bibnamefont {Zheng}}, \bibinfo {author} {\bibfnamefont
  {X.}~\bibnamefont {Zhu}},\ and\ \bibinfo {author} {\bibfnamefont
  {H.}~\bibnamefont {Wang}},\ }\bibfield  {title} {\bibinfo {title}
  {Continuous-variable geometric phase and its manipulation for quantum
  computation in a superconducting circuit},\ }\href
  {https://doi.org/10.1038/s41467-017-01156-5} {\bibfield  {journal} {\bibinfo
  {journal} {Nat. Commun.}\ }\textbf {\bibinfo {volume} {8}},\ \bibinfo {pages}
  {1061} (\bibinfo {year} {2017}{\natexlab{a}})}\BibitemShut {NoStop}%
\bibitem [{\citenamefont {Ning}\ \emph {et~al.}(2019)\citenamefont {Ning},
  \citenamefont {Huang}, \citenamefont {Han}, \citenamefont {Li}, \citenamefont
  {Deng}, \citenamefont {Yang}, \citenamefont {Zhong}, \citenamefont {Xia},
  \citenamefont {Xu}, \citenamefont {Zheng},\ and\ \citenamefont {Zheng}}]{3}%
  \BibitemOpen
  \bibfield  {author} {\bibinfo {author} {\bibfnamefont {W.}~\bibnamefont
  {Ning}}, \bibinfo {author} {\bibfnamefont {X.-J.}\ \bibnamefont {Huang}},
  \bibinfo {author} {\bibfnamefont {P.-R.}\ \bibnamefont {Han}}, \bibinfo
  {author} {\bibfnamefont {H.}~\bibnamefont {Li}}, \bibinfo {author}
  {\bibfnamefont {H.}~\bibnamefont {Deng}}, \bibinfo {author} {\bibfnamefont
  {Z.-B.}\ \bibnamefont {Yang}}, \bibinfo {author} {\bibfnamefont {Z.-R.}\
  \bibnamefont {Zhong}}, \bibinfo {author} {\bibfnamefont {Y.}~\bibnamefont
  {Xia}}, \bibinfo {author} {\bibfnamefont {K.}~\bibnamefont {Xu}}, \bibinfo
  {author} {\bibfnamefont {D.}~\bibnamefont {Zheng}},\ and\ \bibinfo {author}
  {\bibfnamefont {S.-B.}\ \bibnamefont {Zheng}},\ }\bibfield  {title} {\bibinfo
  {title} {Deterministic entanglement swapping in a superconducting circuit},\
  }\href {https://doi.org/10.1103/PhysRevLett.123.060502} {\bibfield  {journal}
  {\bibinfo  {journal} {Phys. Rev. Lett.}\ }\textbf {\bibinfo {volume} {123}},\
  \bibinfo {pages} {060502} (\bibinfo {year} {2019})}\BibitemShut {NoStop}%
\bibitem [{\citenamefont {Zheng}\ \emph {et~al.}(2023)\citenamefont {Zheng},
  \citenamefont {Ning}, \citenamefont {Chen}, \citenamefont {L\"u},
  \citenamefont {Shen}, \citenamefont {Xu}, \citenamefont {Zhang},
  \citenamefont {Xu}, \citenamefont {Li}, \citenamefont {Xia}, \citenamefont
  {Wu}, \citenamefont {Yang}, \citenamefont {Miranowicz}, \citenamefont
  {Lambert}, \citenamefont {Zheng}, \citenamefont {Fan}, \citenamefont {Nori},\
  and\ \citenamefont {Zheng}}]{4}%
  \BibitemOpen
  \bibfield  {author} {\bibinfo {author} {\bibfnamefont {R.-H.}\ \bibnamefont
  {Zheng}}, \bibinfo {author} {\bibfnamefont {W.}~\bibnamefont {Ning}},
  \bibinfo {author} {\bibfnamefont {Y.-H.}\ \bibnamefont {Chen}}, \bibinfo
  {author} {\bibfnamefont {J.-H.}\ \bibnamefont {L\"u}}, \bibinfo {author}
  {\bibfnamefont {L.-T.}\ \bibnamefont {Shen}}, \bibinfo {author}
  {\bibfnamefont {K.}~\bibnamefont {Xu}}, \bibinfo {author} {\bibfnamefont
  {Y.-R.}\ \bibnamefont {Zhang}}, \bibinfo {author} {\bibfnamefont
  {D.}~\bibnamefont {Xu}}, \bibinfo {author} {\bibfnamefont {H.}~\bibnamefont
  {Li}}, \bibinfo {author} {\bibfnamefont {Y.}~\bibnamefont {Xia}}, \bibinfo
  {author} {\bibfnamefont {F.}~\bibnamefont {Wu}}, \bibinfo {author}
  {\bibfnamefont {Z.-B.}\ \bibnamefont {Yang}}, \bibinfo {author}
  {\bibfnamefont {A.}~\bibnamefont {Miranowicz}}, \bibinfo {author}
  {\bibfnamefont {N.}~\bibnamefont {Lambert}}, \bibinfo {author} {\bibfnamefont
  {D.}~\bibnamefont {Zheng}}, \bibinfo {author} {\bibfnamefont
  {H.}~\bibnamefont {Fan}}, \bibinfo {author} {\bibfnamefont {F.}~\bibnamefont
  {Nori}},\ and\ \bibinfo {author} {\bibfnamefont {S.-B.}\ \bibnamefont
  {Zheng}},\ }\bibfield  {title} {\bibinfo {title} {Observation of a
  superradiant phase transition with emergent cat states},\ }\href
  {https://doi.org/10.1103/PhysRevLett.131.113601} {\bibfield  {journal}
  {\bibinfo  {journal} {Phys. Rev. Lett.}\ }\textbf {\bibinfo {volume} {131}},\
  \bibinfo {pages} {113601} (\bibinfo {year} {2023})}\BibitemShut {NoStop}%
\bibitem [{\citenamefont {Xu}\ \emph {et~al.}(2020)\citenamefont {Xu},
  \citenamefont {Sun}, \citenamefont {Liu}, \citenamefont {Zhang},
  \citenamefont {Li}, \citenamefont {Dong}, \citenamefont {Ren}, \citenamefont
  {Zhang}, \citenamefont {Nori}, \citenamefont {Zheng}, \citenamefont {Fan},\
  and\ \citenamefont {Wang}}]{5}%
  \BibitemOpen
  \bibfield  {author} {\bibinfo {author} {\bibfnamefont {K.}~\bibnamefont
  {Xu}}, \bibinfo {author} {\bibfnamefont {Z.-H.}\ \bibnamefont {Sun}},
  \bibinfo {author} {\bibfnamefont {W.}~\bibnamefont {Liu}}, \bibinfo {author}
  {\bibfnamefont {Y.-R.}\ \bibnamefont {Zhang}}, \bibinfo {author}
  {\bibfnamefont {H.}~\bibnamefont {Li}}, \bibinfo {author} {\bibfnamefont
  {H.}~\bibnamefont {Dong}}, \bibinfo {author} {\bibfnamefont {W.}~\bibnamefont
  {Ren}}, \bibinfo {author} {\bibfnamefont {P.}~\bibnamefont {Zhang}}, \bibinfo
  {author} {\bibfnamefont {F.}~\bibnamefont {Nori}}, \bibinfo {author}
  {\bibfnamefont {D.}~\bibnamefont {Zheng}}, \bibinfo {author} {\bibfnamefont
  {H.}~\bibnamefont {Fan}},\ and\ \bibinfo {author} {\bibfnamefont
  {H.}~\bibnamefont {Wang}},\ }\bibfield  {title} {\bibinfo {title} {Probing
  dynamical phase transitions with a superconducting quantum simulator},\
  }\href {https://doi.org/10.1126/sciadv.aba4935} {\bibfield  {journal}
  {\bibinfo  {journal} {Sci. Adv.}\ }\textbf {\bibinfo {volume} {6}},\ \bibinfo
  {pages} {eaba4935} (\bibinfo {year} {2020})}\BibitemShut {NoStop}%
\bibitem [{\citenamefont {Hofheinz}\ \emph {et~al.}(2009)\citenamefont
  {Hofheinz}, \citenamefont {Wang}, \citenamefont {Ansmann}, \citenamefont
  {Bialczak}, \citenamefont {Lucero}, \citenamefont {Neeley}, \citenamefont
  {O'Connell}, \citenamefont {Sank}, \citenamefont {Wenner}, \citenamefont
  {Martinis},\ and\ \citenamefont {Cleland}}]{6}%
  \BibitemOpen
  \bibfield  {author} {\bibinfo {author} {\bibfnamefont {M.}~\bibnamefont
  {Hofheinz}}, \bibinfo {author} {\bibfnamefont {H.}~\bibnamefont {Wang}},
  \bibinfo {author} {\bibfnamefont {M.}~\bibnamefont {Ansmann}}, \bibinfo
  {author} {\bibfnamefont {R.~C.}\ \bibnamefont {Bialczak}}, \bibinfo {author}
  {\bibfnamefont {E.}~\bibnamefont {Lucero}}, \bibinfo {author} {\bibfnamefont
  {M.}~\bibnamefont {Neeley}}, \bibinfo {author} {\bibfnamefont {A.~D.}\
  \bibnamefont {O'Connell}}, \bibinfo {author} {\bibfnamefont {D.}~\bibnamefont
  {Sank}}, \bibinfo {author} {\bibfnamefont {J.}~\bibnamefont {Wenner}},
  \bibinfo {author} {\bibfnamefont {J.~M.}\ \bibnamefont {Martinis}},\ and\
  \bibinfo {author} {\bibfnamefont {A.~N.}\ \bibnamefont {Cleland}},\
  }\bibfield  {title} {\bibinfo {title} {Synthesizing arbitrary quantum states
  in a superconducting resonator},\ }\href
  {https://doi.org/10.1038/nature08005} {\bibfield  {journal} {\bibinfo
  {journal} {Nature}\ }\textbf {\bibinfo {volume} {459}},\ \bibinfo {pages}
  {546} (\bibinfo {year} {2009})}\BibitemShut {NoStop}%
\bibitem [{\citenamefont {Degen}\ \emph {et~al.}(2017)\citenamefont {Degen},
  \citenamefont {Reinhard},\ and\ \citenamefont {Cappellaro}}]{RMP035002}%
  \BibitemOpen
  \bibfield  {author} {\bibinfo {author} {\bibfnamefont {C.~L.}\ \bibnamefont
  {Degen}}, \bibinfo {author} {\bibfnamefont {F.}~\bibnamefont {Reinhard}},\
  and\ \bibinfo {author} {\bibfnamefont {P.}~\bibnamefont {Cappellaro}},\
  }\bibfield  {title} {\bibinfo {title} {Quantum sensing},\ }\href
  {https://doi.org/10.1103/RevModPhys.89.035002} {\bibfield  {journal}
  {\bibinfo  {journal} {Rev. Mod. Phys.}\ }\textbf {\bibinfo {volume} {89}},\
  \bibinfo {pages} {035002} (\bibinfo {year} {2017})}\BibitemShut {NoStop}%
\bibitem [{\citenamefont {Monroe}\ \emph {et~al.}(1995)\citenamefont {Monroe},
  \citenamefont {Meekhof}, \citenamefont {King}, \citenamefont {Itano},\ and\
  \citenamefont {Wineland}}]{PRL4714}%
  \BibitemOpen
  \bibfield  {author} {\bibinfo {author} {\bibfnamefont {C.}~\bibnamefont
  {Monroe}}, \bibinfo {author} {\bibfnamefont {D.~M.}\ \bibnamefont {Meekhof}},
  \bibinfo {author} {\bibfnamefont {B.~E.}\ \bibnamefont {King}}, \bibinfo
  {author} {\bibfnamefont {W.~M.}\ \bibnamefont {Itano}},\ and\ \bibinfo
  {author} {\bibfnamefont {D.~J.}\ \bibnamefont {Wineland}},\ }\bibfield
  {title} {\bibinfo {title} {Demonstration of a fundamental quantum logic
  gate},\ }\href {https://doi.org/10.1103/PhysRevLett.75.4714} {\bibfield
  {journal} {\bibinfo  {journal} {Phys. Rev. Lett.}\ }\textbf {\bibinfo
  {volume} {75}},\ \bibinfo {pages} {4714} (\bibinfo {year}
  {1995})}\BibitemShut {NoStop}%
\bibitem [{\citenamefont {Meekhof}\ \emph {et~al.}(1996)\citenamefont
  {Meekhof}, \citenamefont {Monroe}, \citenamefont {King}, \citenamefont
  {Itano},\ and\ \citenamefont {Wineland}}]{PRL1796}%
  \BibitemOpen
  \bibfield  {author} {\bibinfo {author} {\bibfnamefont {D.~M.}\ \bibnamefont
  {Meekhof}}, \bibinfo {author} {\bibfnamefont {C.}~\bibnamefont {Monroe}},
  \bibinfo {author} {\bibfnamefont {B.~E.}\ \bibnamefont {King}}, \bibinfo
  {author} {\bibfnamefont {W.~M.}\ \bibnamefont {Itano}},\ and\ \bibinfo
  {author} {\bibfnamefont {D.~J.}\ \bibnamefont {Wineland}},\ }\bibfield
  {title} {\bibinfo {title} {Generation of nonclassical motional states of a
  trapped atom},\ }\href {https://doi.org/10.1103/PhysRevLett.76.1796}
  {\bibfield  {journal} {\bibinfo  {journal} {Phys. Rev. Lett.}\ }\textbf
  {\bibinfo {volume} {76}},\ \bibinfo {pages} {1796} (\bibinfo {year}
  {1996})}\BibitemShut {NoStop}%
\bibitem [{\citenamefont {Wineland}(2013)}]{RMP1103}%
  \BibitemOpen
  \bibfield  {author} {\bibinfo {author} {\bibfnamefont {D.~J.}\ \bibnamefont
  {Wineland}},\ }\bibfield  {title} {\bibinfo {title} {Nobel lecture:
  Superposition, entanglement, and raising schr\"odinger's cat},\ }\href
  {https://doi.org/10.1103/RevModPhys.85.1103} {\bibfield  {journal} {\bibinfo
  {journal} {Rev. Mod. Phys.}\ }\textbf {\bibinfo {volume} {85}},\ \bibinfo
  {pages} {1103} (\bibinfo {year} {2013})}\BibitemShut {NoStop}%
\bibitem [{\citenamefont {Mohseni}\ \emph {et~al.}(2008)\citenamefont
  {Mohseni}, \citenamefont {Rezakhani},\ and\ \citenamefont
  {Lidar}}]{PRA032322}%
  \BibitemOpen
  \bibfield  {author} {\bibinfo {author} {\bibfnamefont {M.}~\bibnamefont
  {Mohseni}}, \bibinfo {author} {\bibfnamefont {A.~T.}\ \bibnamefont
  {Rezakhani}},\ and\ \bibinfo {author} {\bibfnamefont {D.~A.}\ \bibnamefont
  {Lidar}},\ }\bibfield  {title} {\bibinfo {title} {Quantum-process tomography:
  Resource analysis of different strategies},\ }\href
  {https://doi.org/10.1103/PhysRevA.77.032322} {\bibfield  {journal} {\bibinfo
  {journal} {Phys. Rev. A}\ }\textbf {\bibinfo {volume} {77}},\ \bibinfo
  {pages} {032322} (\bibinfo {year} {2008})}\BibitemShut {NoStop}%
\bibitem [{\citenamefont {Nielsen}\ and\ \citenamefont {Chuang}(2011)}]{book}%
  \BibitemOpen
  \bibfield  {author} {\bibinfo {author} {\bibfnamefont {M.~A.}\ \bibnamefont
  {Nielsen}}\ and\ \bibinfo {author} {\bibfnamefont {I.~L.}\ \bibnamefont
  {Chuang}},\ }\href@noop {} {\emph {\bibinfo {title} {Quantum Computation and
  Quantum Information: 10th Anniversary Edition}}},\ \bibinfo {edition} {10th}\
  ed.\ (\bibinfo  {publisher} {Cambridge University Press},\ \bibinfo {year}
  {2011})\BibitemShut {NoStop}%
\bibitem [{\citenamefont {Song}\ \emph
  {et~al.}(2017{\natexlab{b}})\citenamefont {Song}, \citenamefont {Xu},
  \citenamefont {Liu}, \citenamefont {Yang}, \citenamefont {Zheng},
  \citenamefont {Deng}, \citenamefont {Xie}, \citenamefont {Huang},
  \citenamefont {Guo}, \citenamefont {Zhang}, \citenamefont {Zhang},
  \citenamefont {Xu}, \citenamefont {Zheng}, \citenamefont {Zhu}, \citenamefont
  {Wang}, \citenamefont {Chen}, \citenamefont {Lu}, \citenamefont {Han},\ and\
  \citenamefont {Pan}}]{PRL180511}%
  \BibitemOpen
  \bibfield  {author} {\bibinfo {author} {\bibfnamefont {C.}~\bibnamefont
  {Song}}, \bibinfo {author} {\bibfnamefont {K.}~\bibnamefont {Xu}}, \bibinfo
  {author} {\bibfnamefont {W.}~\bibnamefont {Liu}}, \bibinfo {author}
  {\bibfnamefont {C.-p.}\ \bibnamefont {Yang}}, \bibinfo {author}
  {\bibfnamefont {S.-B.}\ \bibnamefont {Zheng}}, \bibinfo {author}
  {\bibfnamefont {H.}~\bibnamefont {Deng}}, \bibinfo {author} {\bibfnamefont
  {Q.}~\bibnamefont {Xie}}, \bibinfo {author} {\bibfnamefont {K.}~\bibnamefont
  {Huang}}, \bibinfo {author} {\bibfnamefont {Q.}~\bibnamefont {Guo}}, \bibinfo
  {author} {\bibfnamefont {L.}~\bibnamefont {Zhang}}, \bibinfo {author}
  {\bibfnamefont {P.}~\bibnamefont {Zhang}}, \bibinfo {author} {\bibfnamefont
  {D.}~\bibnamefont {Xu}}, \bibinfo {author} {\bibfnamefont {D.}~\bibnamefont
  {Zheng}}, \bibinfo {author} {\bibfnamefont {X.}~\bibnamefont {Zhu}}, \bibinfo
  {author} {\bibfnamefont {H.}~\bibnamefont {Wang}}, \bibinfo {author}
  {\bibfnamefont {Y.-A.}\ \bibnamefont {Chen}}, \bibinfo {author}
  {\bibfnamefont {C.-Y.}\ \bibnamefont {Lu}}, \bibinfo {author} {\bibfnamefont
  {S.}~\bibnamefont {Han}},\ and\ \bibinfo {author} {\bibfnamefont {J.-W.}\
  \bibnamefont {Pan}},\ }\bibfield  {title} {\bibinfo {title} {10-qubit
  entanglement and parallel logic operations with a superconducting circuit},\
  }\href {https://doi.org/10.1103/PhysRevLett.119.180511} {\bibfield  {journal}
  {\bibinfo  {journal} {Phys. Rev. Lett.}\ }\textbf {\bibinfo {volume} {119}},\
  \bibinfo {pages} {180511} (\bibinfo {year} {2017}{\natexlab{b}})}\BibitemShut
  {NoStop}%
\end{thebibliography}%
\nocite{2,3,PRL113601,5,6,PRL4714,PRL1796,PRA032322}
\begin{acknowledgments}
	\textbf{Funding:} This work was supported by the National Natural Science Foundation of China under (Grants No. 12474356, No. 12475015, No. 12274080, No. 12174058, No. 11875108, No. 12204105), and Innovation Program for Quantum Science and Technology under Grant No. 2021ZD0300200. 
	
	\textbf{Competing interests:} Authors declare that they have no competing interests.
	
	\textbf{Data and materials availability:} All data are available in the main text or the supplementary materials.
	
\end{acknowledgments}

\end{document}


\begin{center}
		\Large \textbf{Supplementary Materials for} \\
		\vspace{0.5em}
		\Large \textbf{Critical quantum metrology robust against dissipation and non-adiabaticity} \\
		\vspace{0.5em}
		Jia-Hao L\"{u} \textit{et al.} \\
		\vspace{1em}
		Corresponding author:Zhen-Biao Yang, zbyang@fzu.edu.cn; Huai-Zhi Wu, huaizhi.wu@fzu.edu.cn; Shi-Biao Zheng, t96034@fzu.edu.cn \\
	\end{center}
	\vspace{10em}
\textbf{This PDF file includes:}
\begin{itemize}
	\item \textbf{Supplementary Text}
	\item \textbf{Figs. S1 to S15}
	\item \textbf{References}
\end{itemize}
\vspace{40em}
\section{Fisher information of the driven Jaynes-Cummings model}

The population of the excited state $|e\rangle$ is defined as the expectation value of the excitation-number operator
$P_{e}=\left\langle
\left\vert e\right\rangle \left\langle e\right\vert \right\rangle $. In the two-level approximation, the operator $\left\vert e\right\rangle
\left\langle e\right\vert $ can be rewritten as
$ \left\vert e\right\rangle \left\langle e\right\vert =\frac{1}{2}(I+\sigma_{z})$,
where $I$ denotes the identity operator and $\sigma _{z}=\left\vert
e\right\rangle \left\langle e\right\vert -\left\vert g\right\rangle
\left\langle g\right\vert $. Then the standard deviation of $P_{e}$ is given
by
\begin{eqnarray}
\Delta P_{e} &=&\frac{1}{2}\sqrt{\left\langle (1+\sigma
_{z})^{2}\right\rangle -\left\langle 1+\sigma _{z}\right\rangle ^{2}} =\frac{1}{2}\sqrt{1-\left\langle \sigma _{z}\right\rangle ^{2}}.
\end{eqnarray}%
Using the expression $\left\langle \sigma _{z}\right\rangle =2P_{e}-1$, we
obtain
\begin{eqnarray}
\Delta P_{e} &=&\frac{1}{2}\sqrt{1-(2P_{e}-1)^{2}} =\sqrt{P_{e}(1-P_{e})}.
\end{eqnarray}%
For the dark state $|\phi_0\rangle$, $P_{e}=|c_-|^2=\frac{1}{2}(1-\sqrt{1-\varepsilon ^{2}})$.  Substituting $P_{e}$ into the above
expression, we obtain $\Delta P_{e}=\varepsilon /2$. The corresponding Fisher information is defined as
\begin{eqnarray}
    {\cal F}=\frac{1}{P_{e}}\left( \frac{dP_{e}}{d\varepsilon }\right) ^{2}+%
\frac{1}{P_{g}}\left( \frac{dP_{g}}{d\varepsilon }\right) ^{2},
\end{eqnarray}
and is adopted in the main text. In the two-level approximation, $P_{e}+P_{g}=1$. Then ${\cal F}$ can be
re-expressed as%
\begin{eqnarray}
{\cal F} &=&\left( \frac{1}{P_{e}}+\frac{1}{1-P_{e}}\right) \left( \frac{%
dP_{e}}{d\varepsilon }\right) ^{2} =\frac{1}{P_{e}(1-P_{e})}\left( \frac{dP_{e}}{d\varepsilon }\right) ^{2}.\label{Fisher}
\end{eqnarray}%
Using $dP_{e}/d\varepsilon =\varepsilon /(2\sqrt{1-\varepsilon ^{2}})$, we finally
have%
\begin{eqnarray}
    {\cal F}=\frac{1}{1-\varepsilon ^{2}}.
\end{eqnarray}
On the other hand, the quantum Fisher information is \cite{1}%
\begin{eqnarray}
{\cal I} &=&4[\left\langle \partial _{\varepsilon }\phi _{0}\right\vert
\left. \partial _{\varepsilon }\phi _{0}\right\rangle +\left\langle \partial
_{\varepsilon }\phi _{0}\right\vert \left.\phi _{0}\right\rangle
^{2}] =\frac{1}{1-\varepsilon ^{2}},
\end{eqnarray}%
which implies that the Cram\'{e}r-Rao bound is saturated for all values of $%
\varepsilon$ for our experimental measurements.

\section{Numerical simulations of the driven Jayness-Cummings model}
\begin{figure}[!hbp]
	\includegraphics[width=1\textwidth]{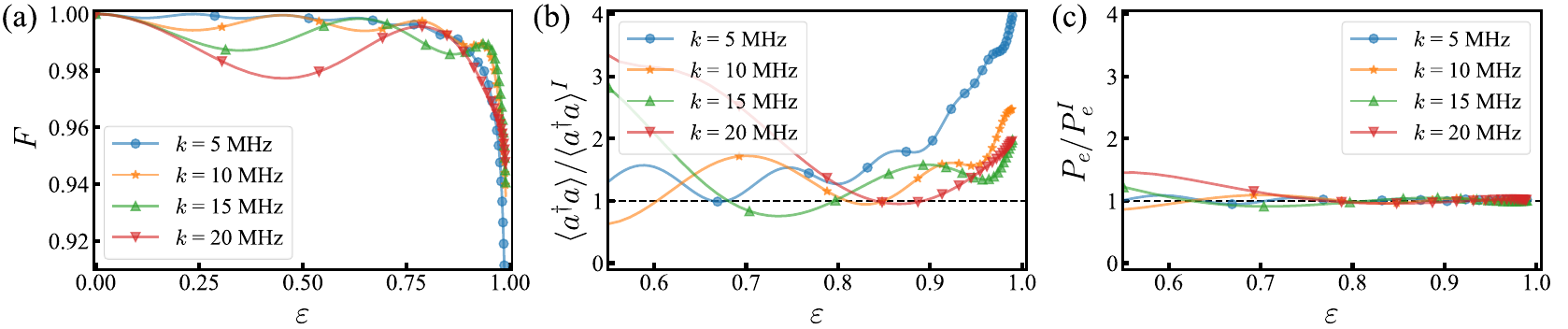}
	\caption{The numerical simulations based on the master equation. (a) Fidelities evolutions of the dark state as a function of $\varepsilon$. The ratio of the measured (b) average photon number as well as the (c) qubit's excitation number to that in the ideal dark state,  versus the control parameter $\varepsilon$ for different $k$, where $\langle a^\dagger a\rangle^I$ and $P^I_e$ denote the average photon number and the $\vert e\rangle$-state population  of the ideal dark state, respectively.}  \label{twolevel_pe}
\end{figure} 
\begin{figure}
	\includegraphics[width=1\textwidth]{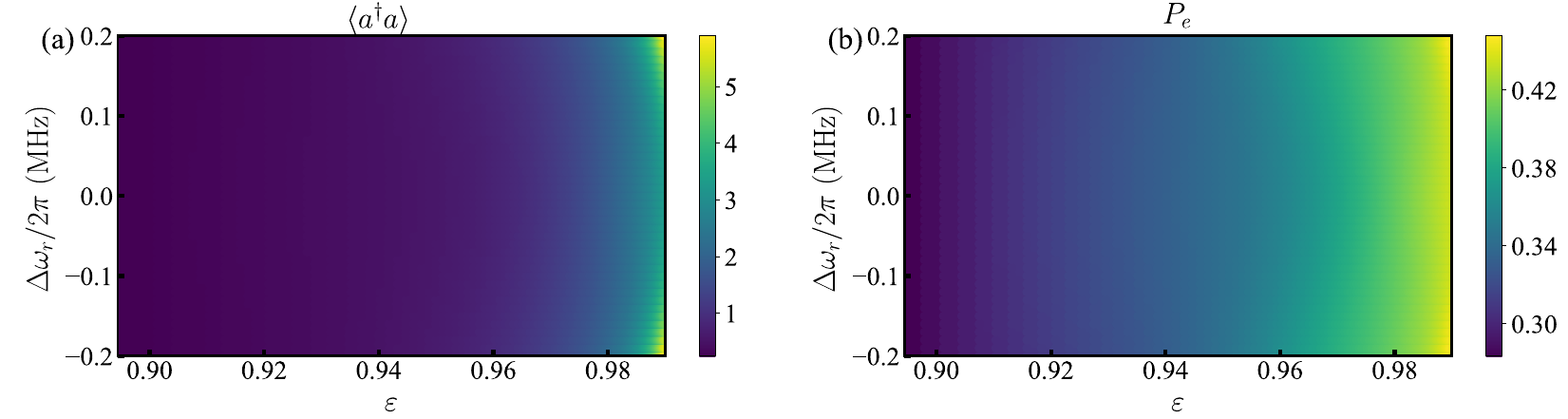}
	\caption{The numerical simulations  of (a) the average photon number and (b) the qubit's excitation number versus the detuning of the signal field with respect to the resonator frequency.}  \label{twolevel_n}
\end{figure} 
In order to quantify the effect of the decoherence and non-adiabatic transitions, we here describe the dynamics by the Lindblad master equation 
\begin{eqnarray}
    \dot{\rho}=-i[H,\rho]+\kappa_q{\cal L}[\sigma_-]+\gamma_q{\cal L}[\vert e\rangle\langle e\vert] + \kappa_r{\cal L}[a],
\end{eqnarray}
where $H$ is given by Eq. (1) of the main text with $\Omega=2\pi\times20.9$ MHz, and the Lindblad superoperators are defined as
\begin{eqnarray}
{\cal L}[{O}]=O\rho O^\dagger-\frac{1}{2}O^\dagger O\rho-\frac{1}{2}\rho O^\dagger O.
\end{eqnarray}
The energy decaying rates for the qubit and the resonator are $\kappa_q=0.05$ MHz and $\kappa_r=0.08$ MHz respectively, and $\gamma_q=0.08$ MHz is the dephasing rate of the qubit.

We then perform the numerical simulations on the ratio of the measured average photon number as well as the qubit's excitation number  to that in the ideal dark state, versus the control parameter $\varepsilon$. These ratios, together with the fidelity evolutions of the dark state, are presented in Fig. \ref{twolevel_pe}, which shows that, as $\varepsilon\to1$, the probability of leakage to the bright state increases dramatically, as a consequence of the decoherence and non-adiabatic effect. Owing to the high distinction of the average photon number between the ideal dark state and the bright state, the measured average photon number is sensitive to the state leakage. In distinct contrast, when approaching the critical point, the qubit's excitation number of the ideal dark state tends to be $1/2$, which is consistent with the bright state. This implies that the qubit's excited-state population ($P_{e}$) is insensitive to leakage to the bright state, making it qualified to serve as a robust indicator for quantum sensing.

Although the dark state is obtained under the resonant condition, $P_{e}$ is insensitive to the frequency fluctuations. To illustrate this point, we perform a numerical simulation to quantify the effect of a frequency deviation $\Delta_{\omega_r}=\omega_r-\omega_s$ of the signal field. As shown in Fig. \ref{twolevel_n}, $P_e$ is robust to the slight frequency detuning of the signal field, which further verifies the robustness of the sensing protocol.

  It should be noted that $P_{e}$ becomes increasingly sensitive to the ramping speed when $\varepsilon$ is approaching the critical point, where the energy gaps are closed. The smaller the ramping speed which is determined by $k$, the better the adiabatic condition being satisfied. However, the increase of the time would cause more serious decoherence. Therefore, there is a trade-off between the non-adiabaticity and decoherence. 

\begin{figure}
	\includegraphics[width=1\textwidth]{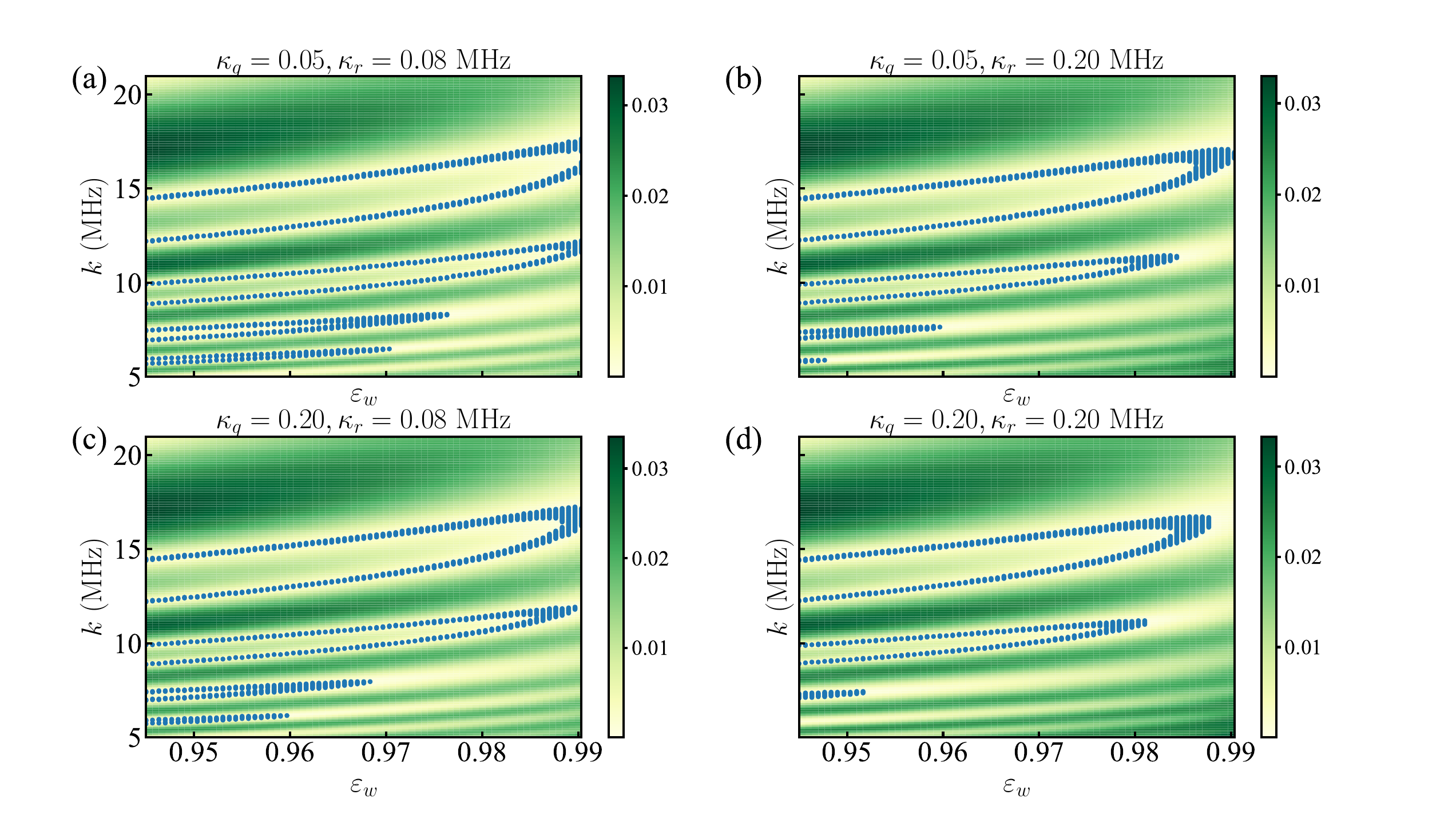}
         \caption{Relative deviation of $P_e$ from the ideal value versus $\varepsilon_w$ and $k$ for different decaying rates. }  \label{section2_decay}
\end{figure}

For certain coherence times, the optimal ramping speed depends on the choice of the working point, $\varepsilon_{w}$, around which the control parameter $\varepsilon$ is assumed to be. In Fig. \ref{section2_decay}(a), we present the relative error of $P_{e}$, defined as $D=|P_{e}- P_{e}^{I}|/P_{e}^{I}$, as a function of $\varepsilon_{w}$ and $k$ for the decaying rates of our systems:  $\kappa_{q}=0.05$ MHz and $\kappa_{r}=0.08$ MHz. Here $k$ determines the ramping speed, as shown in Eq. (5) of the main text. The dots represent the points where the errors are not larger than $0.1\%$. As this upper bound is much smaller than the error caused by the parameter fluctuation in our system, the value of $k$ at each of such points can be reasonably taken as the optimal value for the corresponding working point $\varepsilon_{w}$. In our experiment, we choose $k=10$ MHz, which represents an optimal value for $\varepsilon_{w}=0.98$. For this working point, there exist other optimal values of $k$ larger than $10$ MHz. However, due to the limitation of the anharmonicity of the qubit in our circuit QED system, larger ramping speeds may incur more serious leakage to the qubit's third level. Figure \ref{section2_decay}(b)-(d) show $D$ versus $\varepsilon_{w}$ and $k$ for other values of the decaying rates.

\section{Numerical simulations of the driven Xmon-resonator system}

In this section, we focus mainly on the influence brought about by the energy levels higher than the first excited state $|e\rangle$ of the qubit. In this case, the Lindblad master equation can be rewritten as
\begin{eqnarray}
\dot{\rho}&=&-i[H_{full},\rho]+\kappa_q{\cal L}[q]+\gamma_q{\cal L}[q^\dagger q] + \kappa_r{\cal L}[a], \nonumber\\
H_{full}&=&\omega_ra^\dagger a+\omega_eq^\dagger q-\frac{\chi}{2}q^{\dagger2}q^2+\Omega \lbrack( a^{\dagger }q +aq^\dagger)+\varepsilon (a^{\dagger}e^{-i\omega_st}+ae^{i\omega_st})/2\rbrack,
\end{eqnarray}
where $q^\dagger$ $(q)$ and $a^\dagger$ $(a)$ denote the creation (annihilation) operator for the Xmon qubit and the resonator respectively, and $\chi$ is the anharmonicity for qubit's level ladder; $\omega_e$ is the transition frequency between $|g\rangle$ and $|e\rangle$ and $\omega_r$ represents the resonator's frequency; $\varepsilon $ characterizes the rescaled amplitude of the microwave field coupled to the photonic mode with the frequency $\omega_s$.

\begin{figure}
	\includegraphics[width=1\textwidth]{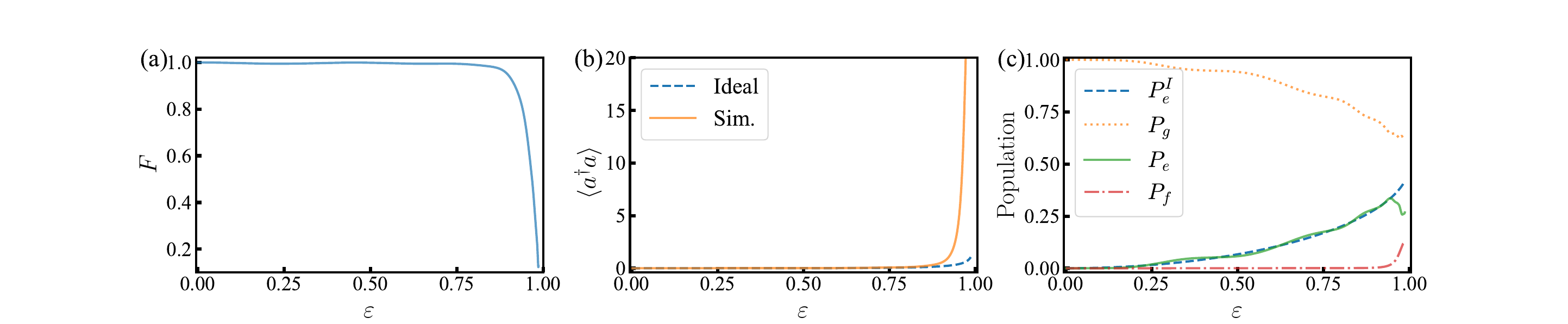}
         \caption{The numerical simulations based on the Hamiltonian of Eq. (\ref{Hami}). (a) Fidelity of the qubit-resonator state to the ideal dark state as a function of $\varepsilon$. (b) The average photon number and (c) the qubit's population distribution  as a function of $\varepsilon$. $P^I_e$ denotes the $\vert e\rangle$-state population of the ideal dark state. In  the simulation, the parameters are the same as that shown in the main text.  }    \label{full}
\end{figure}

For simplicity, we include only the second excited state $|f\rangle$, while we note that the influences of the higher excited states on the system's dynamics are slight and can be neglected compared to $\vert f\rangle$. The Hamiltonian for the driven qutrit-resonator system is given by  (setting $\hbar=1$)
\begin{eqnarray}
H' = \omega_e\vert e\rangle\langle e \vert+\left(2\omega_e-\chi\right)\vert f\rangle\langle f\vert+\omega_ra^\dagger a+\Omega \left[a^\dagger\left(\vert g\rangle\langle e\vert + \sqrt{2}\vert e\rangle\langle f\vert+\frac{\varepsilon}{2}e^{-i\omega_st}\right) + \rm{H.c.}\right], \label{Hami}
\end{eqnarray}

We then perform the numerical simulations based on the Hamiltonian $H'$ in Eq. (\ref{Hami}). As shown in Fig.~\ref{full}, when the resonator couples resonantly with both the qubit and the microwave field, the system's dynamics deviates significantly from the desired one when approaching the critical point, where the infidelity for the ideal dark state increases exponentially. Additionally, the average photon number  exhibits a behavior that is very different from the ideal one near the critical point, owing to the large leakage to the bright states. Although the qubit's $\vert e\rangle$-state population is insensitive to the leakage to the bright states in the two-level approximation, it presents an undesirable behavior near the critical point where the population of the $\vert f\rangle$-state increases to $0.13$. 
As shown in Fig.~\ref{full}(c), the non-negligible influence of the $\vert f\rangle$-state breaks down the monotonous feature of the $\vert e\rangle$-state population near the critical point. 

\section{Suppression of leakage to the second-excited state of the Xmon}
\begin{figure}
	\includegraphics[width=1\textwidth]{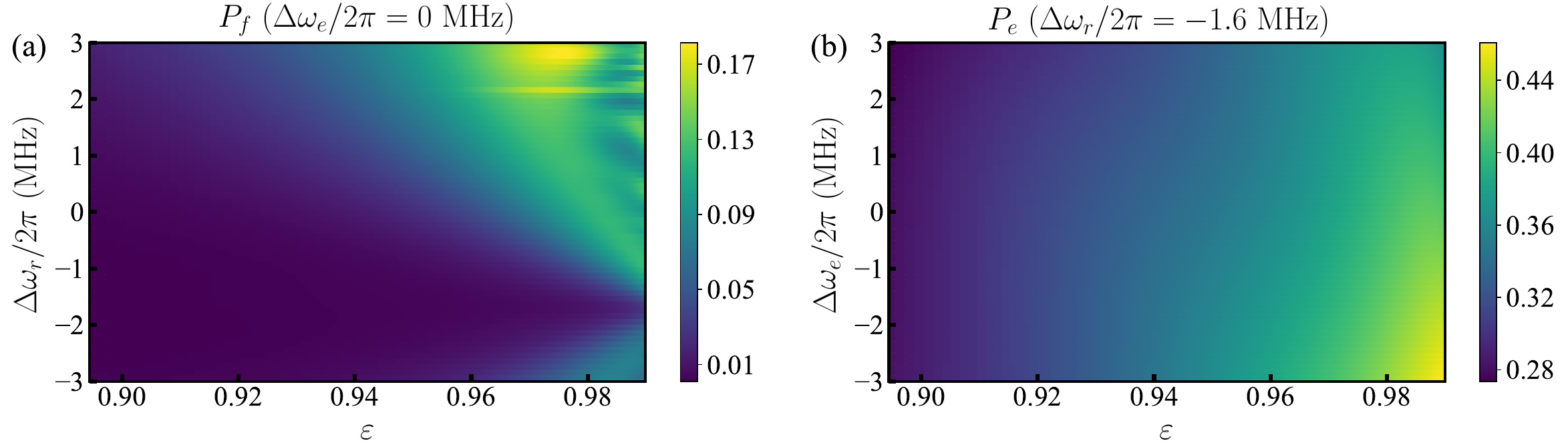}
         \caption{The numerical simulations with the  fine-tuned frequency detunings. (a) The qubit's $\vert f\rangle$-state population versus the detuning of the resonator to the signal field during the evolution. It shows that the probability of leaking to the state $\vert f\rangle $ can be highly suppressed at a specific detuning. We here set the detuning between the qubit and the signal field to zero. (b) The qubit's $\vert e\rangle$-state population versus the detuning of the qubit to the signal field. The detuning between the resonator and the microwave field is chosen as $-2\pi\times 1.6$ MHz. The other parameters are the same as shown above.}  \label{full_unfix}
\end{figure}  
\begin{figure}
	\includegraphics[width=1\textwidth]{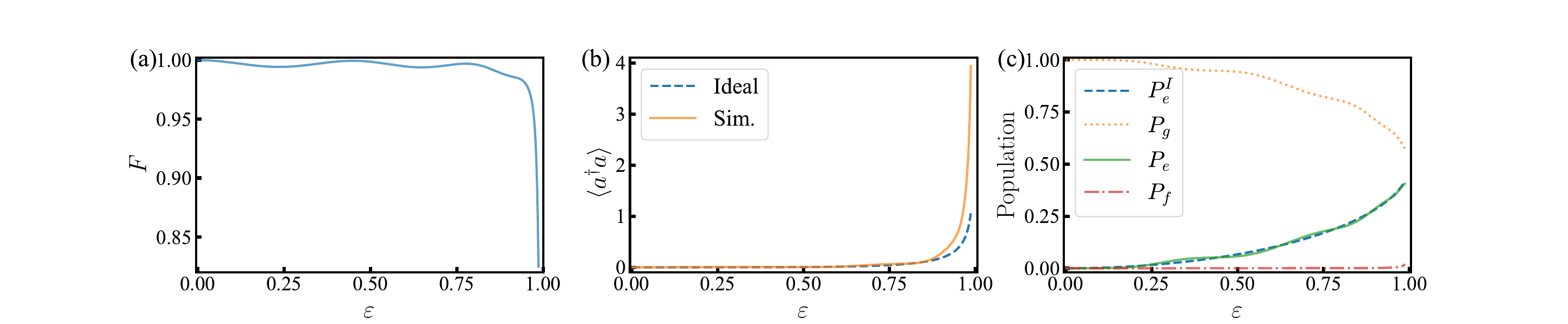}
         \caption{The numerical simulations with  the chosen frequency detunings. (a) Fidelity of the qubit-resonator state to the ideal dark state versus $\varepsilon$. (b) The average photon number and (c) the qubit's population distribution  versus $\varepsilon$. $P^I_e$ denotes the $\vert e\rangle$-state population of the ideal dark state. In  the simulation, the detunings of the resonator and the qubit with respect to the signal field are tuned to $-2\pi \times1.6$ MHz and $2\pi\times 1.0$ MHz, respectively. The other parameters are the same as shown above.}  \label{full_fix}
\end{figure}  
To make the sensor work near the critical point, it is necessary to suppress the population of the second excited state of the Xmon. We find that this can be significantly optimized by fine tuning the detunings of the Xmon and the resonator from the signal field. With the detunings being included, the system Hamiltonian in the framework rotating at the frequency of the signal field is
\begin{eqnarray}
H^{\prime}_I=\Delta\omega_e\vert e\rangle\langle e\vert+(2\Delta\omega_e-\chi)\vert f\rangle\langle f\vert+\Delta\omega_ra^\dagger a+\Omega \left[a^\dagger\left(\vert g\rangle\langle e\vert + \sqrt{2}\vert e\rangle\langle f\vert+\frac{\varepsilon}{2}\right) + \rm{H.c.}\right],
\end{eqnarray}
where $\Delta\omega_r=\omega_r-\omega_s$ and $\Delta\omega_e=\omega_e-\omega_s$ are the detunings  of the resonator and the qubit with respect to the signal field, respectively.

We then perform the numerical simulations by tuning  $\Delta\omega_r$ and $\Delta\omega_e$ successively, as  exhibited in Fig.~\ref{full_unfix}, which shows that the probability of leaking to the $\vert f\rangle$-state can be highly suppressed through the optimization of the frequency detunings, and the $\vert e\rangle$-state population agrees well with that of the on-resonance dynamics. Seen from Fig.~\ref{full_unfix}, we note that the set of optimal detunings of the resonator and the qubit with respect to the signal field are $-2\pi \times1.6$ MHz and $2\pi\times 1.0$ MHz, respectively. With the chosen frequency detunings, when approaching the critical point we find that the fidelity with respect to the effective model remains higher than 0.9 for $\varepsilon=0.98$; the associated average photon number is greatly reduced as compared to the case without such a frequency adjustment; and more importantly, the probability of leaking to the $\vert f\rangle$-state decreases correspondingly such that the qubit's $\vert e\rangle$-state population becomes well  consistent with the ideal situation, as shown in Fig. \ref{full_fix}. In view of experiments, we further fine tune the detunings of the resonator and the qubit around the numerically estimated optimal parameters (i.e. $\Delta\omega_r=-2\pi \times1.6$ MHz and $\Delta\omega_e=2\pi\times 1.0$ MHz), making the overall evolution dynamics match the theoretical predictions (as elaborated in the main text).

\section{Correction of  XY crosstalk}
The superconducting circuit sample used in our experiment consists of five frequency-tunable Xmon qubits interconnected by a bus resonator, as described elsewhere \cite{2,3,4}. Our experiment involves two of the qubits, one of which  serves as the test qubit, labeled as $Q_T$, to establish the critical quantum dynamics together with the bus resonator, and the other, labeled as $Q_A$, to read  out the photon number distribution of the resonator by observing its on-resonance Rabi signal. Additionally, we use the XY-control line of the qubit $Q_R$ to apply a microwave drive to the resonator by crosstalk interactions. The remaining qubits are at their sweep points, so as to effectively decouple from the dynamics.

\begin{figure}
	\includegraphics[width=1\textwidth]{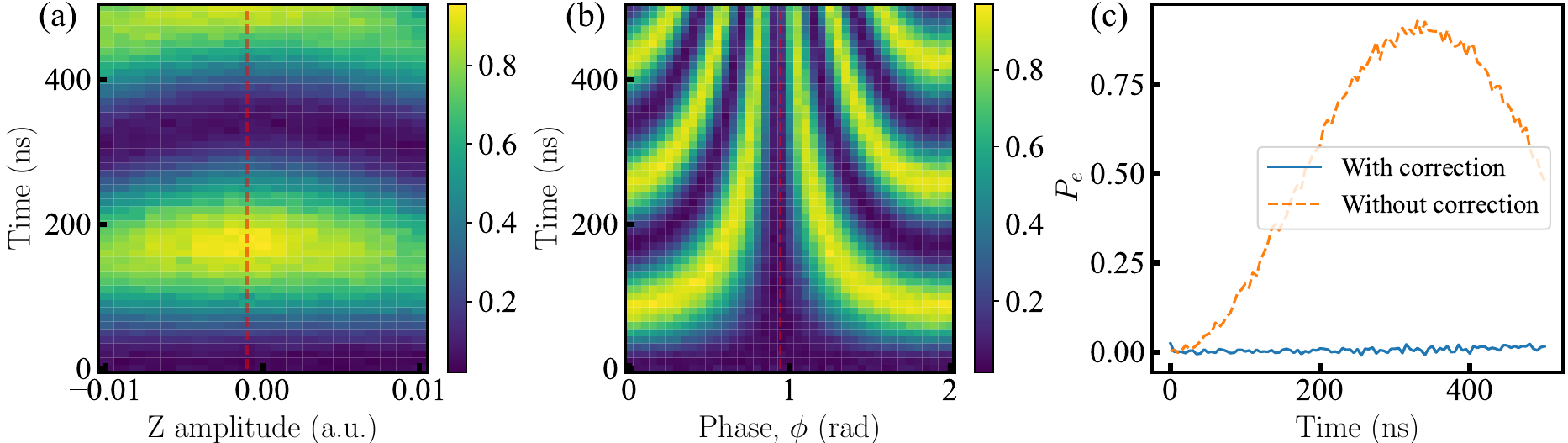}
     \caption{Correction of the XY-crosstalk. (a) The measurement of the XY-crosstalk amplitude. A microwave pulse with frequency $\omega_I$ is applied to $Q_R$'s XY-control line  after tuning $Q_T$ to $\omega_I$, generating a crosstalk Rabi oscillation on $Q_T$. The Rabi oscillations are measured with small deviation in frequency. The slowest Rabi oscillation corresponds to the crosstalk amplitude, as the red vertical dashed line shows. Here the pulse amplitude is set to $2\pi\times 5$ MHz. (b) The measurement of the XY-crosstalk phase. A microwave pulse with amplitude equal to crosstalk amplitude is added to $Q_T$'s XY-control line to cancel the crosstalk effect. The phase in which qubit is rarely excited is the crosstalk phase. (c) Experiment test for the crosstalk correction. With the applications of the XY-crosstalk correction, $Q_T$ is barely excited. The pulse amplitude here is set to $2\pi\times10 $ MHz. }\label{xyct}
\end{figure}

The pulse sequence of our experiment is shown in Fig.~\ref{sequence}. When $Q_T$ is tuned to couple resonantly with the resonator, the phase and the drive strength of the XY-drive crosstalks between $Q_T$ and the resonator must be corrected \cite{5}. Figure~\ref{xyct} illustrates the calibration process for the XY-drive applied to the resonator as an example. 

We start by calibrating the amplitude of the crosstalk, with the results shown in Fig.~\ref{xyct}(a). For this,  $Q_T$ is biased to the  
 frequency $\omega_I/2\pi = 5.53$ GHz, which is close to the resonator's frequency, and the other qubits are decoupled from $Q_T$ and the resonator. It should be emphasised that the reason we are unable to perform the XY crosstalk calibration at the resonator's frequency is that the on-resonance dynamics between $Q_T$ and the resonator complicates the effect of the crosstalk, which makes the crosstalk amplitude extremely difficult to capture. We here assume that a small frequency detuning has no effect on the crosstalk. After that, a strong flat-top-envelope microwave pulse is applied to the $Q_R$'s XY-control line with frequency $\omega_I$, generating a crosstalk excitation on $Q_T$. The amplitude of the crosstalk of the resonator to $Q_T$ can be obtained by fitting the slowest Rabi oscillation of the  evolution of $Q_T$'s excitation, resulting from different frequency deviation  in $\omega_I$.

We characterize the phase of the crosstalk in terms of the frequency corresponding to the slowest Rabi oscillation in Fig.~\ref{xyct} (b). At this point, a microwave pulse  with amplitude equal to the crosstalk amplitude of the resonator to $Q_T$, should be applied to $ Q_T$ to counteract the effect of the crosstalk. We then monitor the evolution of $Q_T$' excitation for different phases of $Q_T$'s microwave pulse, with the $\pi$-phase of $Q_R$'s microwave pulse. As shown in Fig.~\ref{xyct} (b),  $Q_T$ is barely excited during the evolution process at a specific phase, meaning  the phase difference of the XY-control lines of $Q_R$ to $Q_T$.

The calibrations of $Q_T$ to $Q_R$' XY-control line are performed in the same way. To correct the crosstalks, 
we bias $Q_{T}/Q_R$ to $\omega_I$, and simultaneously apply on-resonance microwave pulses to them,  with amplitudes $A_{1}/A_2$ and phases $\phi_{1}/\phi_2$. During the process,  the other qubits are effectively decoupled from these two qubits. However, owing to the effect of the crosstalks, the microwave's effect on each qubit at this point becomes

\begin{eqnarray}
\begin{bmatrix}
A_1e^{i\phi_1}+A_{12}A_2e^{i(\phi_{12}+\phi_2)}  \\
A_2e^{i\phi_2}+A_{21}A_1e^{i(\phi_{21}+\phi_1)}
\end{bmatrix}=
\begin{bmatrix}
1 & A_{12}e^{i\phi_{12}}  \\
A_{21}e^{i\phi_{21}} & 1
\end{bmatrix}
\begin{bmatrix}
A_1e^{i\phi_1}  \\
A_2e^{i\phi_2}
\end{bmatrix},
\end{eqnarray}
where $A_{jk}$ and $\phi_{jk}$ denote the crosstalk amplitude and phase of qubit $k$ to qubit $j$, respectively, quantified by means of the method described above. This means that, to cancel the effect of the crosstalk, a corrected microwave pulse 
\begin{eqnarray}\begin{bmatrix}
1 & A_{12}e^{i\phi_{12}}  \\
A_{21}e^{i\phi_{21}} & 1
\end{bmatrix}^{-1}
\begin{bmatrix}
A_1e^{i\phi_1}  \\
A_2e^{i\phi_2}
\end{bmatrix},\end{eqnarray}
should be applied to get the expected microwave pulse.

To ensure the validity of the XY-crosstalk correction, we bias $Q_T$  to $\omega_I$ and  apply the  corresponding calibrated microwave pulses to $Q_R$ and $Q_T$, respectively. We then observe the evolution of $Q_T$' excitation, as displayed in Fig.~\ref{xyct} (c). The result verifies the validity of such a correction.

\section{ Calibration of the signal field amplitude}
In our experimental protocol, a time-dependent transverse field should be applied to the resonator. It is necessary to calibrate the linear relation between the pulse amplitude of the microwave and the driving field strength of the resonator. 

\begin{figure}
	\includegraphics[width=1\textwidth]{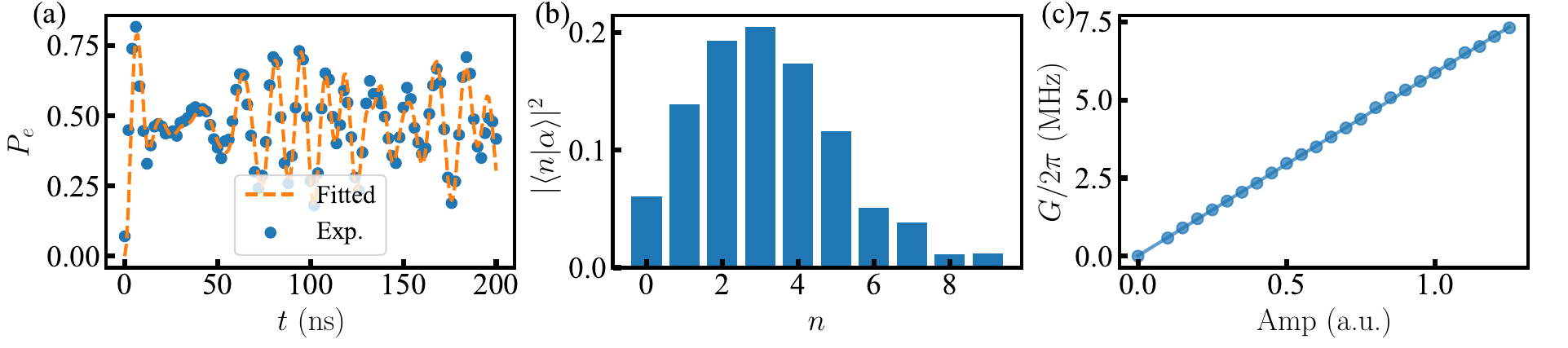}
     \caption{Calibration of the drive strength of the microwave. (a) The experimental (dots), the fitted (solid line) Rabi oscillations and (b) the corresponding photon-number distribution of the resonator. Here, $\tau$ is set to $100$ ns with a specific pulse amplitude of the microwave. (c) The drive strength $G$  as a function of the pulse amplitude of the microwave.}\label{amp2omega}
\end{figure}

\begin{figure}
	\includegraphics[width=0.6\textwidth]{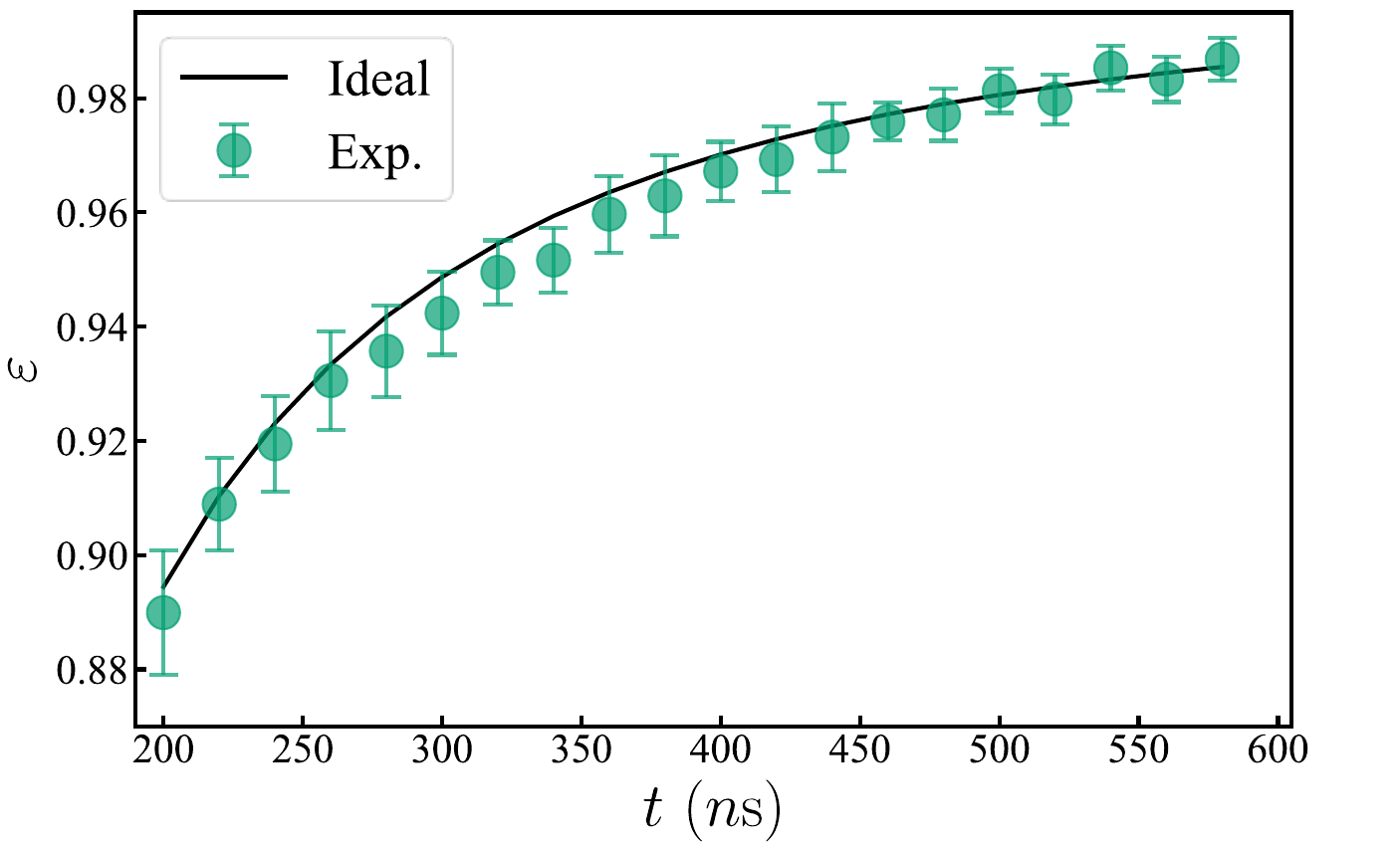}
     \caption{The measurement of the rescaled driving field strength ($\varepsilon$) of the resonator. The values of $\varepsilon$ measured at different times are denoted by the cyan round spots, with the error bars characterizing the fluctuations due to the control imperfections. The black solid line represents the ideal results. }\label{section6}
\end{figure}

To this end, we first use the XY-control line of $Q_R$ to apply a microwave drive with the amplitude $\xi$ to the resonator by the crosstalk interaction, which produces a equivalent driving field strength $G$ to drive the resonator. After a  time $\tau$, the state of the resonator can be described as 
\begin{eqnarray}
\vert R(\tau)\rangle= e^{-iG(a+a^\dagger)\tau}\vert R(0)\rangle=\vert \alpha\rangle,
\end{eqnarray}
where $\vert R(0)\rangle=\vert 0\rangle$ denotes the initial state of the resonator, and $\alpha=-iG\tau$ represents the displacement coefficient. The average photon number $|\alpha|^2$ of such a coherent state can then be read out  by switching off the drive and tuning $Q_A$ to couple resonantly with the resonator (the details are described in the next part). After that, the corresponding driving field strength $G=|\alpha|/\tau$ can be deduced. A linear connection of the pulse amplitude of the microwave to the driving field strength of the resonator can be naturally fitted through the applications of  different pulse amplitude of the microwave, as exhibited in  Fig. \ref{amp2omega}. 
Figure \ref{section6}  presents the rescaled driving amplitude measured with our critical sensing protocol. As shown in the numerical simulation of Fig. \ref{twolevel_pe}(c), the qubit's excited-state population, which is used as the sensing indicator, is insensitive to both the non-adiabaticity and decoherence effects near the critical point. Therefore, the deviations of the measured field amplitudes from the preset values are mainly due to the fluctuations of the driving field itself.

\section{Observed photon-number distribution} 

\begin{figure}[!hbp]
	\includegraphics[width=0.9\textwidth]{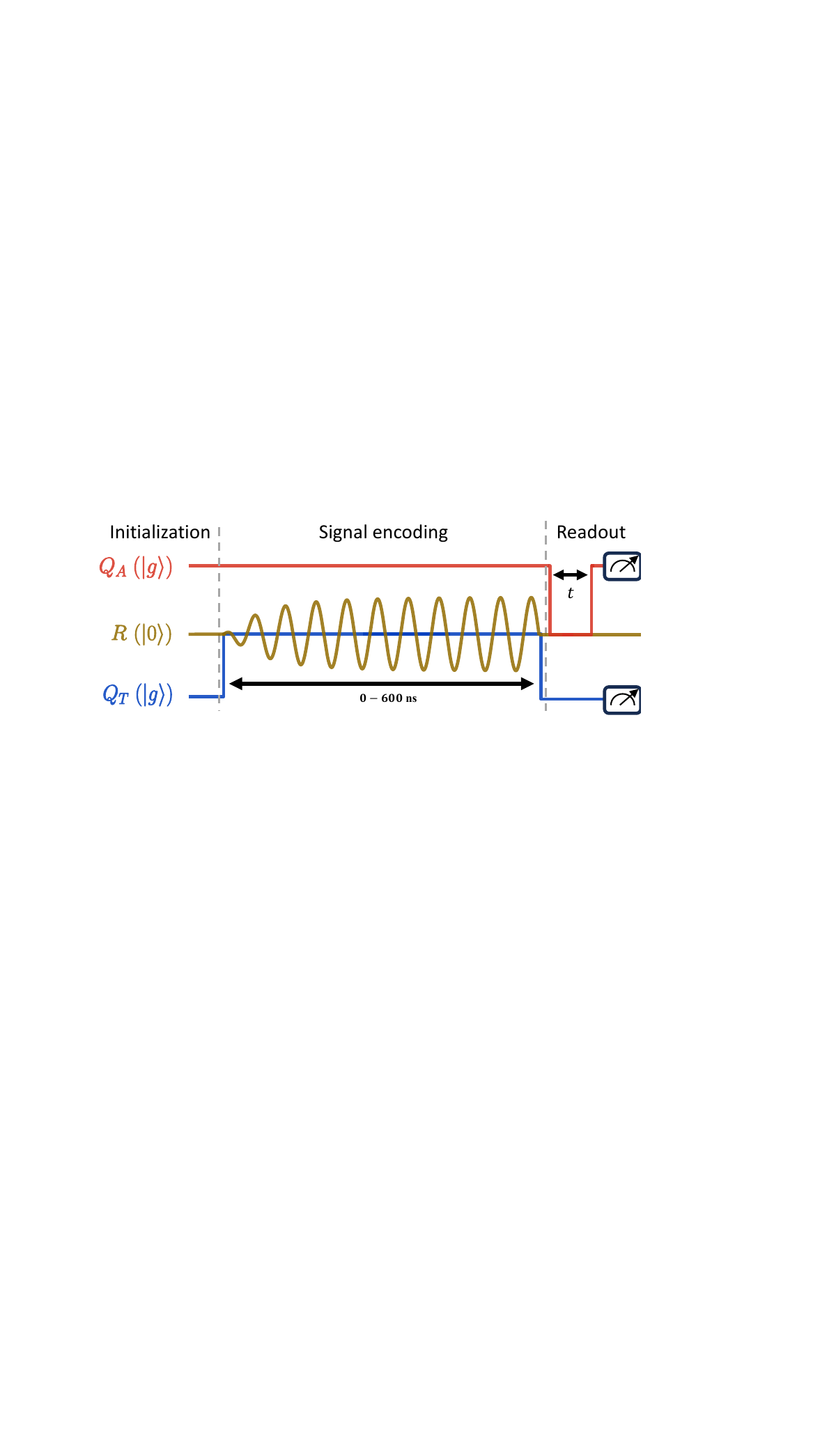}
     \caption{Pulse sequence. The qubit ($Q_T$) is tuned from its frequency to the resonator's
frequency. At the same time the signal field with a rescaled amplitude $%
\varepsilon $ is coupled to the resonator. After the quench process, the qubit
is biased back to its idle frequency for state readout, and the ancilla qubit ($Q_A$) is tuned to couple resonantly with the resonator to measure the photon-number distribution.}\label{sequence}
\end{figure}
After the interaction dynamics between the qubit $Q_T$ and the resonator, the transverse field is switched off, so that the qubit is effectively decoupled from the resonator. The photon number distribution of the resonator is read out by tuning the ancilla qubit $Q_A$, initially in its ground state, to the resonator frequency, and then observing its Rabi oscillations \cite{6}. The excited state population $P^\varepsilon_e(\tau)$ of $Q_A$, after tuning back to its idle frequency, for the rescaled amplitude $\varepsilon$  at the  evolution time $\tau$  is measured, the pulse sequence is shown in Fig.~\ref{sequence}. The measured average photon numbers are calculated by the photon-number distribution, which is fitted by the time-dependent Rabi signal as 
\begin{figure}
	\includegraphics[width=1\textwidth]{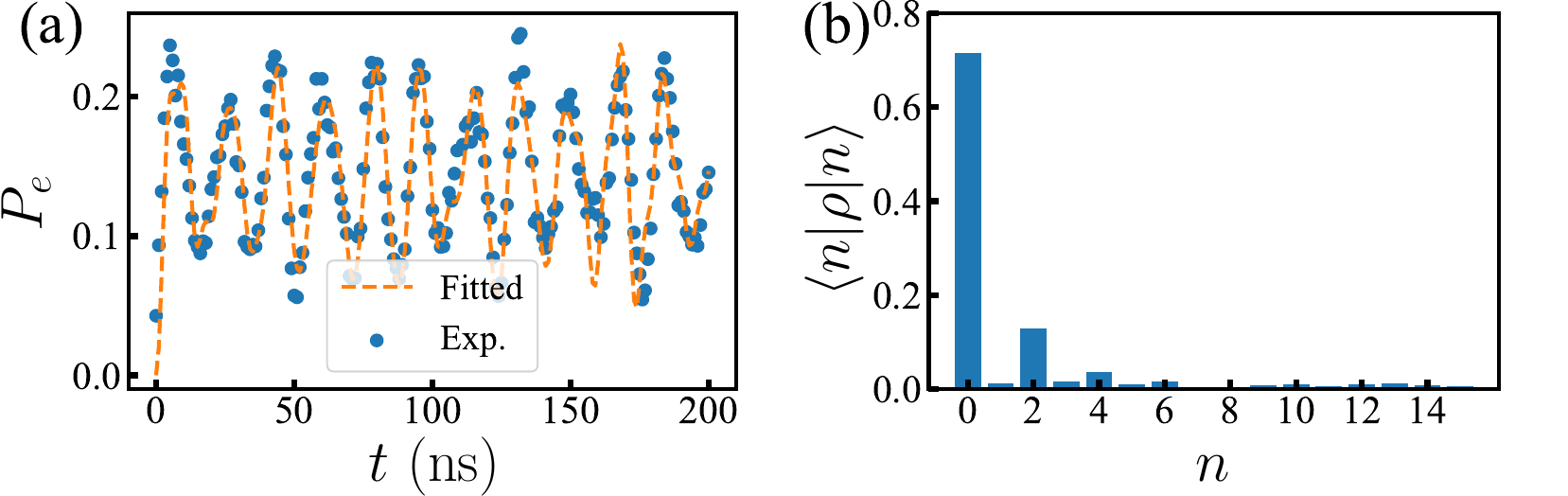}
     \caption{(a) The experimental (dots), the fitted (solid line) Rabi oscillations and (b) the corresponding photon-number distribution of the resonator with $\varepsilon=0.964$.}\label{photon}
\end{figure}
\begin{figure}
	\includegraphics[width=1\textwidth]{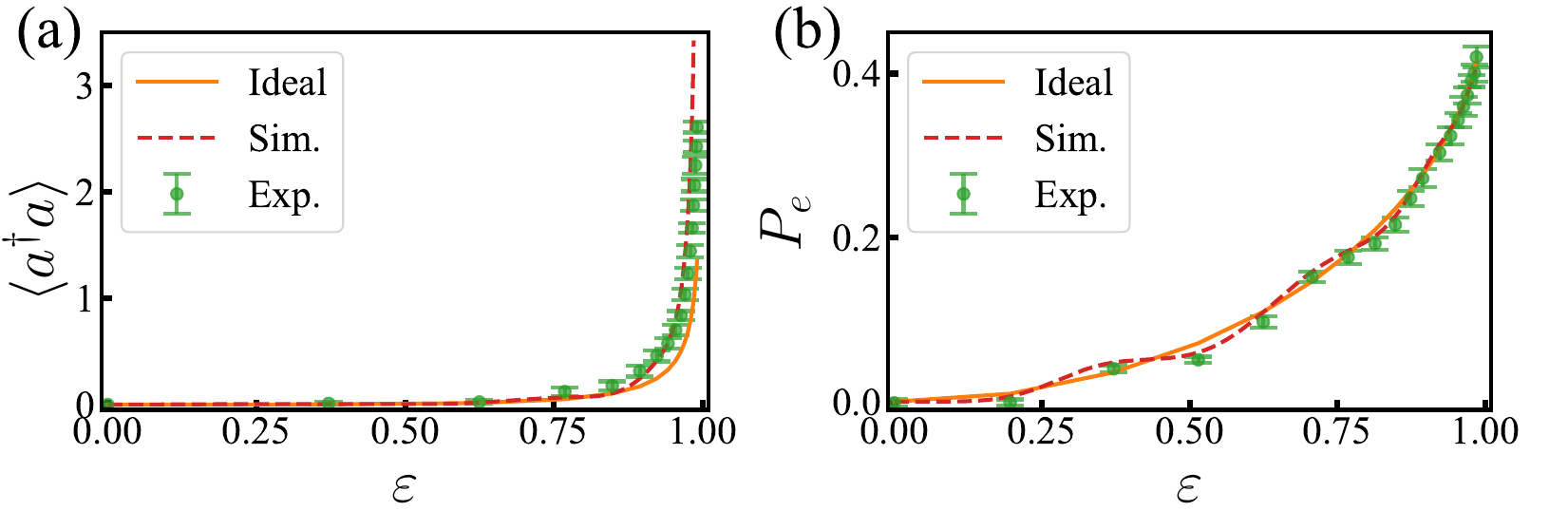}
     \caption{(a) The experimental average photon number and (b) the $\vert e\rangle$-state population as a function of $\varepsilon$ during the whole dynamics. The solid lines denote the result of the ideal dark state, and the dashed lines are that for the numerical simulations. }\label{pe}
\end{figure}

\begin{eqnarray}
P^\varepsilon_e(\tau)=\frac{1}{2}\left[1-P^\varepsilon_g(0)\sum^{n_{max}}_{n=0}{P_n}e^{-\kappa_n\tau}\cos(2\sqrt{n}\Omega_A\tau)\right],
\end{eqnarray}
where $P^\varepsilon_g(0)$ denotes the probability for $Q_A$ initially in its ground state, $P_n$ presents the probability for the $\vert n\rangle$-state photon-number distribution, $n_{max}$ is the fitted cutoff of the photon number, $\kappa_n=n^l/\kappa_r$ 
 $(l=0.7)$ is the empirical decay rate of the $\vert n\rangle$-state Rabi oscillation and $\Omega_A$ represents the on-resonance photonic swapping rate between $Q_A$ and the resonator. Here, the ideal dark state of the resonator is a squeezed vacuum state, which means that the  high-photon state distribution affects the dynamics markedly, increasing the difficulty of the fitting. The measured time-dependent Rabi oscillation signal  and its fitting are shown in Fig. \ref{photon}(a). The corresponding photon-number distribution is presented in Fig. \ref{photon}(b).

In Fig. \ref{pe}, we present the average photon number  of the resonator  and the $\vert e\rangle$-state population of $Q_T$ during the whole dynamics. As expected, the qubit's excitation number is robust against the imperfect system dynamics rather than the average photon number, making $P_e$  serve as an ideal indicator for critical quantum sensing.

\section{Comparison with Rabi measurement}
It is enlightening to compare our critical protocol with the
conventional Rabi measurement method \cite{RMP035002}. The Rabi
measurement is realized by coupling the signal field to a qubit \cite{RMP035002}. The dynamics is described by the Hamiltonian 
\begin{eqnarray}
H_{Rabi}=\frac{\epsilon }{2}(\left\vert g\right\rangle \left\langle
e\right\vert +\left\vert e\right\rangle \left\langle g\right\vert ),
\end{eqnarray}
where $\epsilon $ is the amplitude of the signal field. Under this
Hamiltonian, the qubit state, initially being $\left\vert g\right\rangle $,
evolves as
\begin{eqnarray}
\left\vert \psi (t)\right\rangle =\cos (\theta /2)\left\vert g\right\rangle
-i\sin (\theta /2)\left\vert e\right\rangle,
\end{eqnarray}
where $\theta =\epsilon t$. The $\left\vert e\right\rangle $-state population makes Rabi oscillations, given by
\begin{eqnarray}
P_{e}=\frac{1}{2}(1-\cos \theta ).
\end{eqnarray}
The interferometer works by measuring the deviation of this transition
probability from a well-chosen reference point $P_{0}$, referred to as the
bias point, corresponding to a known value of the field amplitude ($\epsilon
_{0}$) and a well-chosen time $t$. The interferometer is most sensitive to a
small deviation ($\delta \epsilon $) from $\epsilon _{0}$ at the point $%
P_{0}=1/2$, where $\left\vert dP_{e}/d\theta \right\vert $ has the maximum $%
1/2$. The bias point corresponds to $\epsilon _{0}t_{n}=n\pi /2$, where $%
n=1,3,5$.... Around the bias point, the deviation of $P_{e}$ from $P_{0}$
depends on 

\begin{eqnarray}
	\Delta P_{e} =-\frac{1}{2}\cos [(\epsilon _{0}+\delta \epsilon )t_{n}] 
	\simeq \frac{1}{2}(-1)^{(n-1)/2}\delta \epsilon t_{n}.
\end{eqnarray}%
 The precision of
the Rabi measurement method strongly depends upon the accuracy of the
timing. Set the deviation of the real time from the desired value ($t_{n}$)
to be $\delta t_{n}$. The error of the $\left\vert
e\right\rangle $-state population caused by this imperfect timing is given
by   
\begin{eqnarray}
	\delta P_{e} \simeq\frac{1}{2}(-1)^{(n-1)/2}\epsilon _{0}\delta t_{n},
\end{eqnarray}%
which linearly scales with the time control error $\delta t_{n}$.

\begin{figure}
	\includegraphics[width=0.6\textwidth]{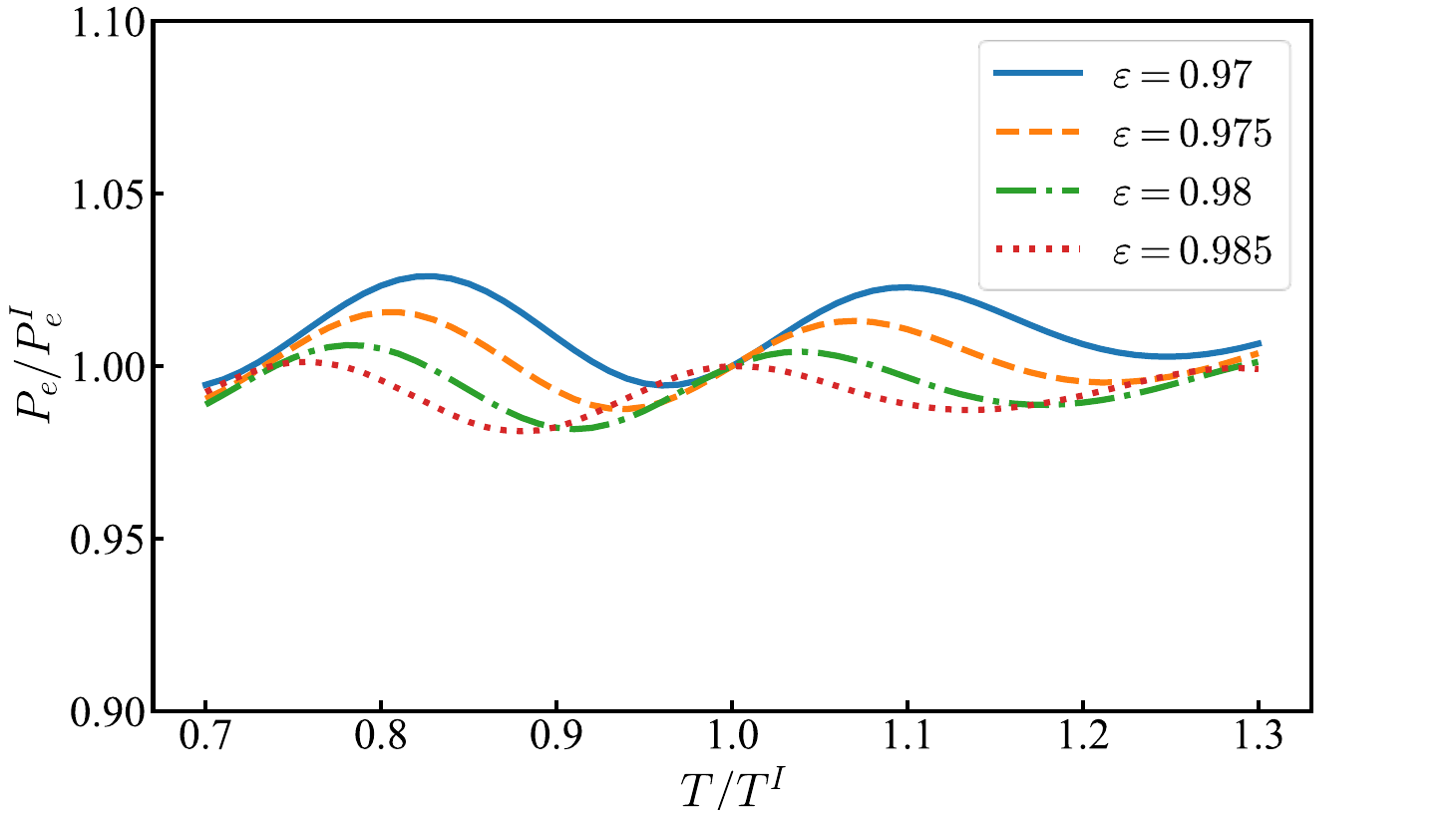}
     \caption{The qubit's $\vert e\rangle$-state population $P_e$ as a function of the ramping time $T$ for specific $\varepsilon(T)$, where $P^I_e$ and $T^I$ denote the simulated results with $k=10$ MHz. In the simulation,  $P_e$ at each point is calculated using the master equation, where the system parameters are the same as those shown  in the main text.}\label{section8}
\end{figure}

The present protocol combines the intrinsic robustness of the adiabatic
evolution with the high sensitivity near the critical point, and thus is
insensitive to the control error. In our adiabatic protocol, the control
parameter $\varepsilon $ depends on the time as
\begin{eqnarray}
\varepsilon (t)=\sqrt{1-(k^{2}t^{2}+1)^{-1}}.
\end{eqnarray}
The time needed to reach the preset value of $\varepsilon $ is determined by 
$k$, which controls the ramping velocity. As shown in Fig. 2(d) of the main
text, within the working regime the qubit's $\left\vert e\right\rangle $%
-state population ($P_{e}$) almost remains the same for a wide range of the
values of $k$. This implies that the scheme is robust against imperfect
timing. To further confirm this point, we perform a simulation of $P_{e}$
for different values of the ramping time $T$. Figure \ref{section8} displays $P_{e}$ versus $T$ for
different values of $\varepsilon (T)$. $P_{e}$ at each point is calculated using the master
equation, where the system parameters are the same as those in the text. The
result unambiguously demonstrates that $P_{e}$ is indeed insensitive to the
ramping time. For example, for $\varepsilon (T)=0.985$, $%
P_{e}$ falls within the narrow regime $[ 0.98P^I_{e}, 1.0P^I_{e} ]$ even when $T$ is ranged from $0.7 T^I$
to $1.3 T^I$, where $P^I_e$ and $T^I$ denote $\vert e\rangle$-state population and total time for the simulated results with $k=10$ MHz, respectively. We note that this robustness is achieved at the price of needing a
resonator to realize the JCM, which is unnecessary in the Rabi measurement
method.


\section{Errors and scaling of the Fisher information
}
\begin{figure}
	\includegraphics[width=1\textwidth]{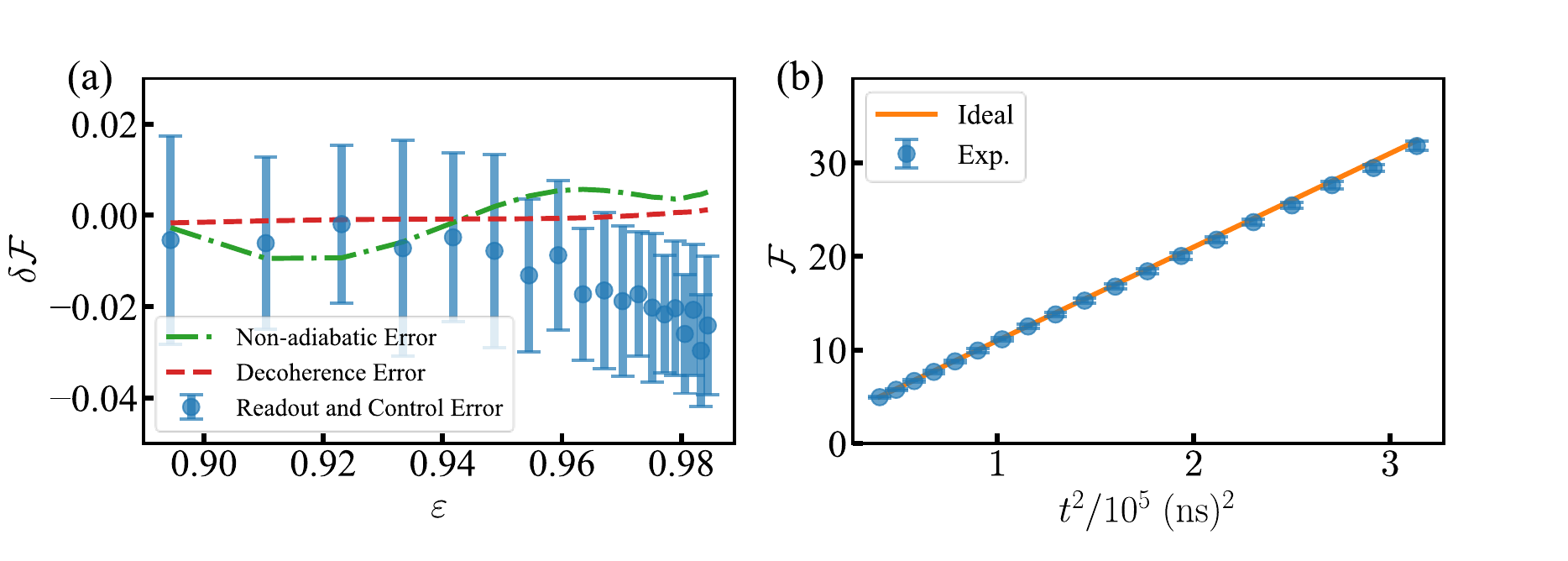}
     \caption{(a) Errors of $\mathcal{F}$ versus the control parameter $\varepsilon$. The error sources include non-adiabaticity, decoherence, and readout and calibration imperfection. Each error source causes a deviation of the measured $\mathcal{F}$ from the ideal value. The effect of each error source is quantified by the ratio between the resulting deviation and the ideal value.  (b)  The Fisher information $\mathcal{F}$ versus $t^2$.    }\label{section9}
\end{figure}
As show in Eq. \ref{Fisher}, the Fisher information depends on both the $\vert e\rangle$-state population ($P_{e}$) of the qubit and its derivative with respect to the control parameter $\varepsilon$. This derivative cannot be directly obtained from the experimental data, which are measured for discrete values of $\varepsilon$. To obtain this derivative at each point, we fit $P_{e}$ as a function of $\varepsilon$ with the measured data, and then calculate the derivative. The error sources of $\mathcal{F}$ include  non-adiabaticity, decoherence, readout infidelity, and calibration error. To quantify these errors, we first calculate $\mathcal{F}$ for different values of $\varepsilon$, based on the Hamiltonian dynamics and the master equation, respectively. The difference between the result obtained with the Hamiltonian dynamics and that for the ideal dark state is due to the non-adiabaticity. The deviation of the result based on the master equation from that with the Hamiltonian dynamics originates from the decoherence. The difference between the experimental data and the simulation with the master equation is attributed to the readout error and the fluctuation of the control parameter. We cannot individually quantify these two errors. Fig. \ref{section9}(a) displays the relative error budgets, defined as the ratios between these deviations and the ideal value of $\mathcal{F}$, versus $\varepsilon$. These errors can be mitigated by improving the coherence times of the system and readout fidelity, and by suppressing the parameter fluctuation.


It is enlightening to explore the scaling of the Fisher information with the
time which can be considered as the sensing resource \cite{1}. 
 When time dependence of the control parameter $\varepsilon $ is
given by Eq. (5), the Fisher information for the ideal dark state linearly
scales with the square of the time, given by 
 \begin{eqnarray}
\mathcal{F}=k^2t^2+1.
\end{eqnarray} 
This corresponds to the optimal Heisenberg scaling precision \cite{1}. Figure \ref{section9}(b) shows the time dependence of the Fisher information obtained from the measured $P_e$, which well agrees with the ideal result (solid line).

\section{Implementation of the metrological protocol in ion traps}
The JC model can be realized in ion trap with the vibrational frequency $\nu 
$, where the photonic mode is replaced by the phononic mode \cite{PRL4714,PRL1796,RMP1103}. The
qubit is formed by two hyperfine ground states $\left\vert g\right\rangle $
and $\left\vert e\right\rangle $, separated by the energy $\omega _{0}$. The
qubit and the phononic mode are coupled through virtual excitation of an
excited state ($\left\vert f\right\rangle $) with a pair of laser beams with
the frequencies  $\omega +\omega _{0}+\Delta _{0}-\nu $ and $\omega
+\Delta _{0}$, where $\omega $ is the energy difference between $\left\vert
e\right\rangle $ and $\left\vert f\right\rangle $. When the detuning $\Delta
_{0}$ is much larger than $\nu $ and the amplitude of the laser beams, $%
E_{0} $, the two laser beams couple the two ground states $\left\vert
g\right\rangle $ and $\left\vert e\right\rangle $ in a Raman manner, with
the strength $\chi _{0}=E_{0}^{2}/\Delta _{0}$. Consider the system dynamics
in the sideband-resolved regime $\chi _{0}\ll \nu $ and in the Lamb-Dicke
limit $\eta _{0}=\delta k_{0}/\sqrt{2\nu M}\ll 1$, where $\delta k_{0}$
denotes the wavevector difference between the laser fields and $M$ is the
mass of the ion. Then the coupling between the internal and external degrees
of freedom is described by the JC Hamiltonian 
\begin{eqnarray}
H_{JC}=\Omega ( a^{\dagger }\left\vert g\right\rangle \left\langle
e\right\vert +a\left\vert e\right\rangle \left\langle g\right\vert ), 
\end{eqnarray}
where $\Omega =\eta _{0}\chi _{0}$, and $a^{\dagger }$ and $a$ denote
creation and annihilation operators for the phononic mode. The phononic mode
can be driven by two pairs of laser beams with the frequencies ($\omega
+\omega _{0}+\Delta _{1}-\nu ,\omega +\omega _{0}+\Delta _{1}$) and ($\omega
+\Delta _{1}-\nu ,\omega +\Delta _{1}$). With suitable choice of the
polarizations, the two pairs of laser beams solely drive the transitions $%
\left\vert g\right\rangle \longleftrightarrow \left\vert f\right\rangle $
and $\left\vert e\right\rangle \longleftrightarrow \left\vert f\right\rangle 
$, respectively. When the detuning $\Delta _{1}$ is much larger than $\nu $
and the amplitude of the laser beams, $E_{1}$, these two pairs of laser
beams couple $\left\vert g\right\rangle $ and $\left\vert e\right\rangle $
to the phononic mode, respectively. In the sideband-resolved regime $\chi
_{1}=E_{1}^{2}/\Delta _{1}\ll \nu $ and in the Lamb-Dicke limit $\eta
_{1}=\delta k_{1}/\sqrt{2\nu M}\ll 1$, the resulting interactions are
described by the Hamiltonian 
\begin{eqnarray}
H_{d}=\lambda (\left\vert g\right\rangle \left\langle g\right\vert
+\left\vert e\right\rangle \left\langle e\right\vert )(a^{\dagger }+a), 
\end{eqnarray}
where $\lambda =\eta _{1}\chi _{1}$. We here have assumed that the two pair
of laser beams have the same wavevector difference $\delta k_{1}$. Since the
identity operator $I=\left\vert g\right\rangle \left\langle g\right\vert
+\left\vert e\right\rangle \left\langle e\right\vert $ can be discarded, the
combination of $H_{JC}$ and $H_{d}$ is equivalent to the driven JC
Hamiltonian.

Under the condition $\delta k_{0}\gg \delta k_{1}$, we have $\eta _{0}\gg
\eta _{1}$ and $d\lambda /d\chi _{1}\ll d\Omega /d\chi _{0}$. This implies
that $\lambda $ is much more insensitive to the fluctuation of $\chi _{1}$
than $\Omega $ to the fluctuation of $\chi _{0}$. Therefore, we can set $%
\lambda $ to be a known fixed value and use this model to estimate $\chi _{0}
$. Below the critical point, the system has a unique dark state, where $P_{e}
$ is directly related to $\chi _{0}$ by
\begin{eqnarray}
P_{e}=\frac{1}{2}\left[ 1-\sqrt{1-\left( 2\lambda /\eta _{0}\chi _{0}\right)
	^{2}}\right] .
\end{eqnarray}
As the dynamics of a trapped ion can be manipulated in a well controlled
manner, it is expected that the driven JCM and the associated critical
metrological protocol can be readily realized in ion traps. Compared to the
circuit QED architecture, the ion trap has the following advantages.
Firstly, the qubit can be encoded in two hyperfine ground states of a
trapped ion so that it has a long lifetime. Secondly, the electronic state
of a trapped ion can be well restricted within a two-dimensional Hilbert
space, so that the error due to leakage out of the qubit space is highly
suppressed. Thirdly, the detection efficiency of the state of an ionic qubit
can nearly reach 100\% \cite{RMP1103}, which is much higher than that achievable for
superconducting qubits.

\begin{figure}
	\includegraphics[width=0.6\textwidth]{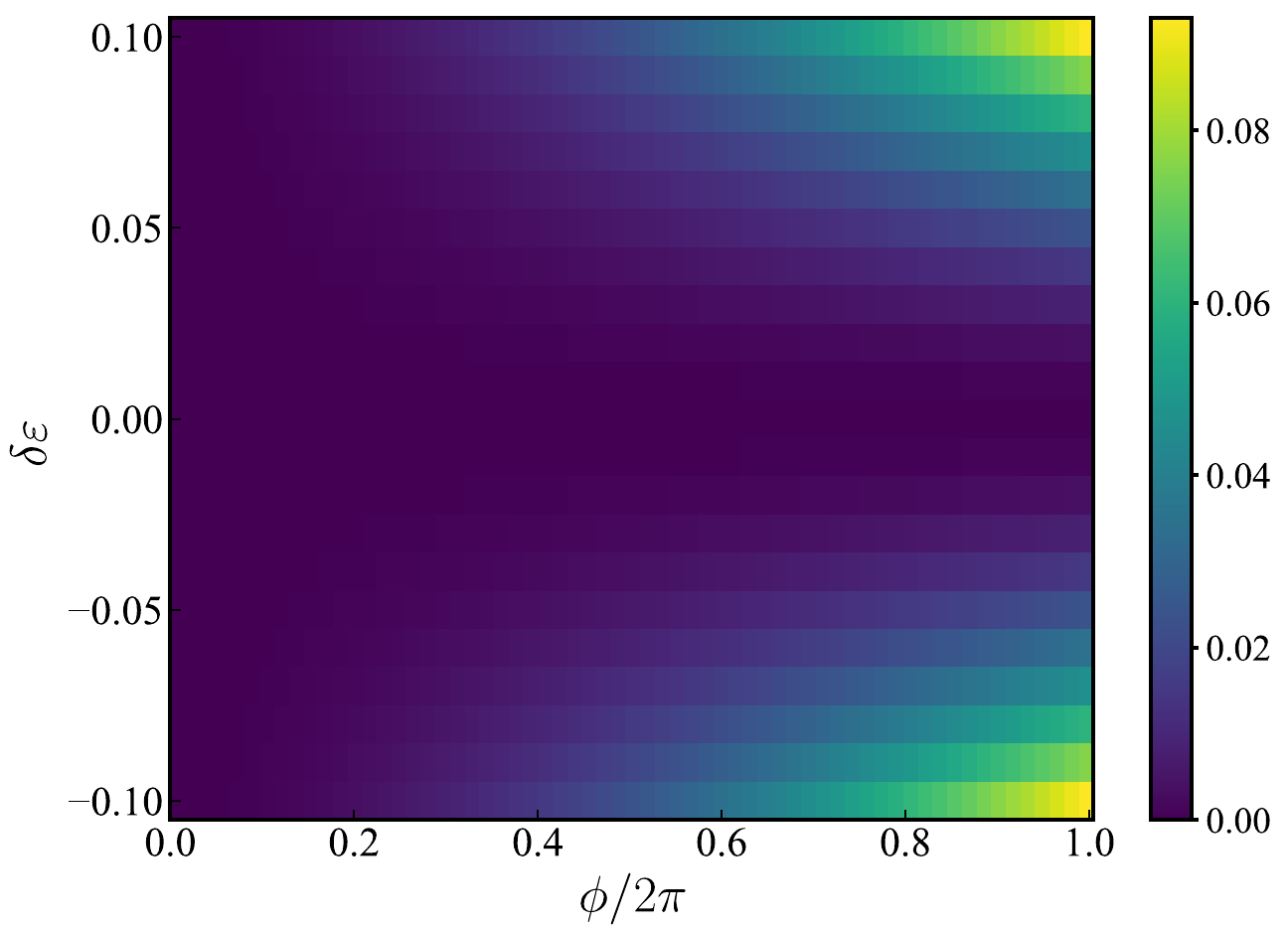}
     \caption{Single-qubit gate infidelities  versus the rotation angle $\phi$ and the relative deviation of the amplitude $\delta\varepsilon$ of the driving field.   }\label{section11}
\end{figure}

\section{Single-qubit gate infidelities due to inaccurate calibration of driving strengths}

A single-qubit rotation around any axis on the equatorial plane of
the Bloch sphere can be realized by resonantly driving the qubit with an
external field. The qubit dynamics is given by the Hamiltonian 
\begin{eqnarray}
H_{dr}=\frac{\epsilon }{2}(e^{i\theta }\left\vert g\right\rangle
\left\langle e\right\vert +e^{-i\theta }\left\vert e\right\rangle
\left\langle g\right\vert ),
\end{eqnarray}
where $\epsilon $ is the amplitude of the driving field and $\theta $ is the
phase. After a time t, this Hamiltonian produces a rotation around the axis (%
${\bf n}$) that has an angle $\theta $ to the x axis on the equatorial plane
of the Bloch sphere, given by
\begin{eqnarray}
R_{{\bf n}}(\phi )=\exp (-i\phi \sigma _{{\bf n}}/2),
\end{eqnarray}
where $\phi =\epsilon t$ and $\sigma _{{\bf n}}=\cos \theta \sigma _{x}+\sin
\theta \sigma _{y}$, with $\sigma _{x}=\left\vert g\right\rangle
\left\langle e\right\vert +\left\vert e\right\rangle \left\langle
g\right\vert $ and $\sigma _{y}=i(\left\vert g\right\rangle \left\langle
e\right\vert -\left\vert e\right\rangle \left\langle g\right\vert )$. A
deviation of the driving amplitude from the desired value $\epsilon $, given
by $\epsilon \delta \varepsilon $ would result in an error of the rotation
angle, $\delta \phi =\phi \delta \varepsilon $. The resulting gate error can
be quantified by $1-{\cal F}_{p}$, where ${\cal F}_{p}$ denotes the process fidelity, defined by ${\rm tr}(\chi_{I}\chi_{\delta})$, with $\chi$ being the quantum process tomography matrix of single-qubit gate \cite{PRA032322}. Fig. \ref{section11} shows $1-{\cal F}_{p}$ versus $\phi $ and $\delta \varepsilon $. The result implies that implementation of a
high-fidelity single-qubit gate requires high-precision calibration of the
amplitude of the driving field. Such gates are crucial for realization of
quantum algorithms \cite{book}, as well as for
multi-qubit entangled state tomography \cite{PRL180511}. Our protocol may
be used to improve the calibration precision of the amplitudes of the
driving fields used for implementation of these gates on different quantum
computing platforms, including superconducting circuits and ion traps.

\bibliography{siref}

